\documentclass{amsart}

\usepackage[margin = 1.3in]{geometry}

\usepackage[norelsize,ruled]{algorithm2e}
\usepackage{amssymb}
\usepackage{epsfig}  		
\usepackage{epic,eepic}       
\usepackage{amsmath}
\usepackage{amsmath,amsfonts,amssymb}
\usepackage{graphicx}
\usepackage{amscd}
\usepackage{pb-diagram}
\usepackage{url}
\usepackage{caption}
\usepackage{lipsum}

\usepackage[usenames,dvipsnames]{xcolor}
\usepackage{mathtools}
\usepackage{hyperref}

\newtheorem{remark}{Remark}

\newcommand{\norm}[1]{\left\lVert#1\right\rVert}
\newcommand{\abs}[1]{\left\lvert#1\right\rvert}

\newcommand{\R}{\mathbb{R}}

\newcommand{\bbL}{\mathbb{L}}
\newcommand{\bbX}{\mathbb{X}}
\newcommand{\bbY}{\mathbb{Y}}

\DeclareMathOperator{\last}{last}

\usepackage{chngcntr}
\counterwithout{counter}{figure}

\usepackage{enumitem}
\newlist{examples}{enumerate}{1}
\setlist[examples]{label=(\thetable.\arabic*)}

\begin{document}


\title{A persistence landscapes toolbox for topological statistics}

\thanks{P.B is supported by AFOSR grant FA9550-13-1-0115. P.D is supported by the Advanced Grant of the European Research Council GUDHI (Geometric Understanding in Higher Dimensions), DARPA grant FA9550-12-1-0416 and AFOSR grant FA9550-14-1-0012.}

\author{Peter Bubenik}
\address{ Department of Mathematics, University of Florida, Gainesville, FL, USA.}
\email{peter.bubenik@gmail.com }

\author{Pawe\l\ D\l otko}
\address{Geometrica, INRIA Saclay, Ile-de-France, France and Institute of Computer Science, Jagiellonian University, Krakow, Poland.}
\email{pawel.dlotko@inria.fr}

\begin{abstract}

Topological data analysis provides a multiscale description of the geometry and topology of quantitative data. 
The persistence landscape is a topological summary that can be easily combined with tools from statistics and machine learning. 
We give efficient algorithms for calculating persistence landscapes, their averages, and distances between such averages.
We discuss an implementation of these algorithms and some related procedures.
These are intended to facilitate the combination of statistics and machine learning with topological data analysis.
We present an experiment showing that the low-dimensional persistence landscapes of points sampled from spheres (and boxes) of varying dimensions differ.
\end{abstract}

\keywords{%
topological data analysis, 
persistent homology, 
statistical topology, 
topological machine learning, intrinsic dimension.
}


\maketitle

\section{Introduction} \label{sec:intro}

We provide some algorithms and computational tools for statistical topological data analysis.
In particular, we give algorithms for calculating the persistence landscape, a functional summary of persistence modules. 
We also give algorithms for calculating the averages of such summaries, and for calculating distances between such averages.
These tools also provide an alternative computational approach for calculating distances between topological summaries that may be useful when other methods are computational prohibitive.
In addition, we specify an implementation of these algorithms and some related tools that we have made publicly available.
 
We are motivated by \emph{topological data analysis}~\cite{ghrist:survey,carlsson:topologyAndData}.
Its main tool, \emph{persistent homology} provides a multiscale description of the topology of the data of interest, called either a \emph{barcode} or a \emph{persistence diagram}.
Unfortunately this summary is difficult to work with from the point of view of statistics and machine learning. 
For example, it is not feasible to calculate averages.
For these purposes, it is convenient to replace these summaries with a linear summary, that is, a finite- or infinite-dimensional vector. In a linear space it is easy to calculate averages.
 One such vector which does not lose any information is the functional summary called the \emph{persistence landscape}~\cite{peter}.
Since this summary may be thought of as lying in a Hilbert space, in the language of machine learning, it is a \emph{feature map}.
There is an associated \emph{kernel}~\cite{reininghaus:2014} to which standard machine learning tools may be applied.

\subsection{Background}

In the simplest computational setting for topological data analysis, the data of interest is encoded in a 
finite filtered complex,
\begin{equation} \label{eq:sc}
   \mathcal{K}_0 \subset \mathcal{K}_1 \subset \ldots \subset \mathcal{K}_n.
\end{equation}
This is a filtration of the complex $K=K_n$ and it is sometime convenient to add $K_{-1}=\emptyset$.
\emph{Persistent homology}~\cite{elz:tPaS,zomorodianCarlsson:computingPH} gives a multiscale representation of the topology of this complex.
To be precise, one applies homology in some degree with coefficients in some field to \eqref{eq:sc} to obtain a sequence of finite-dimensional vector spaces and linear maps,
\begin{equation} \label{eq:ph}
H(\mathcal{K}_0) \rightarrow
H(\mathcal{K}_1) \rightarrow \ldots \rightarrow
H(\mathcal{K}_n),
\end{equation}
called a \emph{persistence module}.
It turns out that the persistence module can be completely described by a finite sequence of pairs $\{(b_i,d_i)\}$, with $b_i < d_i$. For each such pair $(b_i,d_i)$ there is a choice of a nonzero homology class $\alpha_i \in H(\mathcal{K}_{b_i})$ that is not in the image of $H(\mathcal{K}_{b_i-1})$ and whose image is nonzero in $H(\mathcal{K}_{d_i-1})$ but is zero in $H(\mathcal{K}_{d_i})$. One sometimes says that $\alpha_i$ is born at $b_i$ and dies at $d_i$. Furthermore, the homology classes $\{\alpha_i\}$ and their nonzero images under the maps in \eqref{eq:ph} give a basis for the vector spaces in \eqref{eq:ph}.
Considering these pairs as points in the plane, one obtains the \emph{persistence diagram}. 
Considering them as intervals $[b_i,d_i)$ on obtains the \emph{barcode}.
We will often refer to them as \emph{birth-death pairs}.
In the simple setting of \eqref{eq:sc}, we have $b_i,d_i \in \{0,1,\ldots,n\}$. However we can generalize to $b_i,d_i \in \mathbb{R}$ by associated a corresponding increasing sequence of real numbers with \eqref{eq:sc}.
This summary is stable~\cite{stability,csehm:lipschitz,csgo:persistenceModules} in that small perturbations of the data will lead to small perturbations of these pairs, under suitable choices of distance.
Successful applications of topological data analysis include breast cancer data~\cite{nlc:topologyBreastCancer}, sensor networks~\cite{deSilvaGhrist:coverageInSNvPH}, orthodontic data~\cite{gambleHeo}, signal analysis~\cite{pereaHarer:swipers}, target tracking~\cite{bendich:tracking}, and brain artery data~\cite{bendich:brainArtery}.

Now let us define the persistence landscape~\cite{peter}. 
First, for a birth-death pair $(b,d)$, let us define the piecewise linear function $f_{(b,d)} : \mathbb{R} \rightarrow
[0,\infty]$.
\begin{equation}
 f_{(b,d)} =
  \begin{cases}
   0 & \text{if } x \not \in (b,d) \\
   x - b & \text{if } x \in (b , \frac{b+d}{2}] \\
   -x + d       & \text{if } x \in (\frac{b+d}{2} , d) 
  \end{cases}
  \label{eq:basicLand}
\end{equation}
The \emph{persistence landscape} of the birth-death pairs
$\{ (b_i , d_i) \}_{i=1}^n$ is the sequence of functions $\lambda_k :
\mathbb{R} \rightarrow [0,\infty]$, $k=1,2,3,\ldots$ where $\lambda_k(x)$ is the $k$-th largest value of $\{ f_{(b_i,d_i)}(x) \}_{i=1}^n$. We
set $\lambda_k(x) = 0$ if the $k$-th largest value does not
exist; so $\lambda_k = 0$ for $k>n$.  
Equivalently, the persistence landscape is a function $\lambda: \mathbb{N} \times \mathbb{R} \to [0,\infty]$, where $\lambda(k,t) = \lambda_k(t)$.
In this definition we have assumed that $b$ and $d$ are finite.
In the appendix we show that this definition extends to
the cases where $b$ and/or $d$ are infinite.

Given a set of persistence landscapes, $\lambda^{(1)},\ldots,\lambda^{(N)}$, their average, $\bar{\lambda}$, is defined pointwise, $\bar{\lambda}_k(t) = \frac{1}{N} \sum_{i=1}^N \lambda^{(i)}_k(t)$.
Distances between persistence landscapes and between average persistence landscapes can be given using the $L^{\infty}$ norm,
\begin{equation*}
  \norm{\lambda-\lambda'}_{\infty} = \sup_{k,t} \abs{\lambda_k(t)-\lambda'_k(t)},
\end{equation*}
or the $L^p$ norm, for $1 \leq p < \infty$,
\begin{equation*}
  \norm{\lambda-\lambda'}_p = \left[ \sum_{k=1}^{\infty} \int \abs{\lambda_k(t)-\lambda'_k(t)}^p dt \right]^{\frac{1}{p}}.
\end{equation*}
In~\cite{peter} it is shown that the persistence landscape is stable
 with respect to the $L^p$ distance for $1 \leq p \leq \infty$.
That is, under suitable hypotheses, sufficiently small perturbations of a function under the supremum norm lead to small changes of the persistence landscape of the persistent homology of the sublevel sets of that function under the $L^p$ norm.

 In addition to the persistence landscape, other functional summaries of the persistence module can also be easily averaged and used for statistics and machine learning. 
The simplest of these is the ranks of the maps in the persistence module (called the persistent Betti number function or the rank function).
To help keep distance finite, one can 
smooth the persistence diagram~\cite{donatini:1998,ferri:1998,chepushtanova:2015}
or 
integrate with respect to a weight function~\cite{robins:2015}.
The most sophisticated such smoothing is given in \cite{reininghaus:2014}. 
Variants of the persistence landscape such as silhouettes~\cite{Chazal:2014a} also work.
The representation of persistence diagrams as complex polynomials~\cite{diFabioFerri} has been used for shape classification. In~\cite{carriereOudotOvsjanikov} the authors construct stable (with respect to the Bottleneck distance) feature vectors based on the persistence diagram. Those vectors are used later to compare points in $3$d shapes.
 In addition, one can use the lengths of the $N$ longest bars~\cite{bendich:brainArtery} or algebraic functions on barcodes~\cite{adcock:2013}.

\subsection{Relation to other software}
\label{sec:other-software}

The main aim of most persistent homology software is to start with a finite filtered complex (or perhaps something used to generate such as a complex, such as a finite set of points in Euclidean space) and to produce a set of birth-death pairs. 
Examples include JavaPlex~\cite{plex}, Dionysus~\cite{dionysous}, Perseus~\cite{perseus1}, PHAT~\cite{phat}, and GUDHI~\cite{gudhi}.
These birth-death pairs are the input for our software.

Other recent software developed concurrently to ours is the R package TDA~\cite{fasy:tda}. This package provides R users with tools for topological data analysis. For example, it provides an interface to GUDHI, Dionysus and PHAT. It calculates persistence landscapes using grids and also calculates confidence intervals.  Future versions of TDA may include an interface to our code.

The software described in this paper 
performs exact computations of persistence landscapes. It also implements grid-based computations, such as the ones provided in the TDA~\cite{fasy:tda} package. Since the grid-based computations are straightforward, they are not discussed in this paper. One may switch from the (default) exact computations to the grid-based ones, by making a small change in the self-explaining \emph{config} file in the library's main folder. Pros and cons of grid and exact computations of landscapes along with some description of the limitations of both approaches are provided in Section~\ref{sec:rigorousVsGrid}.

\subsection{Prior work and related work}

It has been shown~\cite{csehm:lipschitz} that the set of persistence diagrams with the Wasserstein distance is a complete and separable metric space, and thus provides a suitable setting for probability and statistics. 
Unfortunately the Fr\'echet mean is not necessarily unique.
For a slightly adjusted metric, there is an algorithm~\cite{tmmh:frechet-means} that converges to an element of the Fr\'echet mean set, though it does not have good computational properties. 
The discontinuity of this procedure can be remedied by using a probabilistic approach~\cite{munch:probabilistic-f}.

Persistence diagrams can be used for statistical inference.
Hypothesis testing using persistence diagrams has been considered for brain MRI data in~\cite{cbk:ipmi2009} and more abstractly in~\cite{robinsonTurner:2013}.
Furthermore, confidence sets for persistence diagrams have been obtained in~\cite{Fasy:2014}.

The persistence landscape allows the use of more statistical machinery.
The bootstrap has been applied to obtain confidence bands for the persistence landscape~\cite{Chazal:2014a} and the average persistence landscape of subsamples has been studied~\cite{Chazal:2014c}.
The persistence landscape has been used to study protein binding~\cite{giseon:maltose} and as a kernel for topological machine learning and compared to the recent multi-scale kernel for persistence diagrams~\cite{reininghaus:2014}.

\subsection{Our work}

We present an algorithm for calculating the persistence landscape corresponding to
$n$ birth-death pairs in time $O(n^2)$ and show that this time
complexity is optimal.
Given $N$ persistence landscapes, each obtained from $n$ birth-death pairs, we 
calculate their average persistence landscape in time $O(n^2N\log N)$.
We show how to calculate the $L^p$ distances for $1 \leq p \leq \infty$ between two such averages in time 
$O(n^2N\log N)$.

If we are willing to slightly perturb our birth and death times, then we give algorithms that are much faster for large $n$ or large $N$. 
To be precise, instead of associating an arbitrary increasing sequence of real numbers with each of our filtered simplicial complex, we round these numbers so that they lie on an equally-spaced grid of size $m$. By the stability theorem of~\cite{stability}, this will perturb the resulting persistence diagram by at most $\frac{\delta}{2}$ in the bottleneck distance, where $\delta$ is the spacing of the grid.
Under this assumption, we give an algorithm for calculating the persistence landscape in time $O(mn\log n)$ and calculating the average landscape in time $O(mnN)$. We can also calculate the distance between two such averages in time $O(mnN)$.

We describe an implementation of these algorithms that we have made publicly available. 
In addition to the above algorithms, we have also implemented a number of procedures that we hope will be helpful to practitioners interested in using these methods for topological data analysis.
For example, we have procedures for plotting persistence landscapes and their averages. Given a number of classes of birth-death pairs, we allow the user to calculate the distance matrix of the corresponding average persistence landscapes, and also to perform pairwise permutation tests for these classes, using the $L^p$ distance, with $1 \leq p \leq \infty$, between their respective average persistence landscapes.
In addition, we provide two nearest-neighbor classifiers, one using the persistence landscape in a single degree, and the other using multiple degrees.
Also, one can compute the inner product of landscapes with the presented software, thus allowing them to be used with kernel methods.

In Section~\ref{sec:data-structures} we describe the input and output for our algorithms.
In Section~\ref{sec:algorithms} we give our main algorithms and calculate their time complexities. We also show why calculating the persistence landscape is not the same as the n-th envelope problem.
In Section~\ref{sec:experiments} we give a few simple numerical experiments demonstrating our implementation of our algorithms.
In Section~\ref{sec:specification} we describe our implementation of our algorithms and some related procedures.

\section{Data structures}
\label{sec:data-structures}

In this section, we describe the inputs and outputs of our main algorithms.

\subsection{Input}
\label{sec:input}

The initial input to our algorithms consists of a list of $n$ pairs of 
numbers $(b,d)$ with $b < d$.  Each of these pairs represents the
birth and death times of a persistent homology class.  Thinking of
these pairs as points we obtain a persistence diagram~\cite{herbert},
and considering them to be intervals we obtain a barcode~\cite{ghrist:survey}.
We will give two algorithms for calculating persistence landscapes from these birth-death pairs.

In Algorithm~\ref{alg:landscapePoints}, for simplicity, we assume that $b$ and $d$ are
finite.  In the appendix, we extend this algorithm to
Algorithm~\ref{alg:landscapePointsInfinite} which also accommodates
infinite intervals.
Reduction to the finite case can be achieved by
removing or truncating infinite intervals or by using extended
persistence~\cite{extendedPersistence},
where extended persistence is used to obtain a persistence module which is eventually zero~\cite{bubenikScott:1}. 

In our implementation, the user is asked to define a
 number $i$ (which can be set to the maximal representable double number) at which infinite intervals will be truncated. 
This choice will depend on the application and/or the persistent homology calculation. 
For example, the persistent homology calculation of a filtered Vietoris-Rips complex is often truncated at some filtration value, because of the exponential growth in size of the complex. It is then sensible to truncate infinite intervals at this maximum filtration value.

In Algorithm~\ref{alg:landscapePointsGrid},
we assume that each $b$ and $d$
is an element of a finite, evenly-spaced grid, 
$a,a+d,a+2d,\ldots,a+md$. 
Such birth-death pairs are often the output of
Perseus~\cite{perseus,perseus1} or Plex~\cite{plex}.
In fact, by rescaling, we assume without loss of generality that values of $b$ and $d$ are elements of $\{0,2,4,\ldots,2m\}$.

\subsection{Output}
\label{sec:output}

In this section, we describe our encoding of persistence landscapes and linear combinations of persistence landscapes.

As defined in Section~\ref{sec:intro}, a \emph{persistence landscape} is a function $\lambda :
\mathbb{N} \times \mathbb{R} \rightarrow [0,\infty]$, or equivalently,
a sequence of functions $\lambda_k:\R \to [0,\infty]$ where $k\geq 1$. For
every fixed $k$, $\lambda_k$ is a piecewise-linear function. 

Since the input consists of $n$ birth-death pairs, $\lambda_k=0$ for $k$ greater than some
fixed $K \leq n$.
We will represent $\lambda_k$ by a vector $\mathbb{L}_k$ of the points
$(x,\lambda_k(x))$ such that $\lambda_k$ is not differentiable in
$x$, which is sorted to have increasing values of $x$. For clarity, we include in $\mathbb{L}_k$ the points $(-\infty,0)$ and $(\infty,0)$.
The projection of $\bbL_k$ onto its first coordinate is the vector of
\emph{critical numbers} and the projection on the second coordinate is
the vector of \emph{critical values}.
We refer to the elements of $\bbL_k$
as \emph{critical points}. Clearly, $\lambda_k$ can be recovered from $\mathbb{L}_k$ by linearly interpolating
consecutive points. This is therefore an exact representation of a persistence landscape. Therefore from now on these two objects will be used interchangeably. Note that when a landscape is represented on a discrete grid, this property do not hold in general. 

Let $P$ denote the total number of critical points in the $\bbL_k$, not including the points $(\pm\infty,0)$, which are not strictly necessary.
In Section~\ref{sec:landscapesAndEnvelopes} we show that $P=O(n^2)$.

In Algorithm~\ref{alg:landscapePointsGrid}, the input numbers are in the set
$\{0,2,4,\ldots,2m\}$.
It follows that the critical numbers and critical values are elements of $\{0,1,2,\ldots,2m\}$.
We represent the persistent landscape by a two-dimensional array $V$ satisfying $V[k][i] = \lambda_k(i)$, with $i \in \{0,1,2,\ldots,2m\}$ and $k \in \{1,\ldots,K\}$, where $K$ is the largest $k$ such that $\lambda_k$ is not identically equal to $0$.
Note that the persistence landscape can be obtained from $V$ by linear interpolation.
It is also fruitful to consider $V$ as a vector of size $K(2m+1)$.

Notice that a linear combination of persistence landscapes is also a sequence of piecewise linear functions. So we encode it in the same way as we do a persistence landscape.

\begin{remark}
  We remark that $K$ is the largest rank of the linear maps in \eqref{eq:ph}.
\end{remark}

\section{Algorithms}
\label{sec:algorithms}

In this section, we describe our main algorithms, which compute persistence landscapes, linear combinations of such persistence landscapes, and distances between such linear combinations.

\subsection{Persistence landscape}
\label{sec:landscapes}

In this section, we present two algorithms to construct the persistence
landscape from a list of birth-death pairs.  For simplicity, Algorithm~\ref{alg:landscapePoints} assumes the input consists of finite numbers. For a variation without this assumption and with the same complexity, see Algorithm~\ref{alg:landscapePointsInfinite} in the appendix.  The computational complexity is $O(n\log(n)+nK)$, where
$n$ denotes the number of input pairs and $K$ the number of nonzero
landscapes. Since $K \leq n$, this algorithm is $O(n^2)$. If we do not need all of the persistence landscape, then there are faster variations, described below.
Algorithm~\ref{alg:landscapePointsGrid} assumes that the input
coordinates 
are elements of a finite, evenly-spaced grid of size $m+1$. 
Its computational complexity is $O(mn + mK\log K)$ which is $O(m n \log n)$.

\medskip

\begin{figure}[h!tb]
\centering
\includegraphics[scale=0.32]{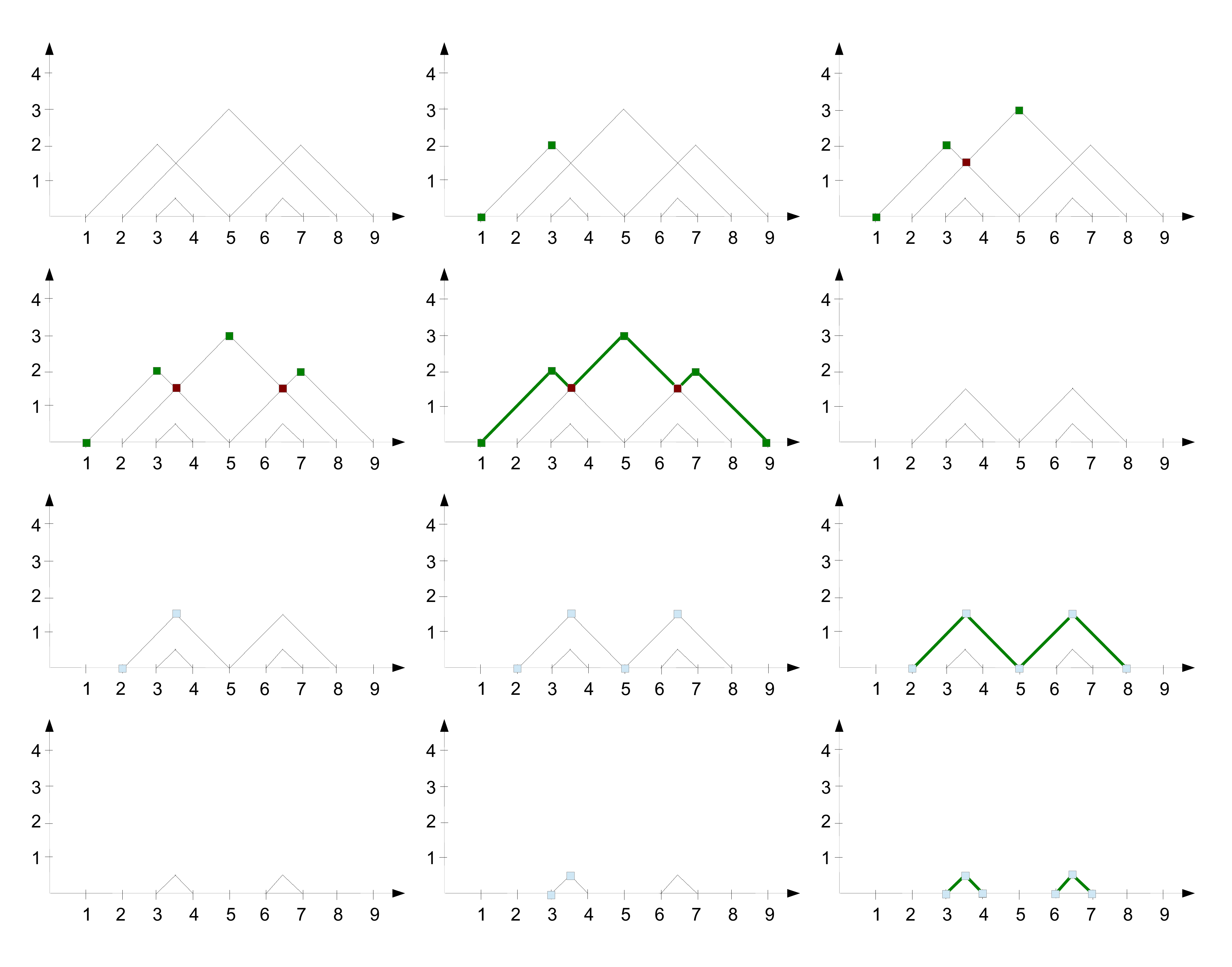}
\caption{Algorithm~\ref{alg:landscapePoints} is used to
  construct the persistence landscape corresponding to the birth-death pairs $\{ (1,5), (2,8), (3,4), (5,9), (6,7) \}$. 
  (a) The functions~\eqref{eq:basicLand} corresponding to the
  birth-death pairs and their corresponding critical points. 
  (b-d) Steps through the \textbf{while} loop to construct $\bbL_1$.
  (e) The graph of $\lambda_1$.
  (f) The graph of the functions corresponding to the remaining pairs
  in the list $A$.
  (g-i) The second iteration of the \textbf{while} loop constructs
  $\bbL_2$.
  (j) The graph of $\lambda_2$.
  (k) The graphs of the functions corresponding to the remaining pairs
  on the list $A$.
  (l) The graph of $\lambda_3$.
}
\label{fig:exLanscapePints}
\end{figure}

An example of the steps in Algorithm~\ref{alg:landscapePoints} is given in Figure~\ref{fig:exLanscapePints}.

\begin{algorithm}[t]
\SetAlgoNoLine
\small
\KwIn{$A = \{(b_i,d_i)\}_{i=1}^n$ -- a list of birth-death pairs, $-\infty < b_i < d_i < \infty$.}
\KwOut{$\{\mathbb{L}_k\}$ -- the persistence landscape, a list of lists of critical points $(x,y)$.}
Sort $A$ first according to increasing $b$ and second according to decreasing $d$\;
$k \leftarrow 1$\;
\While {$A \neq \emptyset$}
{
  Initialize $\bbL_k$\;
  Pop the first term $(b,d)$ from $A$; Let $p$ point to the next term\;
  Add $(-\infty,0),(b,0),(\frac{b+d}{2},\frac{d-b}{2})$ to $\bbL_k$\;
  \While {$\bbL_k.\last \neq (0,\infty)$} 
  {
    \eIf { $d$ maximal among remaining terms in $A$ starting at $p$}
    {
	Add $(d,0),(\infty,0)$ to $\bbL_k$\;
    }
    {
        Let $(b',d')$ be the first of the terms starting at $p$ with $d'>d$\;
	Pop $(b',d')$ from $A$; Let $p$ point to the next term\;
	\If { $b'>d$ }
	{
          Add $(d,0)$ to $\bbL_k$\;
	}
	\eIf { $b'\geq d$ }
	{
          Add $(b',0)$ to $\bbL_k$\;
	}
	{
          Add $(\frac{b'+d}{2},\frac{d-b'}{2})$ to $\bbL_k$\;
          Push $(b',d)$ into $A$ in order, starting at $p$; Let $p$ point to the next term\;
        }
        Add $(\frac{b'+d'}{2},\frac{d'-b'}{2})$ to $\bbL_k$\;
	$(b,d) \leftarrow (b',d')$\;
    }
  }
  ++k\;
}
Return $\{\bbL_k\}$\;
\caption{Compute the persistence landscape.}
\label{alg:landscapePoints}
\end{algorithm}

In Algorithm~\ref{alg:landscapePoints}, we first sort the input list, $A$, in a way that we will be able to pass through $A$ exactly once in order to construct each list $\bbL_k$. This takes time $O(n\log n)$. 
We denote the last pair in $\bbL_k$ by $\bbL_k.\last$.
Each iteration of the outer \textbf{while} loop constructs one of the lists $\bbL_k$. Denote the number of these by $K$. 
The length of $A$ is initially $n$. 
At the start of the outer \textbf{while} loop, the length of $A$ is decreased by one. 
The inner \textbf{while} loop does not increase the length of $A$.
So $K \leq n$.
In each non-terminal iteration of the inner \textbf{while} loop, the position of $p$ in the list $A$ advances, so it repeats at most $n$ times.
Thus the algorithm terminates.
Furthermore, in each iteration of the outer \textbf{while} loop, each pair in $A$ is only considered once, so it takes time $O(n)$. 
Therefore, Algorithm~\ref{alg:landscapePoints} takes time $O(n\log n + Kn)$.

The following faster variations of Algorithm~\ref{alg:landscapePoints} might be of interest for some applications.
If we only want $\bbL_1,\ldots,\bbL_{\lceil \log n \rceil}$, then the algorithm takes time $O(n\log n)$. 
If we only want $\bbL_1,\ldots,\bbL_{\kappa}$, for some fixed $\kappa$, then the algorithm also takes $O(n\log n)$.
If in addition $A$ is already sorted, then the algorithm takes $O(n)$.

\medskip

In Algorithm~\ref{alg:landscapePointsGrid}, we add each birth-death pair's contribution to the persistence landscape to an array of lists $W$. This takes time $O(mn)$. Next we sort each of the lists, which takes time $O(mK\log K)$. Finally we  copy this data to $V$, which takes time $O(mK)$. 
So the algorithm has time complexity $O(mn + mK\log K)$.
Since $K \leq n$, this is $O(mn\log n)$.

\begin{algorithm}[t]
  \SetAlgoNoLine
  \small
  \KwIn{$\{(b_i,d_i)\}_{i=1}^n$ -- a list of pairs $b_i < d_i$ which are elements of $0,2,4,\ldots,2m$\;}
  \KwOut{$V$ -- the persistence landscape, a two-dimensional array of size $K\times (2m+1)$\;}
  Initialize $W$, an array of size $2m+1$ of empty lists of integers\;
  \For { $i = 1$ to $n$ }
  {
		\For { $j = 1$ to $\frac{d_i-b_i}{2}$ }
		{
			Append $j$ to $W[b_i+j]$\;
		}
		\For { $j = 1$ to $\frac{d_i-b_i}{2}-1$ }
		{
			Append $j$ to $W[d_i-j]$\;
		}
  }
  \For { $i = 0$ to $2m$ }
  {
		Sort $W[i]$ in decreasing order\;
  }
  $K \leftarrow \max_{i\in\{0,\ldots,2m\}} \text{length of } W[i]$\;
  Initialize $V$ as a $K\times (2m+1)$ zero matrix\;
  \For { $i=0$ to $2m$ }
  {
	    \For { $k = 0$ to length of $W[i]$ }	
	    {
                  V[k][i] = W[k][i]\;
            }
  }
  \caption{Compute the persistence landscape using a grid.}
  \label{alg:landscapePointsGrid}
\end{algorithm}

\subsection{Persistence landscapes and the $n$-th envelope problem}
\label{sec:landscapesAndEnvelopes}

The problem discussed in this paper, may at first glance may look like the problem of finding envelopes of a set of line segments, which is well known in computational geometry. In this section, we  discuss the finding-envelopes problem from computational geometry and show why the problem of computing the persistence landscape is different. Also we will give a worst case estimate of the spatial complexity of the persistence landscape which will later guarantee the optimality of the algorithms used in this paper.

The upper envelope of a set of line segments $\{L_i\}_{i=1}^k$ is defined as those parts of the line segments which are visible from the point $(0,+\infty)$. Equivalently if we consider $L_i$ to be a piecewise-linear functions set to $-\infty$ outside the line segments, the upper envelope in the point $x \in \mathbb{R}$ is defined as $max_{i\in \{1,\ldots,k\}}L_{i}(x)$. 
Due to the practical importance of this problem, there are many efficient algorithms available to compute upper envelopes~\cite{Hershberger2} practically in linear time. 
Let us have a family of persistence intervals $\{(a_i,b_i)\}_{i=1}^n$. For every interval let us define the line segment $L_{2i-1}$ starting at $(a_i,0)$ and ending at $(\frac{a_i+b_i}{2} , \frac{b_i-a_i}{2})$ and the line segment $L_{2i}$ starting at $(\frac{a_i+b_i}{2} , \frac{b_i-a_i}{2})$ and ending at $(b_i,0)$.
The first landscape $\lambda_1$ is the upper envelope of the family $\{L_i\}_{i=1}^{2n}$. However, from the classical upper envelope computations we are
 not able to get the $\lambda_k$'s for $k > 1$. There are algorithms available in which successive upper envelopes are computed, as presented in~\cite{Hershberger}. However in this case,
 once part of a line segment $L_i$ belongs to the $k$-th envelope, then the whole of $L_i$ is removed from the search of $(k+1)$-st envelope, which is not what is required in the calculation of the
 persistence landscape. Alternatively, one can find all the intersections between the mentioned lines. However, in this case, we will get $O(n^2)$ lines segment in the worst case.


Even though standard results on the n-th envelope problem do not apply
directly,
Michael Kerber has recently pointed out to us
that an adaptation of a line sweep algorithm can indeed be
applied to calculate persistence landscapes. 
The time complexity of this adaptation may be a bit worse than the one
presented in this paper, but there are examples for which it is faster. 
This approach will be explored and compared to one presented here in a
subsequent paper.

\begin{figure}[h!tb]
\centering
\includegraphics[scale=0.35]{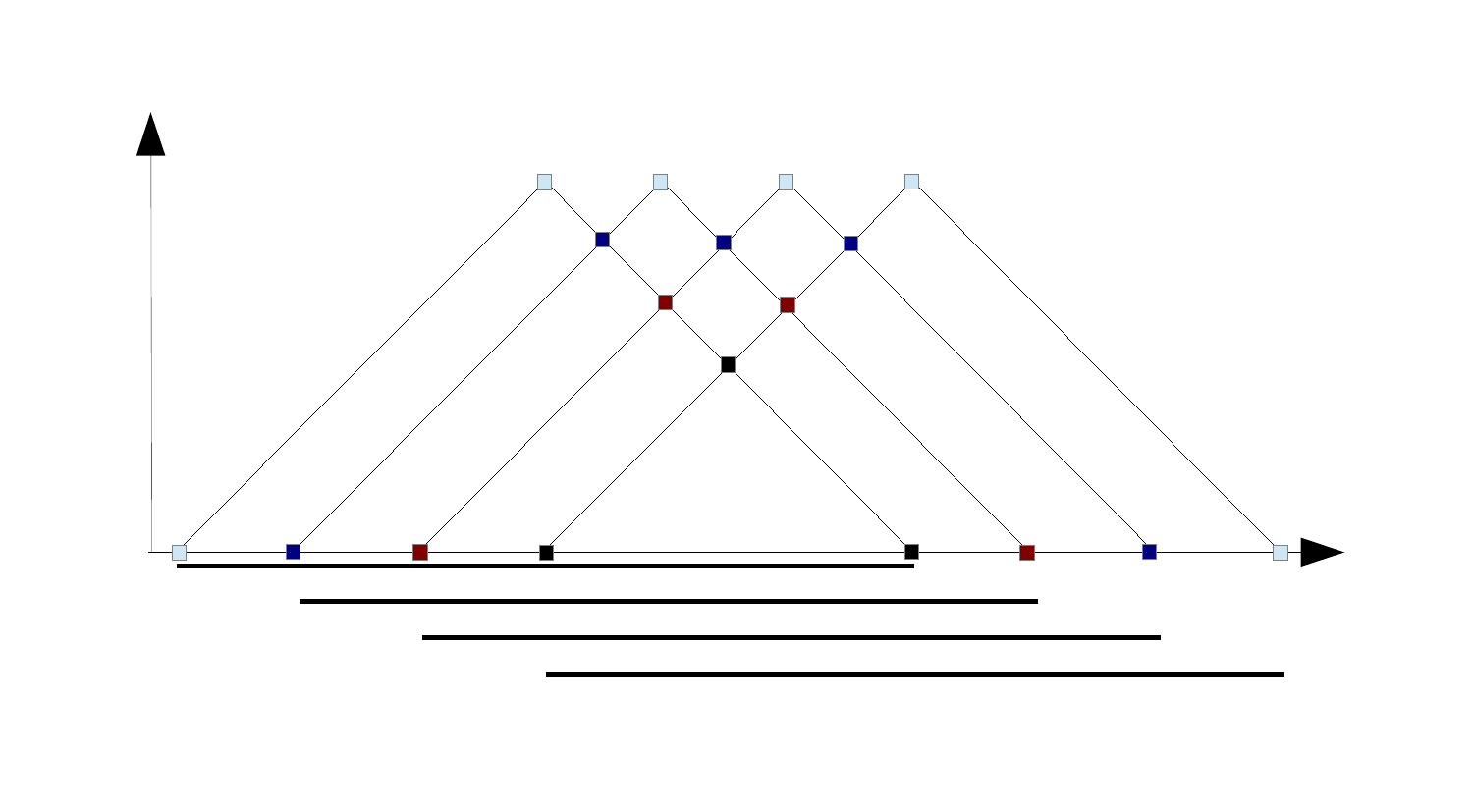}
\caption{The worst case scenario for the complexity of the persistence landscapes occurs when all the pairs of persistence intervals have nonempty intersection. Suppose $n$ persistence intervals having nonempty intersection are given. Then there are $n$ nonzero persistence landscapes and the landscape $\lambda_k$ has $2n+3-2k$ critical points. That gives $n^2+2n = O(n^2)$ critical points in total.}
\label{fig:worstCaseComplexityOfPersistenceLandscapes}
\end{figure}

To end this section, let us determine the complexity of the persistence landscape structure. Clearly the size of the structure and therefore the computational time required to construct the structure is bounded from below by $P$, the number of critical points of persistence landscape. This number can be quadratic in $n$ in the worst case as shown in the Figure~\ref{fig:worstCaseComplexityOfPersistenceLandscapes}. This analysis shows that the algorithms presented in this section are optimal in the worst case. In fact, it is easy to see that if $n$ is the number of intervals given by the birth-death pairs, and $p$ is the number of pairwise non-empty intersections of these intervals, then $P=3n+2p$. Since $p \leq \binom{n}{2}$, we have $P \leq n^2+2n$.

\subsection{Averages and linear combinations}
\label{sec:averages}

In this section, we present algorithms for calculating linear combinations of persistence landscapes.

In Section~\ref{sec:landscapes} we encoded a persistence landscape
$\lambda= \{\lambda_k: \R \to \R\}$ by lists $\{\bbL_k\}$ of pairs of extended real
numbers. 
Fix $k$.
Define $\bbX_k$ and $\bbY_k$ to be the vectors of critical
numbers and critical values obtained from the first and second
coordinates of elements of $\bbL_k$.
Then $\bbY_k = \lambda_k(\bbX_k)$, and
$\lambda_k$ can be obtained from $\bbX_k$ and $\bbY_k$ by linear interpolation. 

Now suppose that we have persistence landscapes
$\lambda^{1},\ldots,\lambda^{N}$ and we wish to calculate the
  linear combination $f = \sum_{j=1}^N a_j \lambda^{j}$, where $a_j
  \in \R$.
Important special cases are the average
of a list of persistence landscapes, $\overline{\lambda} =
\sum_{j=1}^N \frac{1}{N} \lambda^j$, and the difference between the
averages of two groups of persistence landscapes, $\overline{\lambda}
- \overline{\lambda'} = \sum_{j=1}^N \frac{1}{N} \lambda^j +
\sum_{j=1}^{N'} \frac{-1}{N'} \lambda'^j$.

Let $f_k(t) = f(k,t)$. Then $f_k = \sum_{j=1}^N a_j \lambda^{j}_k$.
First we give a naive algorithm for calculating a representation of $f_k$
from the representations
$(\bbX_k^1,\bbY_k^1),\ldots,(\bbX_k^N,\bbY_k^N)$ of
$\lambda_k^1,\ldots,\lambda_k^N$.
First we sort the union of the elements of
$\bbX_k^{1},\ldots,\bbX^{N}_k$, removing repetitions.
Call this vector $\bbX_k$. For each $1\leq j \leq N$, define
$\bar{\bbY}_k^{j} = \lambda_k^{j}(\bbX_k)$.
Now we represent $\lambda_k^{j}$ by $\bbX_k$ and $\bar{\bbY}_k^{j}$.
Again, 
$\lambda^{j}_k$ can be obtained from $\bbX_k$ and $\bar{\bbY}_k^{j}$ by
linear interpolation.
By definition, $f_k(\bbX_k) = \sum_{j=1}^N a_j \lambda_k^{j}(\bbX_k) =
\sum_{j=1}^N a_j \bar{\bbY}_k^{j}$.
Also, $f_k$ can be recovered from $\bbX_k$ and $f_k(\bbX_k)$ by linear
interpolation. 
See Algorithm~\ref{alg:linear-combination}.
In summary, vector space operations on
$\lambda_k^{1},\ldots,\lambda_k^{N}$ are obtained from vector
space operations on $\bar{\bbY}_k^{1},\ldots,\bar{\bbY}_k^{N}$.

\begin{algorithm}[t]
  \SetAlgoNoLine
  \small
  \KwIn{$(\bbX_k^1,\bbY_k^1),\ldots,(\bbX_k^N,\bbY_k^N)$, for some fixed $k$, and $a_1,\ldots,a_N$\;}
  \KwOut{$(\bbX_k,\bbY_k)$ \;}
  Merge the sorted lists $\bbX_k^j$, removing duplicates. Call this vector $\bbX_k$\;
  \For {$j = 1$ to $N$ }
  {
		Calculate $\bar{\bbY}_k^j = \lambda_k^j(\bbX_k)$ by linear interpolation;\
  }
  $\bbY_k \leftarrow \sum_{j=1}^N a_j \bar{\bbY}_k^j$;\
  \Return $(\bbX_k,\bbY_k)$
\caption{Linear combination of persistence landscapes.}
\label{alg:linear-combination}
\end{algorithm} 

Let us consider the time complexity of this algorithm.
Let $n$ be the maximum number of birth-death pairs used to construct each of $\lambda^1,\ldots,\lambda^N$.
Let $P_k$ be the sum of the number of the critical points of the $\lambda_k^1,\ldots,\lambda_k^N$. Note that $P_k = O(Nn)$.
Then 
$f_k$ has at most $P_k$ critical points, and constructing $\bbX_k$ takes $O(P_k)$.
Since the length of $\bbX_k$ may be at most $P_k$, calculating each
$\bar{\bbY}_k^j$ takes $O(P_k)$ and calculating $\bbY_k = f_x(\bbX_k)$ takes
$O(NP_k)$.
So Algorithm~\ref{alg:linear-combination} has time complexity 
$O(nN^2)$.
 
Now if we repeat Algorithm~\ref{alg:linear-combination} for all $k$ we
obtain a linear combination of the full persistence landscape. 
Let $P$ be the sum of the number of critical points of $\lambda^1,\ldots,\lambda^N$.
Then constructing $\{\bbL_k\}$ has time complexity $O(NP)=O(n^2N^2)$.


%

The complexity of the naive algorithm can be improved to $O(n^2N\log N)$ by merging landscapes in a binary tree fashion (this is sometimes referred to as divide-and-conquer). Here we describe this for the case of calculating the average landscape\footnote{In the general case of computations of weighted sums of landscapes, when merging in the first level we compute weighted sums  $(\mathcal{M}_1 \bbX_k^i,\mathcal{M}_1 \bbY_k^i) = a_i(\bbX_k^i,\bbY_k^i) + a_{i+1}(\bbX_k^{i+1},
 \bbY_k^{i+1})$ instead of unweighted ones  $(\mathcal{M}_1 \bbX_k^i,\mathcal{M}_1 \bbY_k^i) = (\bbX_k^i,\bbY_k^i) + (\bbX_k^{i+1},
 \bbY_k^{i+1})$. Also, at the end of computations we do not multiply the last landscape by $\frac{1}{N}$.}. 
Suppose a collection of persistence
 landscapes $(\bbX_k^1,\bbY_k^1), \ldots, (\bbX_k^N,\bbY_k^N)$ is given. Let us assume that $N$ is even. Then this collection is transformed to a new collection $ (\mathcal{M}_1 \bbX_k^1,\mathcal{M}_1 \bbY_k^1),\ldots,(\mathcal{M}_1 \bbX_k^{\frac{N}{2}},\mathcal{M}_1 \bbY_k^{\frac{N}{2}})$ 
where
 $(\mathcal{M}_1 \bbX_k^i,\mathcal{M}_1 \bbY_k^i) = (\bbX_k^i,\bbY_k^i) + (\bbX_k^{i+1},
 \bbY_k^{i+1})$ for $i \in \{1,\ldots,\frac{N}{2}\}$. 
In the case that $N$ is odd, additionally set $(\mathcal{M}_1 \bbX_k^{\frac{N+1}{2}},\mathcal{M}_1 \bbY_k^{\frac{N+1}{2}}) = (\bbX_k^{N},\bbY_k^{N})$. This merging procedure is repeated for the obtained sequence until the sequence contains only
   one element $(\bbX_k^f,\bbY_k^f)$. The average persistence landscape is then $\frac{1}{N}(\bbX_k^f,\bbY_k^f)$. Note that 
the number of landscapes at level $i$ is of the order $\frac{N}{2^i}$. 
The complexity of each of these is $2^i n^2$. 
Therefore, the cost of merging
     neighboring persistence landscapes at level $i$ is $2^i n^2 \frac{N}{2^i} = n^2 N$, which does not depend on $i$. Number of levels is $log_{2}(N)$ , and
      therefore the total cost of computing average persistence landscape is $O( n^2 N log(N) )$.

\medskip

For the case where all of the birth-death pairs have endpoints on an evenly-spaced grid of size $m+1$, we can obtain a linear combination of the persistence landscapes by simply taking a linear combination of the $N$ corresponding vectors of size $K\times(2m+1)$, which has time complexity $O(KmN)$ which is $O(mnN)$.

\subsection{Distances}
\label{sec:distances}

In this section, we compute the $L^p$ and
$L^{\infty}$ distances between two linear combinations of persistence landscapes $\bbL = \{\bbL_k\}$ and $\bbL' = \{\bbL'_k\}$.
Let $P$ be the maximum number of critical points of $\bbL$ and $\bbL'$.
Let $\{(\bbX_k,\bbY_k)\}$ be the representation of the difference between these two persistence landscapes as described in 
Section~\ref{sec:averages}.

The $L^{\infty}$ distance between $\bbL$ and $\bbL'$ is 
\begin{equation*}
  \norm{\bbL-\bbL'}_{\infty} = \max_k \abs{\bbY_k}.
\end{equation*}
This calculation has time complexity $O(P)$.

The $L^p$ distance between $\bbL$ and $\bbL'$ is given by the formula:
\[ \norm{\bbL-\bbL'}_p =  \left[ {\sum_{k=1}^{K} \int \norm{\bbL_k-\bbL'_k}_p^p } \right]^{\frac{1}{p}}   \]
The norm $\norm{\bbL_k-\bbL'_k}_p^p$ can be computed from $(\bbX_k,\bbY_k)$
by summing integrals over intervals given by consecutive elements of
$\bbX_k$.
These have the form $\int_c^d \abs{ax+b}^p dx$ which can be written as one or two integrals of the form
 $\int_c^d (ax+b)^p dx = 
 \frac{(ax+b)^{p+1}}{a(p+1)} |_c^d$.
This calculation also has time complexity $O(P)$.

Now assume that we start with two persistence
diagrams, each of which has at most~$n$ birth-death pairs. 
Since $P=O(n^2)$ we can calculate the $L^{\infty}$ and
$L^p$ distances between the corresponding persistence landscapes in $O(n^2)$.

\medskip

For the case where the points in the birth-death pairs lie on an evenly-spaced grid of size $m$, 
the calculation of distance between linear combinations of persistence landscapes is $O(Km)$,
where $K$ be the maximum $k$ for which either $\{\bbL_k\}$ or $\{\bbL'_k\}$ is nontrivial.
So starting with two persistence diagrams, each of which has at most $n$ birth-death pairs which lie on a grid of size $m$, we can calculate the $L^{\infty}$ and $L^p$ distances between their persistence landscapes in $O(Km)$ which is $O(mn)$. 

\medskip
 
We can combine the results of this section with those of the previous section to calculate the $L^{\infty}$ and $L^{p}$ distances between $f$ and $g$, two linear combinations of at most $N$ persistence landscapes, each of which in obtained from at most $n$ birth-death pairs.
An important special case of this is the calculation of the distance between two average persistence landscapes.
Let $P$ be the total number of critical points in the two sets of persistence landscapes.
Then the representation of $f-g$,  $\{(\bbX_k,\bbY_k)\}$, has at most $P$ critical points, and can be constructed in time 
$O(n^2N\log N)$.

From this, the distance between $f$ and $g$ can be calculated in 
$O(P) = O(n^2N)$
so the total computation time is 
$O(n^2N\log N)$.
%
%
For the case were the endpoints of the intervals lie on a grid of size $m+1$, constructing $f$ and $g$ is $O(mnN)$, and calculating the distance between $f$ and $g$ is $O(mn)$, so the total computation has time $O(mnN)$.

\section{Experiments}
\label{sec:experiments}

In this section, we present the results of some experiments used to test our implementation of our algorithms.

\subsection{ Points sampled from $S^d$}
\label{sec:pointsFromSpheres}

Our first experiment is motivated by the following questions. Suppose we are
given a set of points in $\mathbb{R}^D$ that lie on a lower dimensional sphere $S^d$. How well can we determine $d$ using persistent homology?

For $d \in \{2,3,\ldots,10\}$, we sampled $100$ points from $S^d$ using the uniform distribution. This was done by sampling $100$ points from
the $(d+1)-$dimensional Gaussian distribution and then projecting those points to $S^{d}$.  For an example of such
normalization in topological data analysis, see~\cite{cidsz:mumford}.
This was repeated $1000$ times.
To refute the charge that our detected differences are only due 
to the increase of the average distance between points in higher dimensions, we rescale the sphere on which the points has been projected so that the average distance between points is one. 

For each point cloud generated as described above we computed the
persistent homology of the corresponding Vietoris-Rips complex. 
The parameter values for the radius ranged from $0$ to a radius
for which all inessential $0$, $1$, and $2$ dimensional cycles are
killed, which was $0.7$ for this range of dimensions. 

\begin{remark}
  In this example and all subsequent examples, the filtration values have been rescaled to the range from $0$ to $100$. For simplicity, we have chosen to leave all subsequent calculation in this new scale. If desired, it is easy to rescale the results back to the original scale.
\end{remark}

\begin{figure}
\begin{tabular}{ccc}
\includegraphics[scale=0.34]{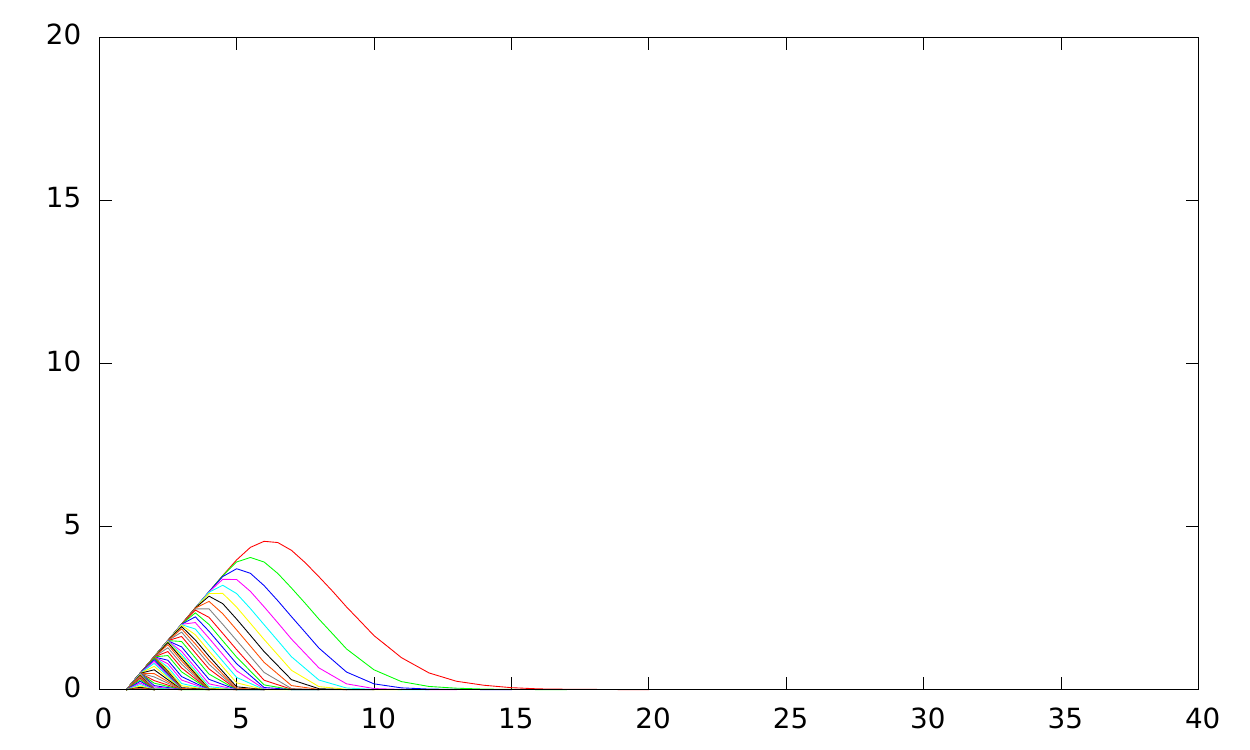}  & \includegraphics[scale=0.34]{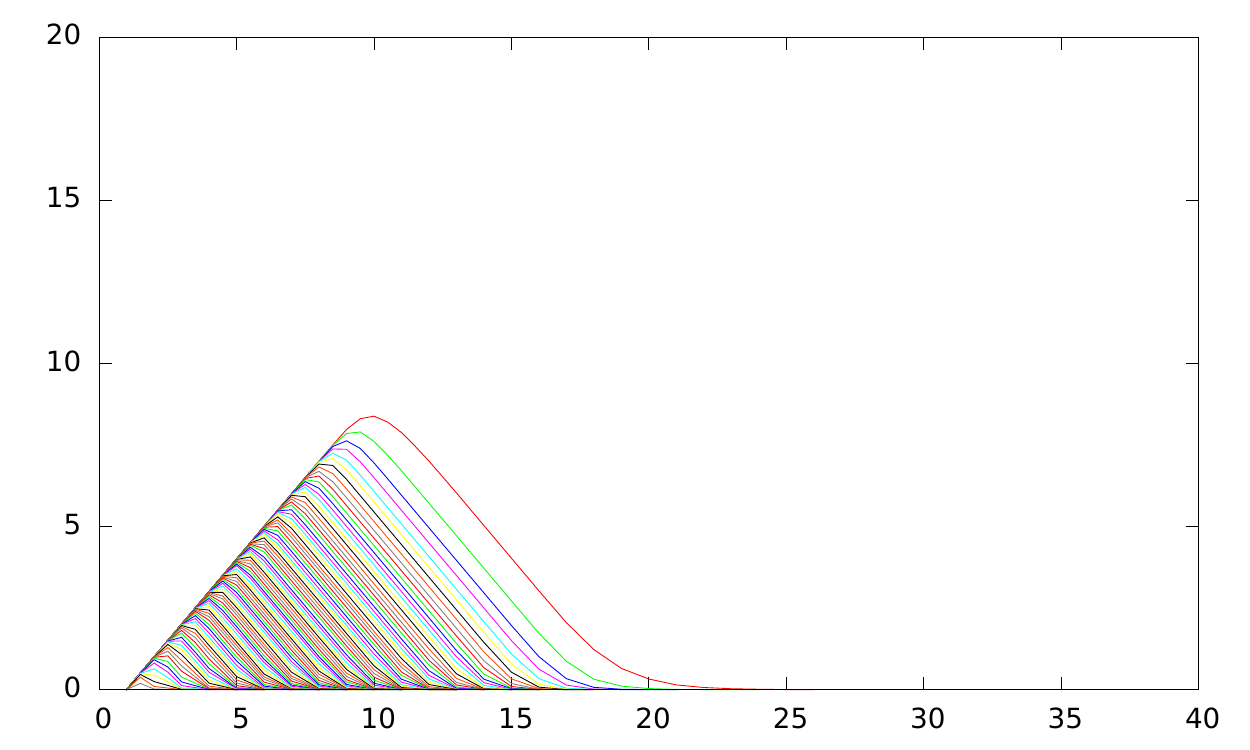} & \includegraphics[scale=0.34]{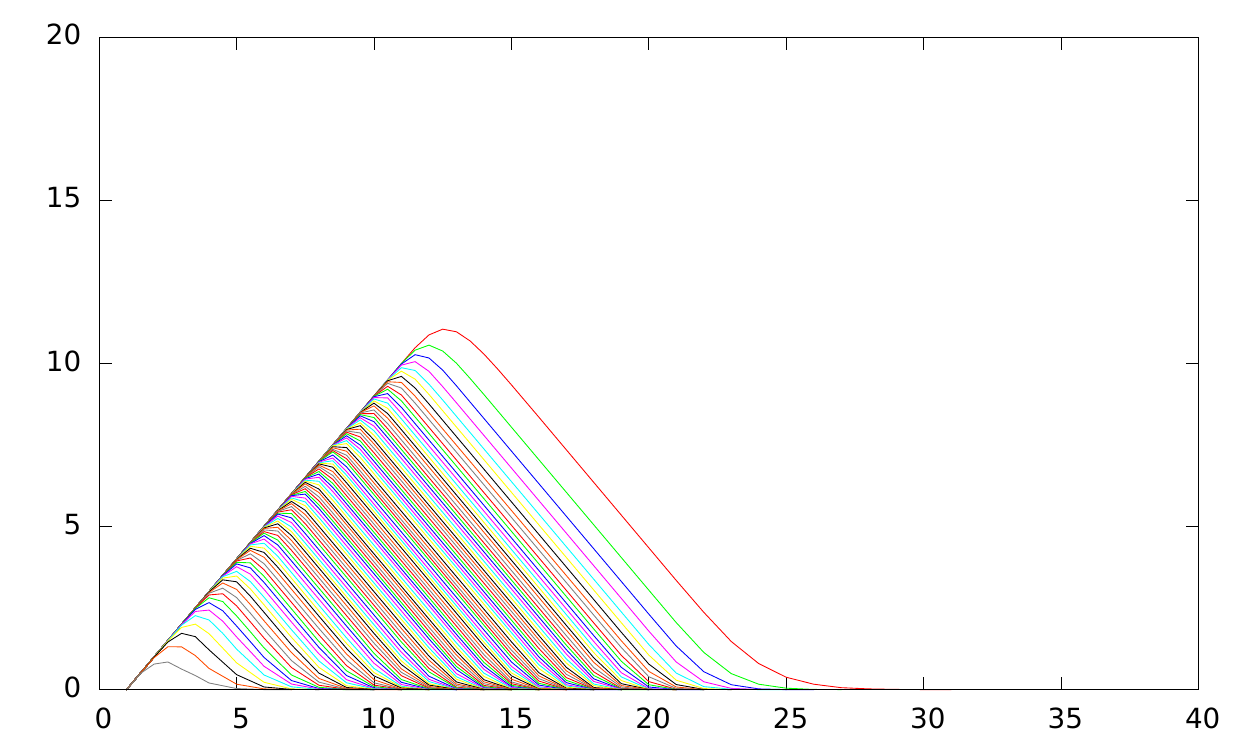} \\ 
dim 2. & dim 3. & dim 4. \\   
\includegraphics[scale=0.34]{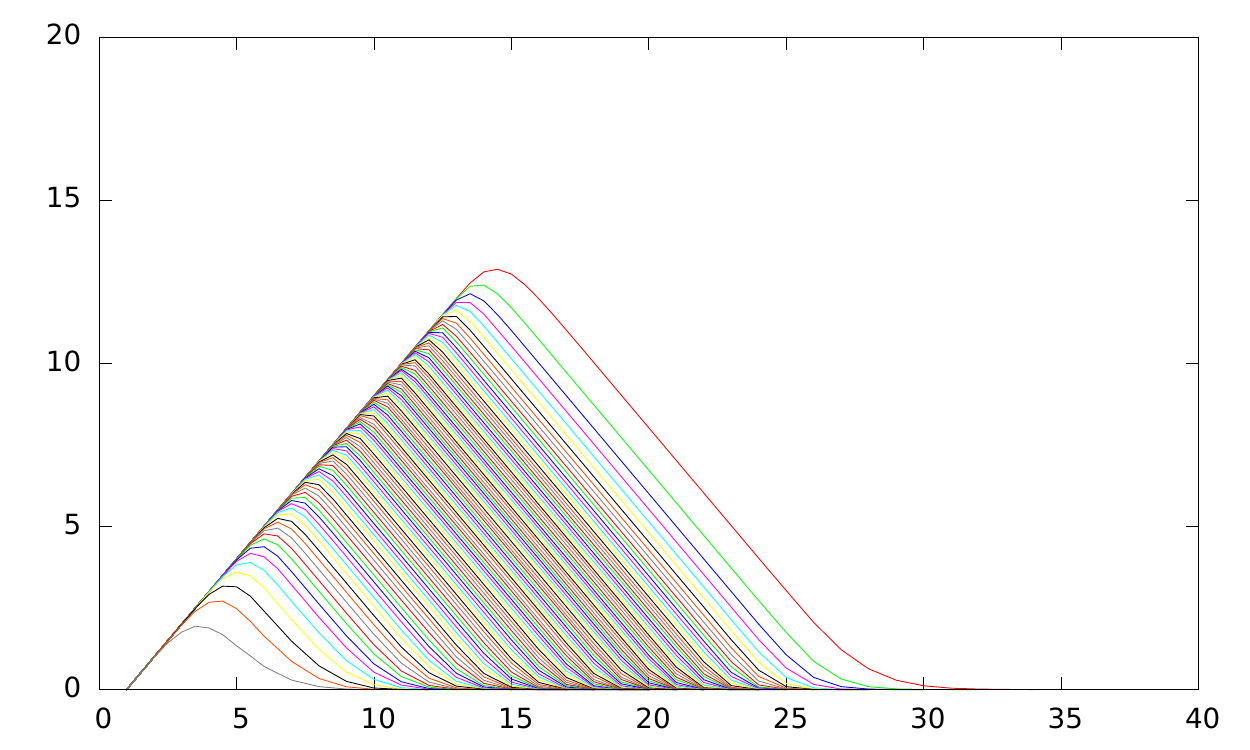} & \includegraphics[scale=0.34]{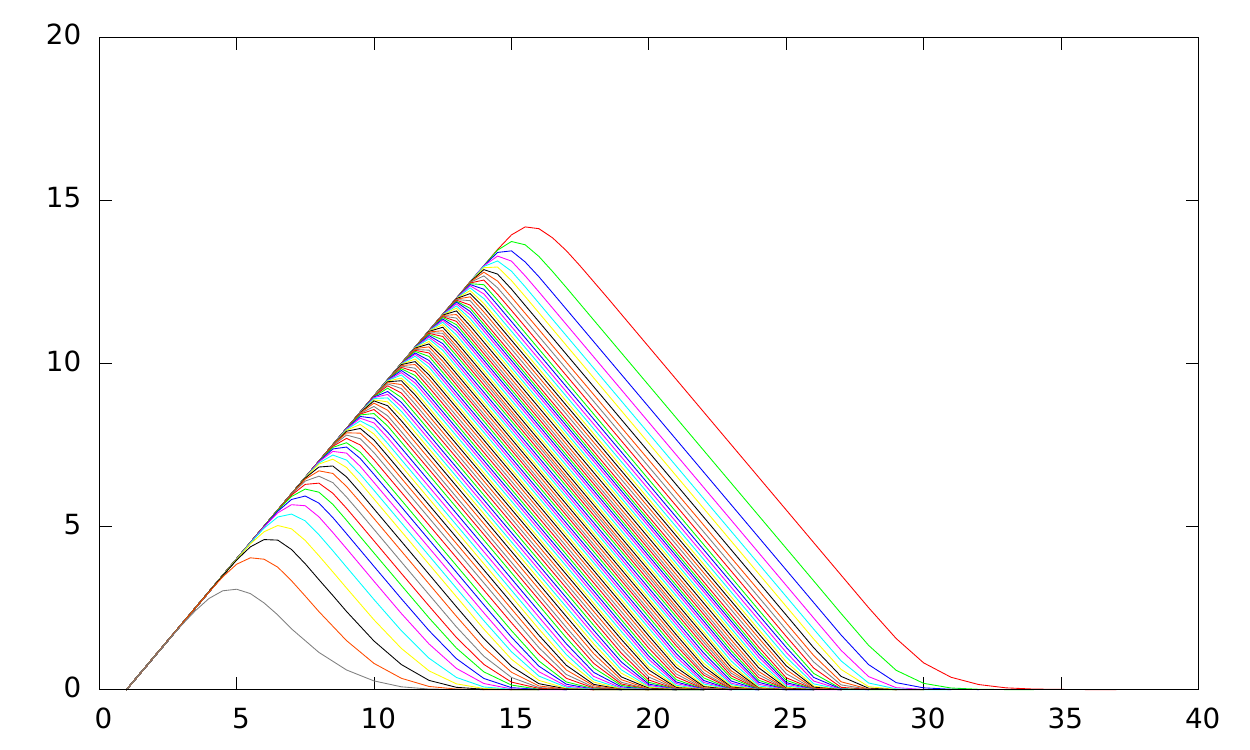} & \includegraphics[scale=0.34]{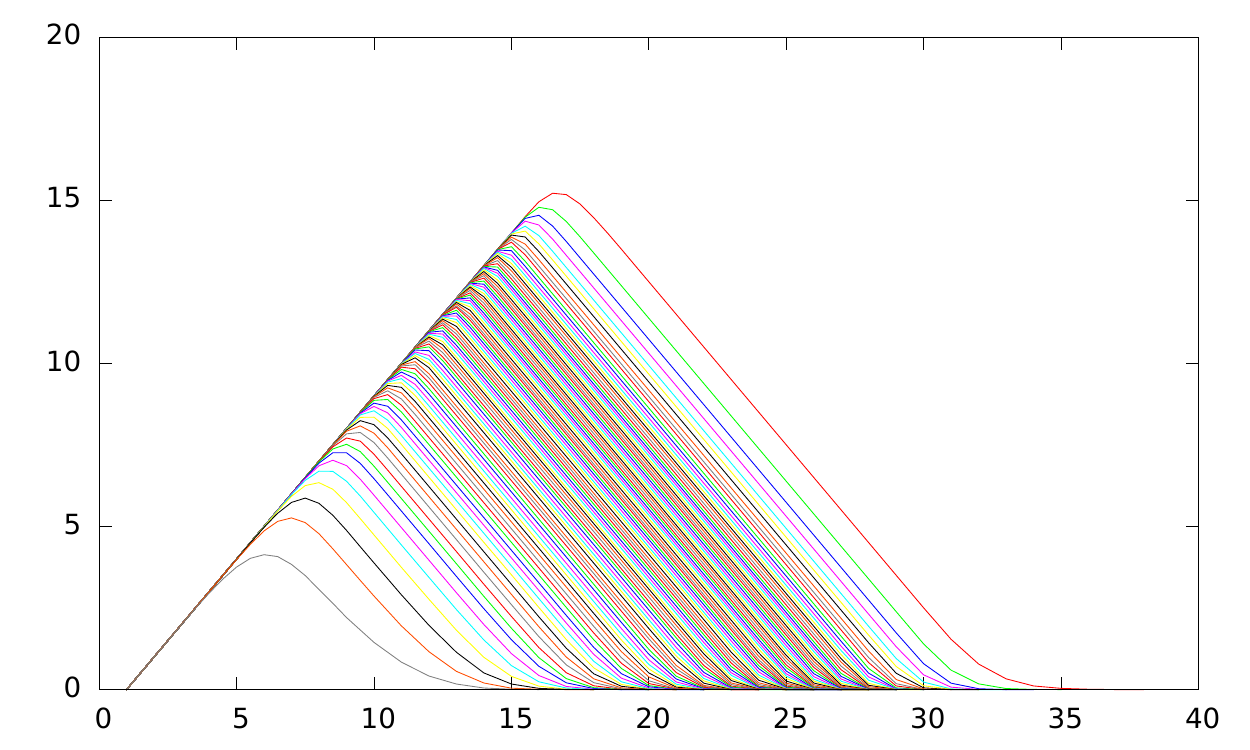} \\ 
dim 5. & dim 6. & dim 7. \\   
\includegraphics[scale=0.34]{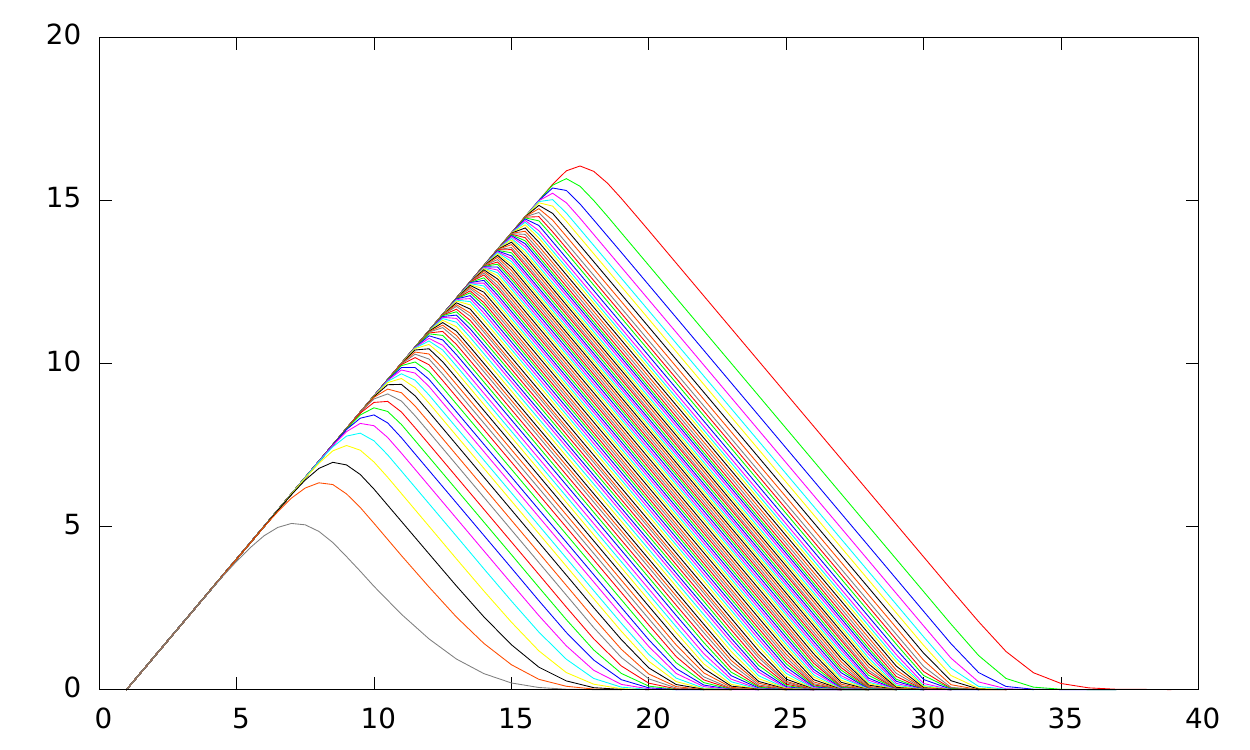} & \includegraphics[scale=0.34]{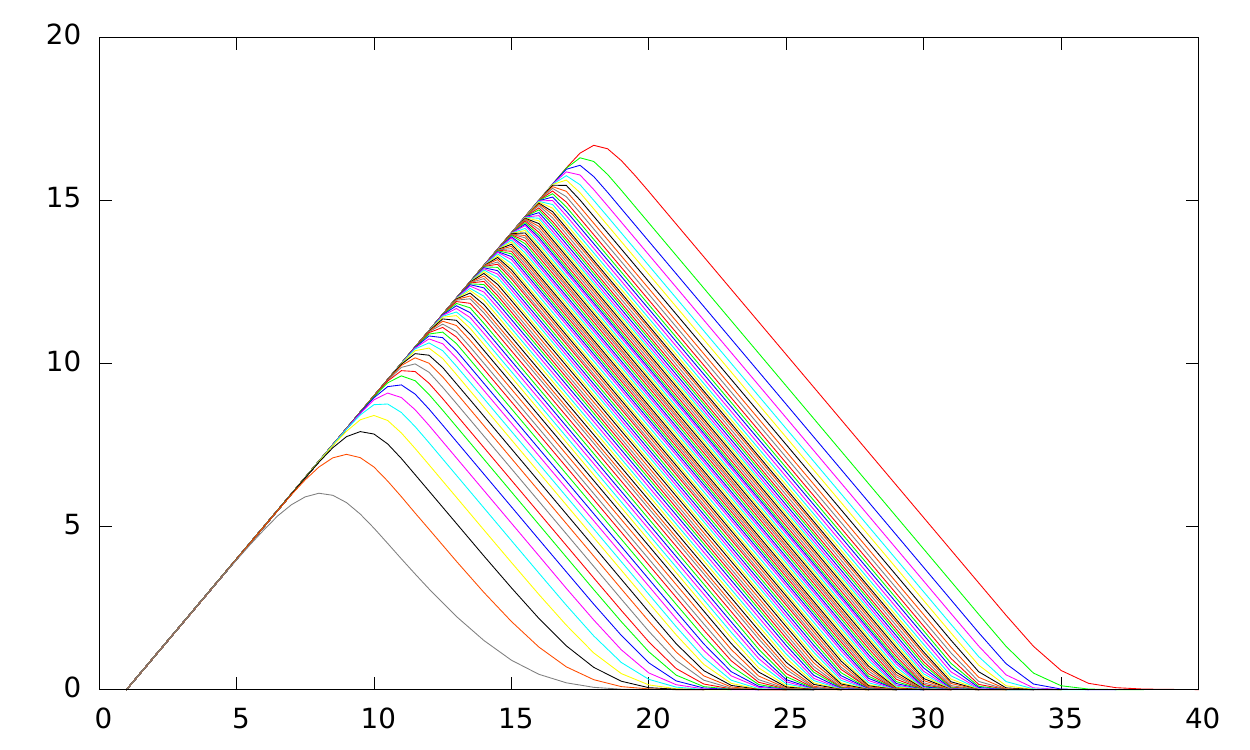} & \includegraphics[scale=0.34]{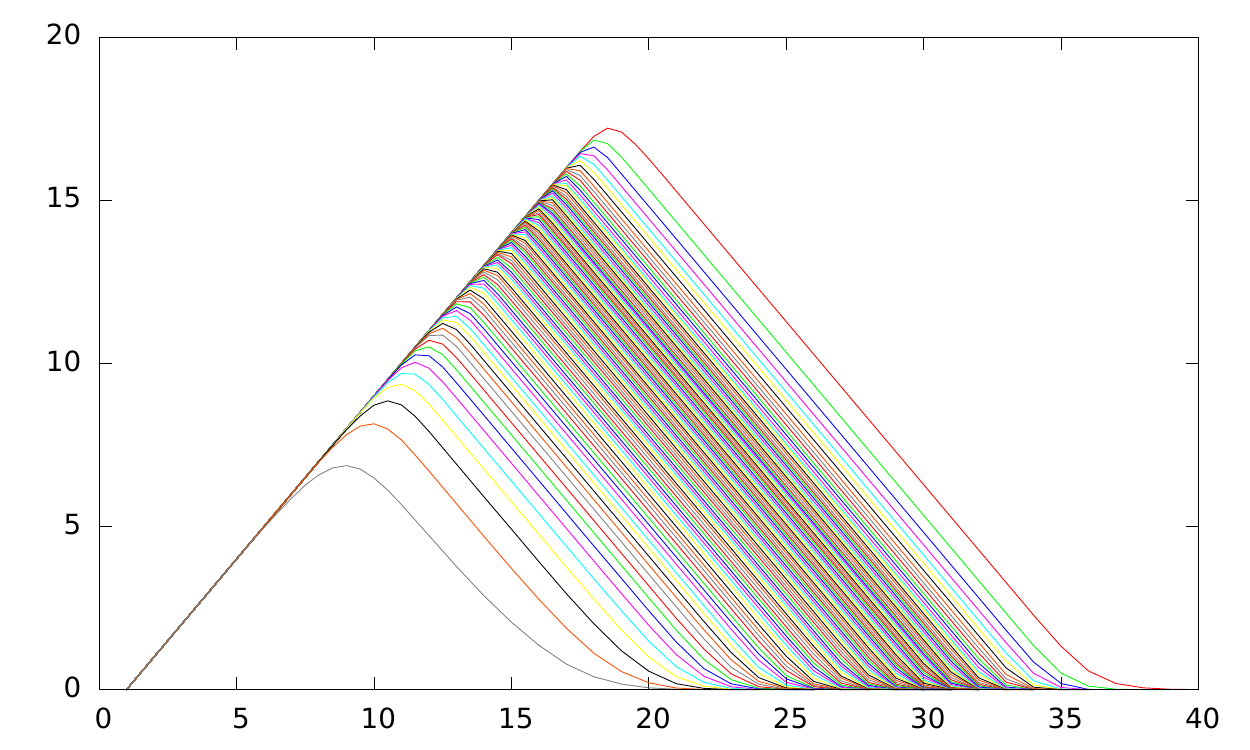} \\ 
dim 8. & dim 9. & dim 10. \\   
\end{tabular}
\caption{Average persistence landscapes in degree $0$ of points sampled from $S^{d}$ for $d \in \{2,\ldots,10\}$. The spheres have been scaled so that the average distance between points is one.}
\label{fig:avLandDim0}
\end{figure}

\begin{figure}
\begin{tabular}{ccc}
  & \includegraphics[scale=0.34]{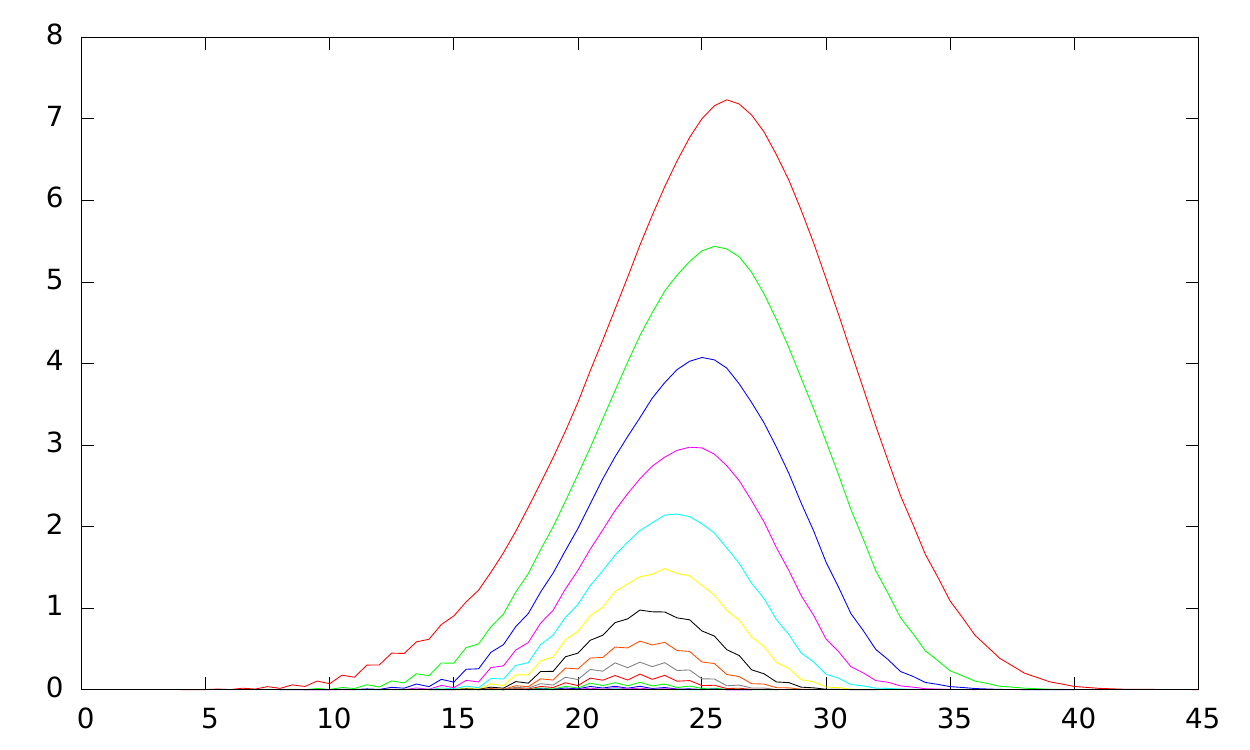} & \includegraphics[scale=0.34]{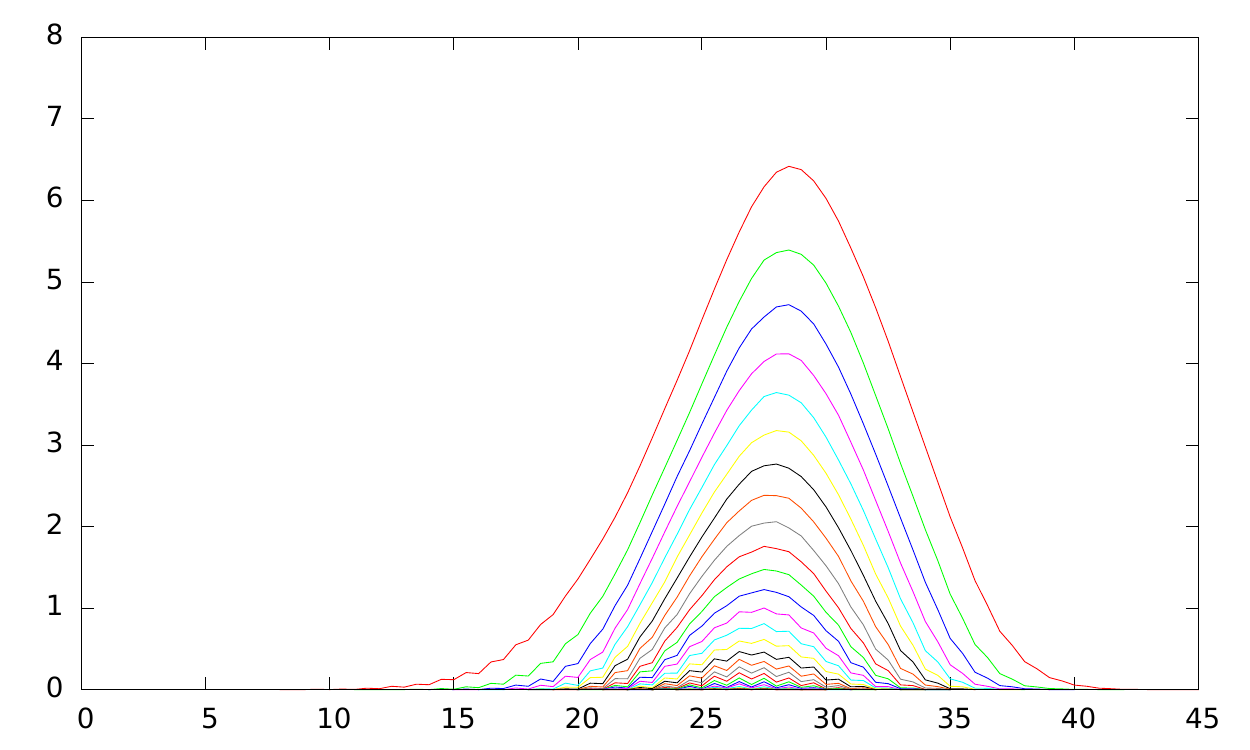} \\ 
dim 2. & dim 3. & dim 4. \\   
\includegraphics[scale=0.34]{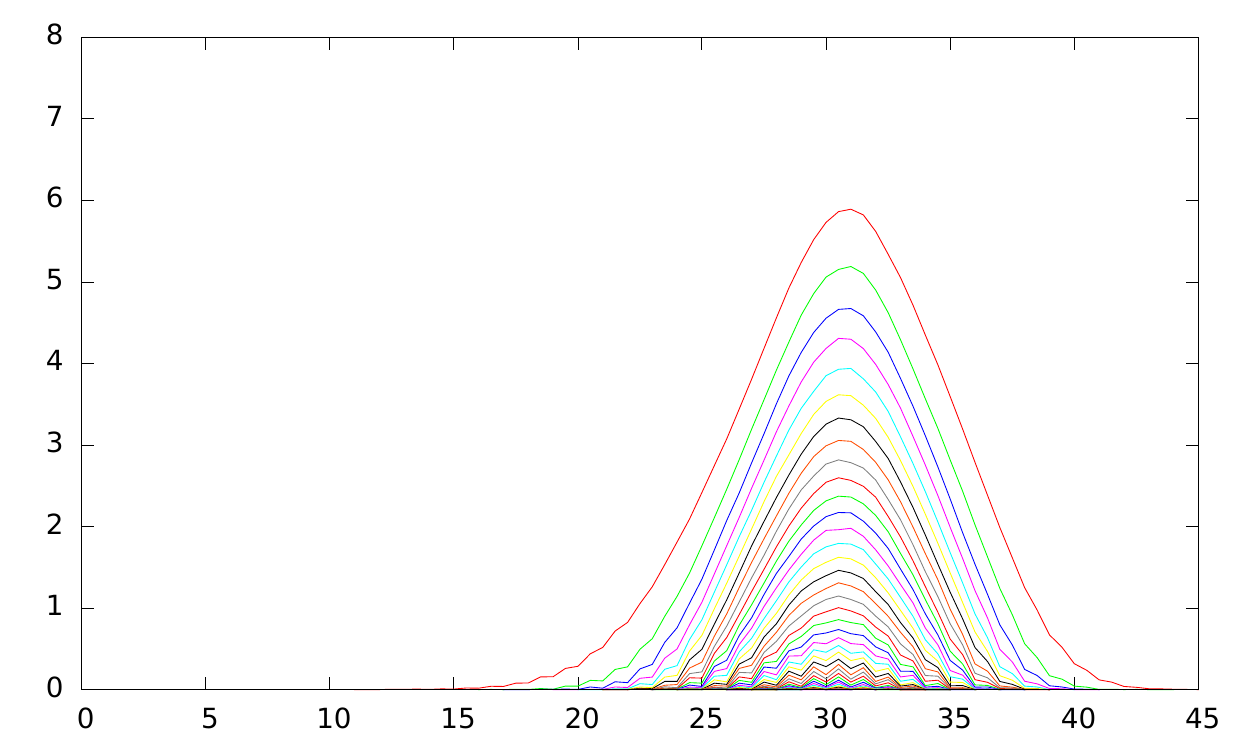} & \includegraphics[scale=0.34]{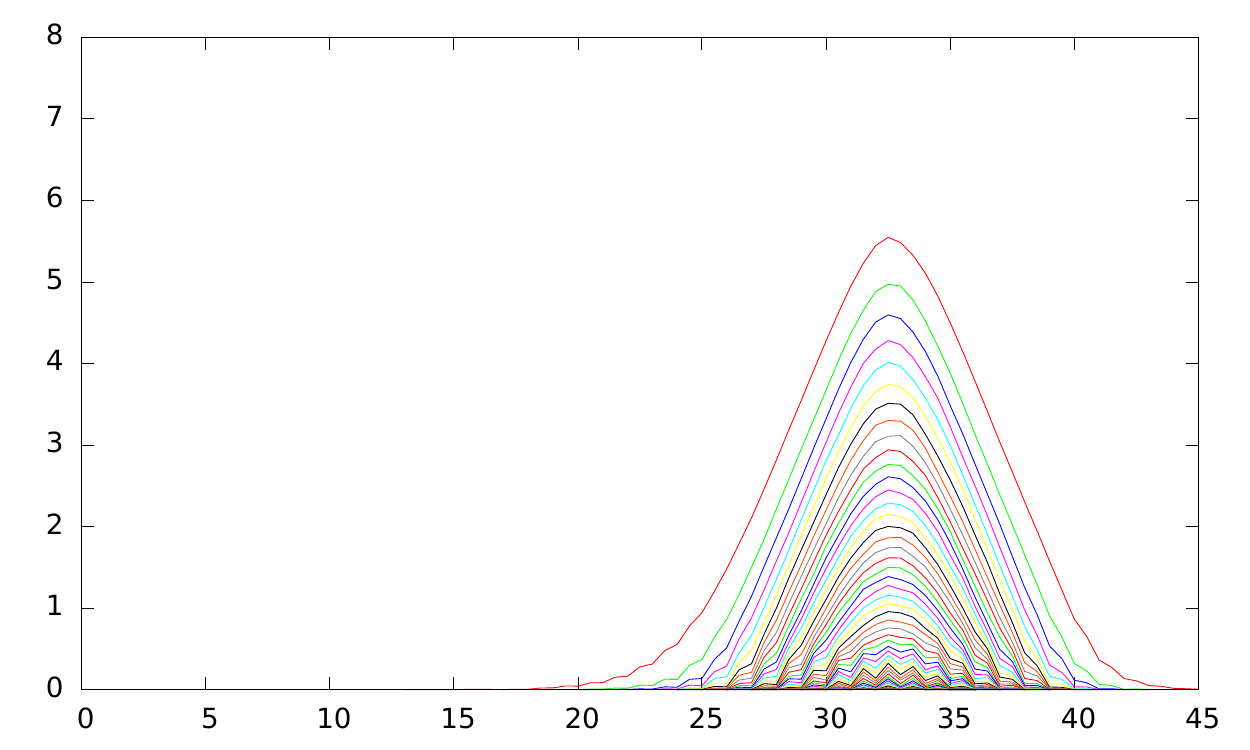} & \includegraphics[scale=0.34]{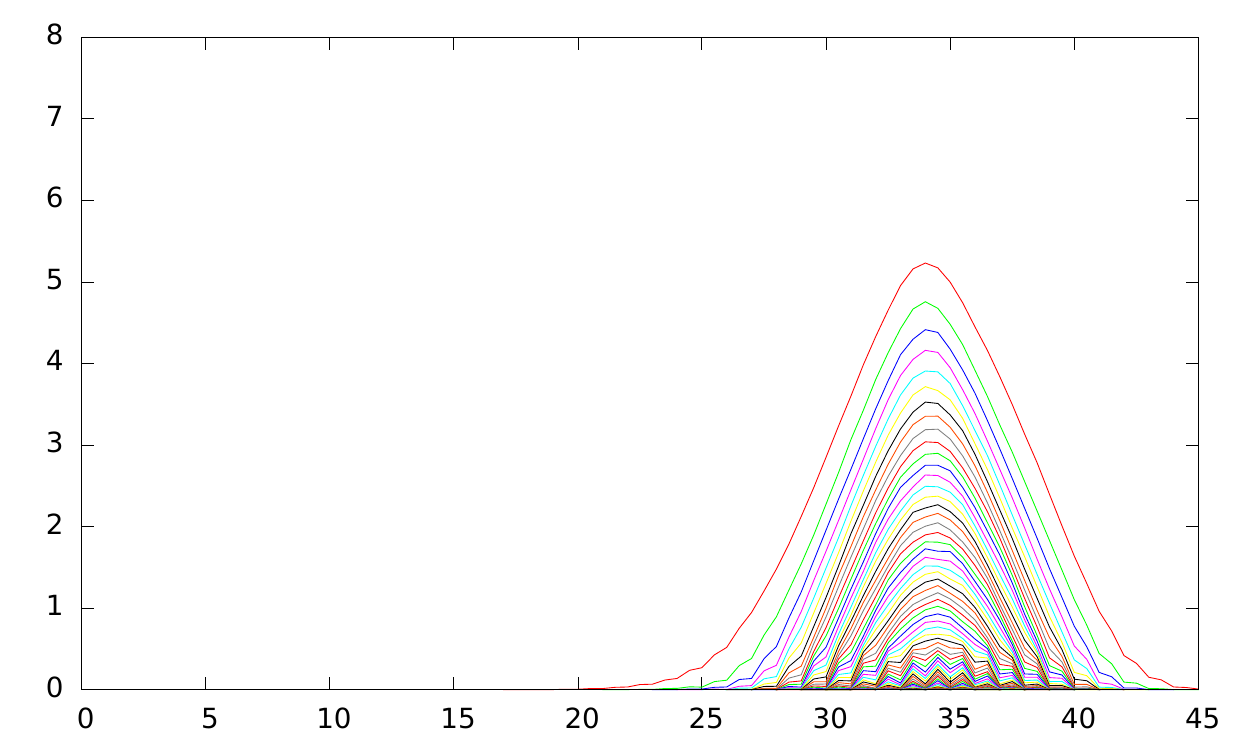} \\ 
dim 5. & dim 6. & dim 7. \\   
\includegraphics[scale=0.34]{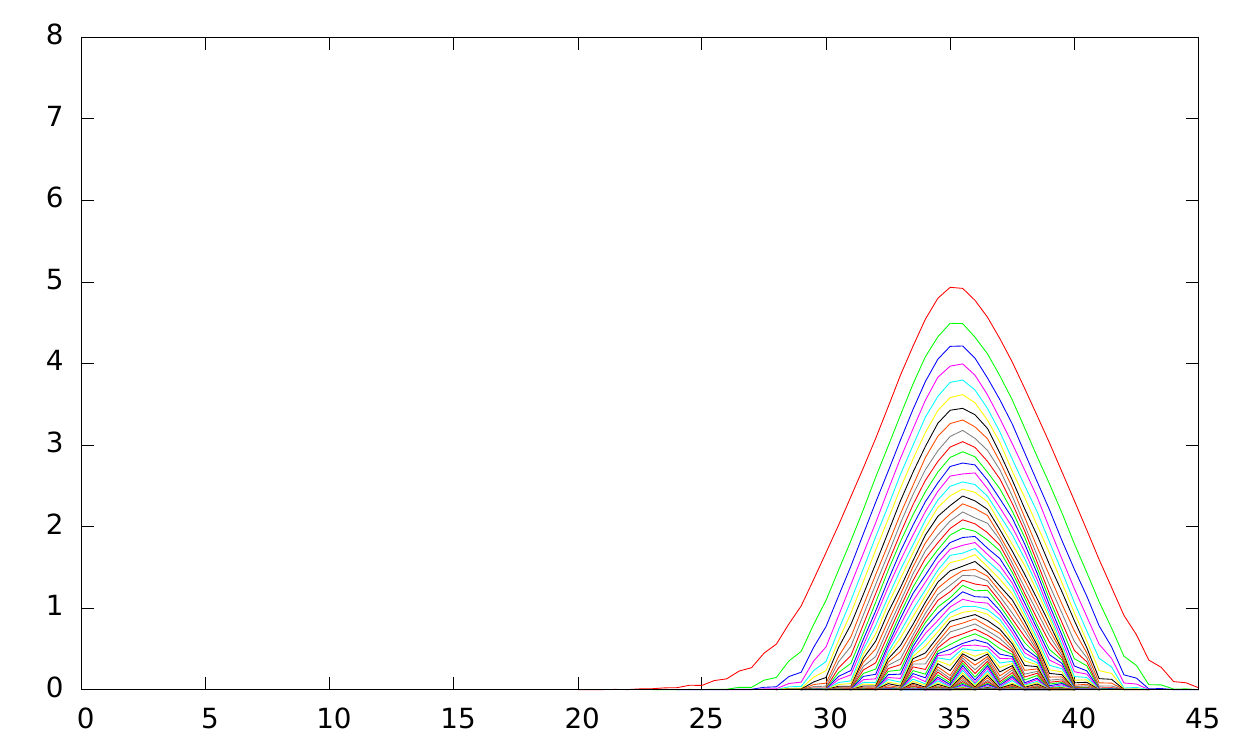} & \includegraphics[scale=0.34]{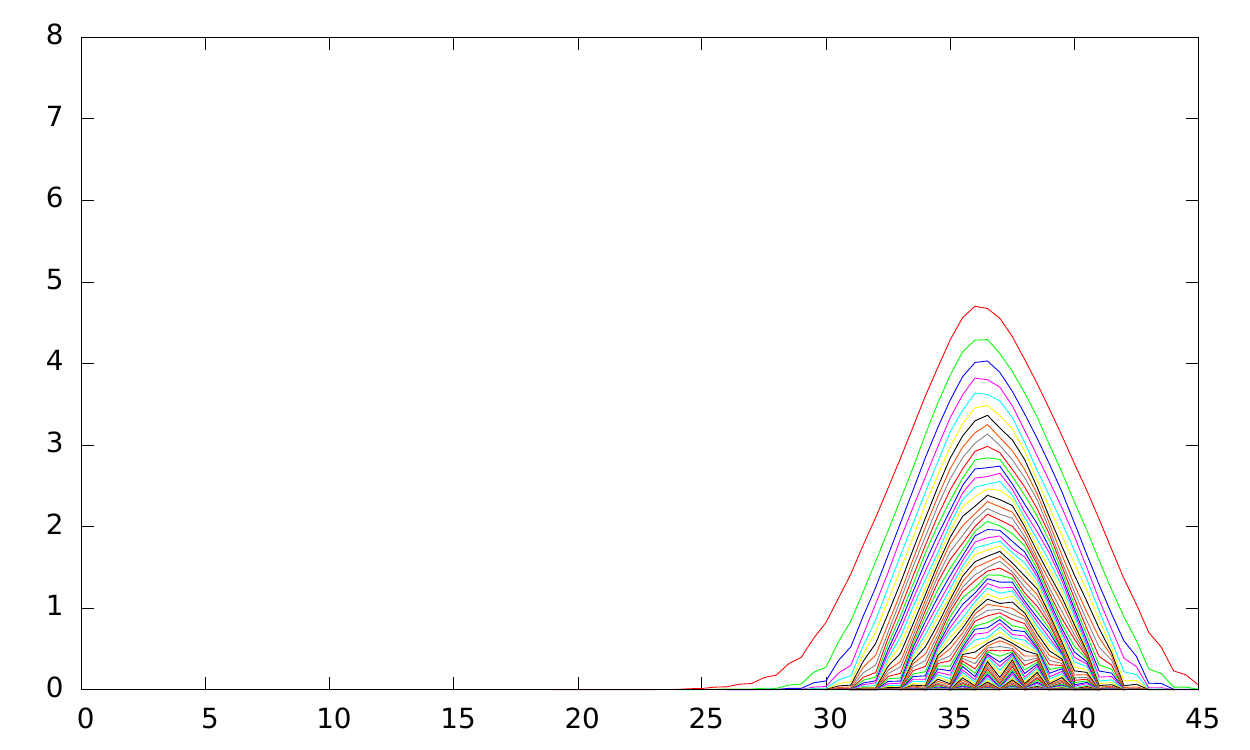} & \includegraphics[scale=0.34]{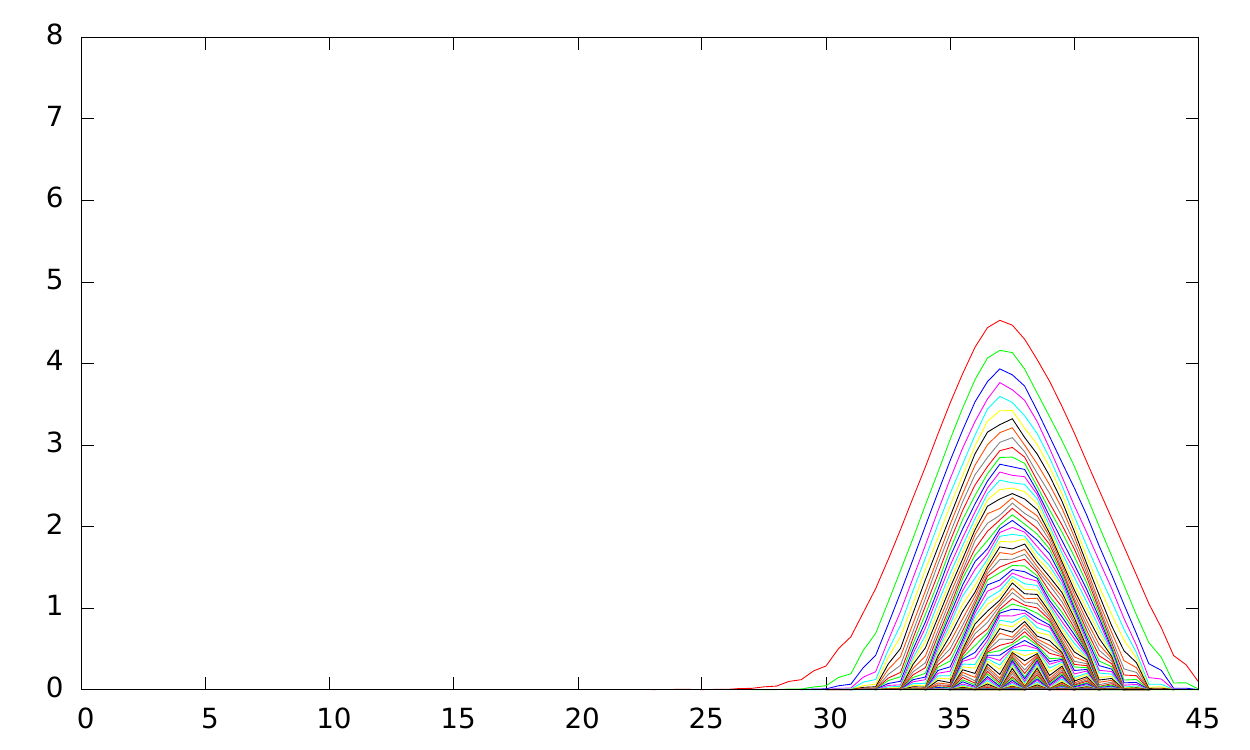} \\ 
dim 8. & dim 9. & dim 10. \\   
\end{tabular}
\caption{Average persistence landscapes in degree $1$ of points sampled from $S^{d}$ for $d \in \{2,\ldots,10\}$. The spheres have been scaled so that the average distance between points is one.}
\label{fig:avLandDim1}
\end{figure}

\begin{figure}
\begin{tabular}{ccc}
  &  & \includegraphics[scale=0.34]{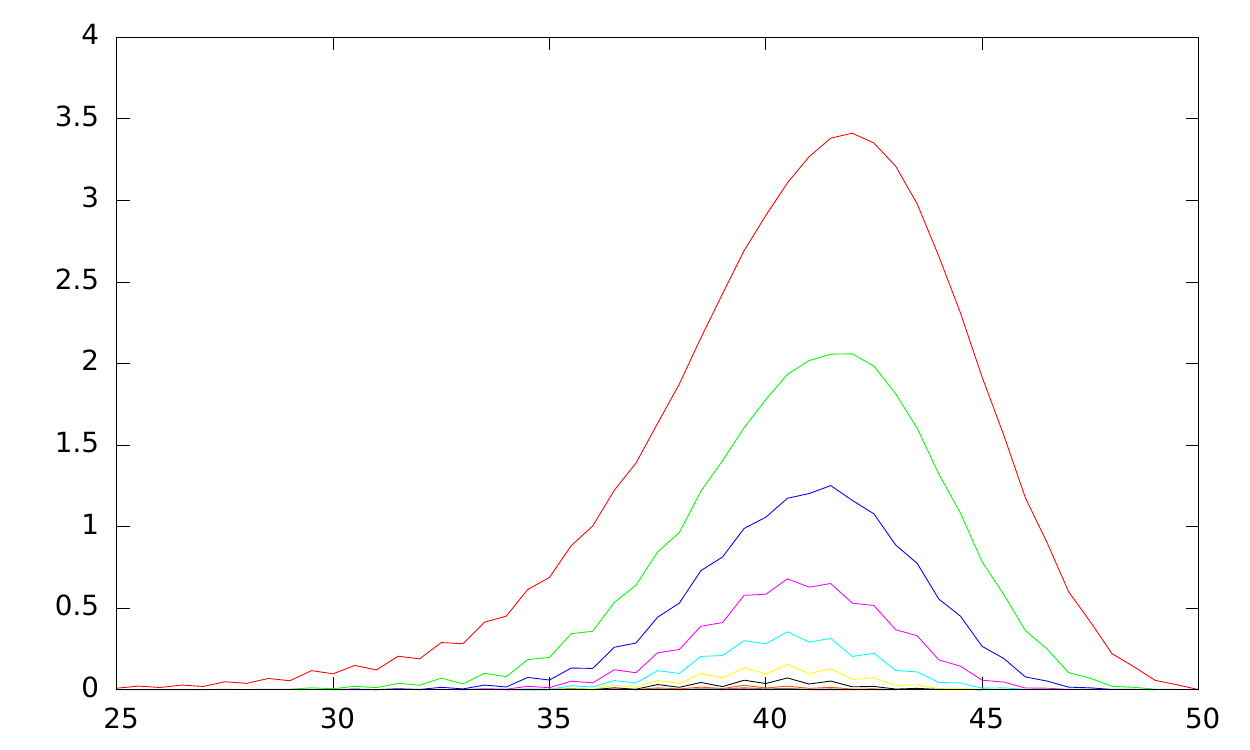} \\ 
dim 2. & dim 3. & dim 4. \\   
\includegraphics[scale=0.34]{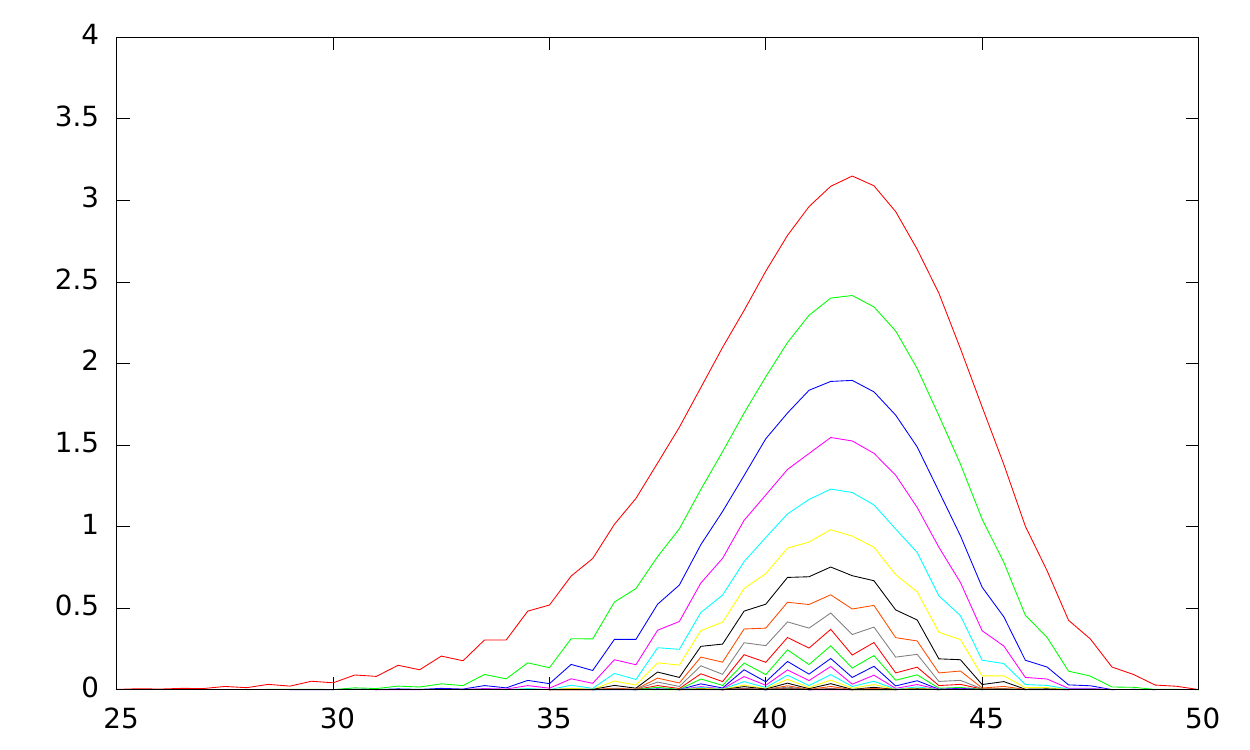} & \includegraphics[scale=0.34]{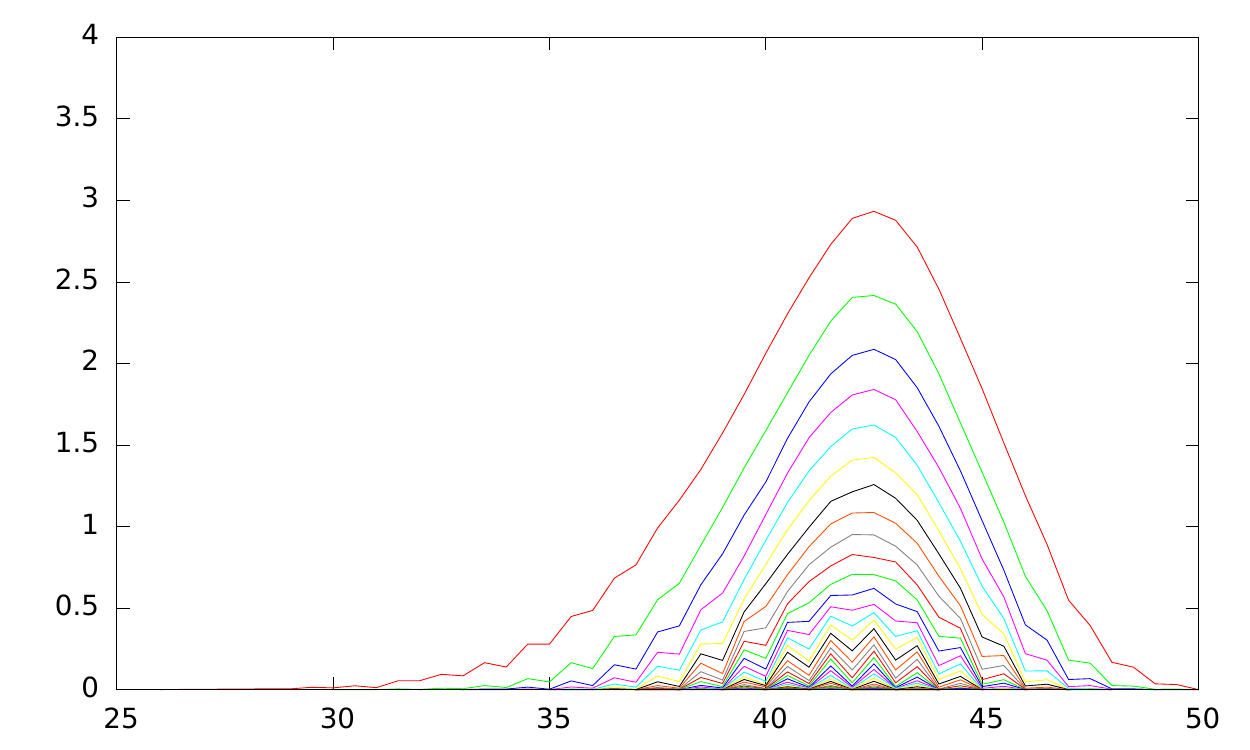} & \includegraphics[scale=0.34]{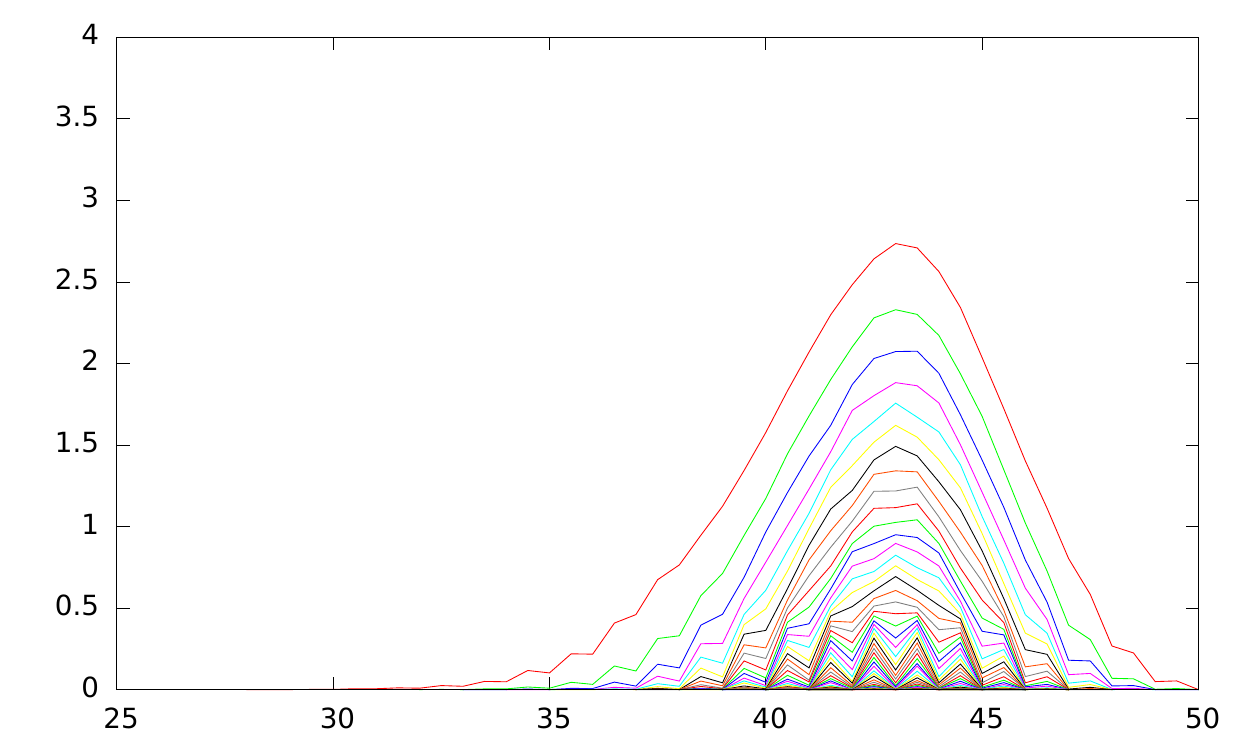} \\ 
dim 5. & dim 6. & dim 7. \\   
\includegraphics[scale=0.34]{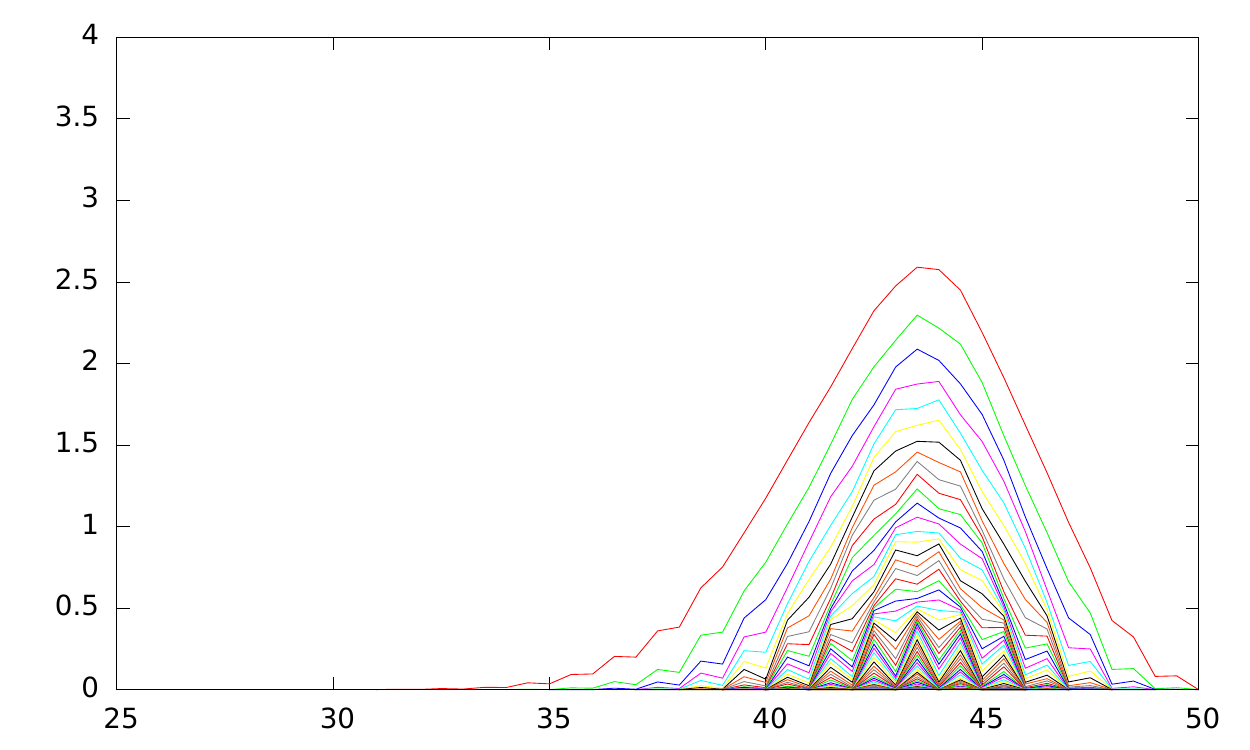} & \includegraphics[scale=0.34]{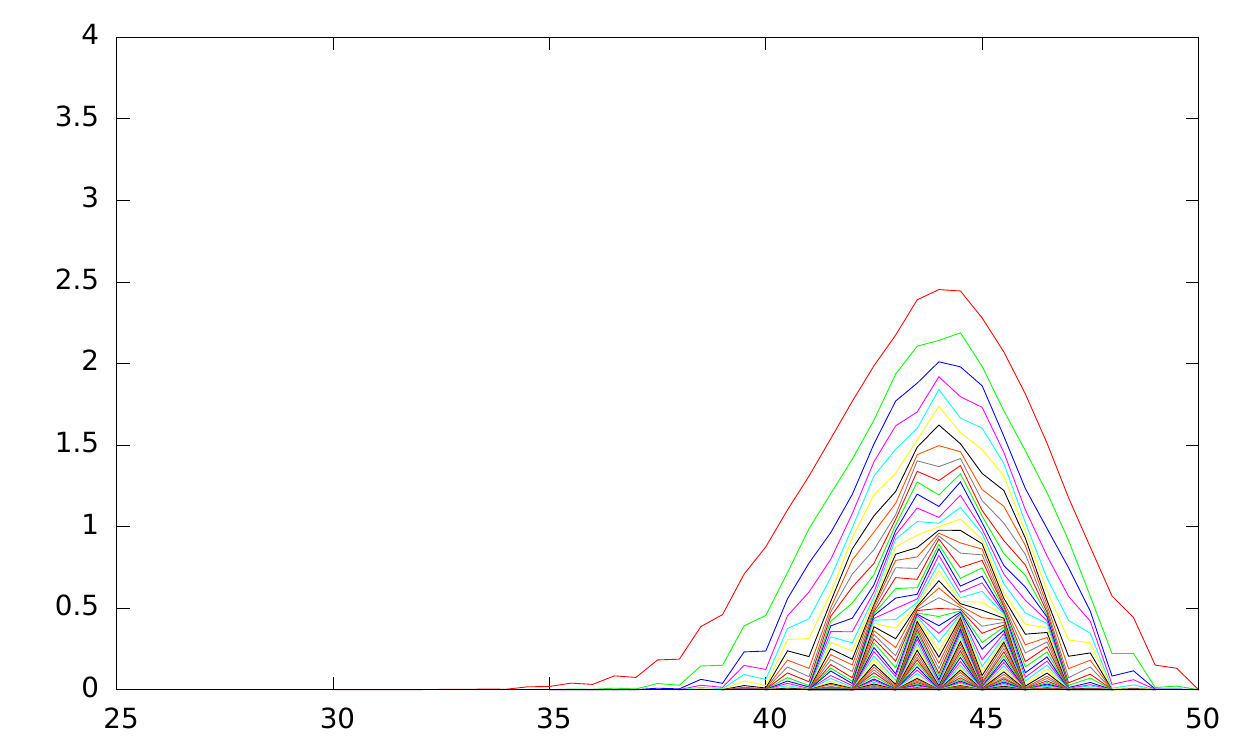} & \includegraphics[scale=0.34]{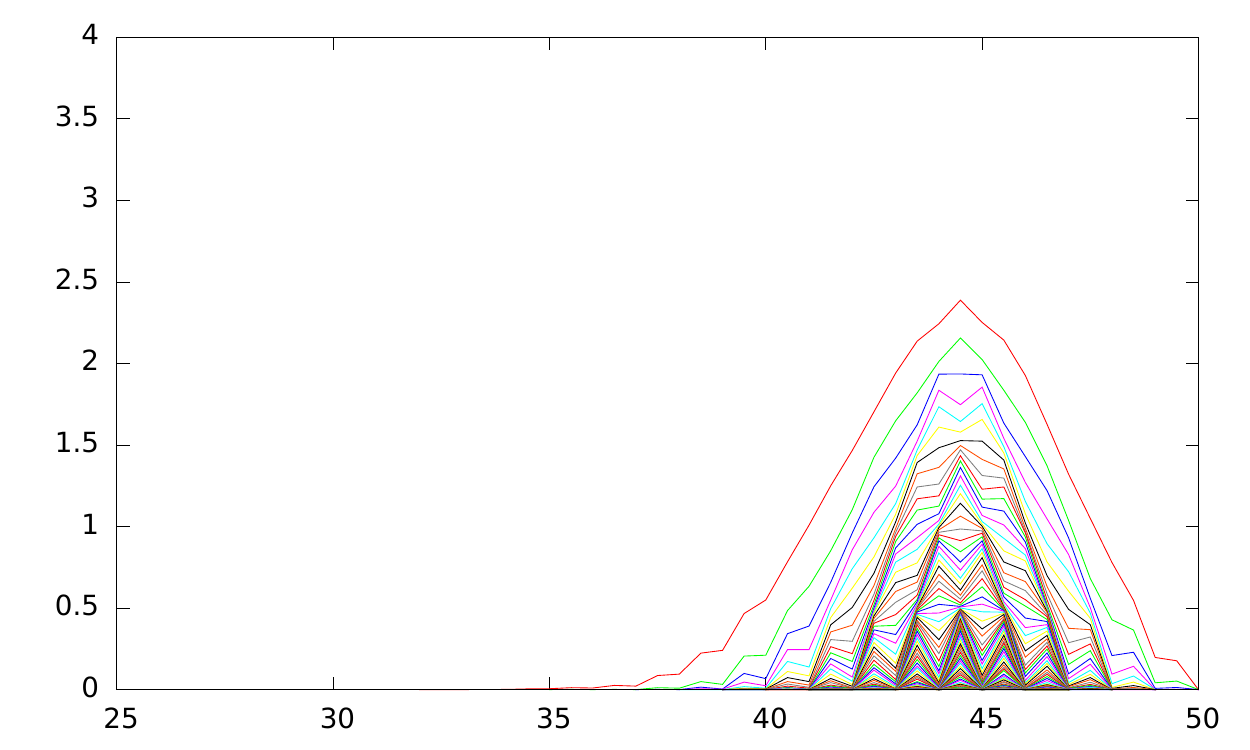} \\ 
dim 8. & dim 9. & dim 10. \\   
\end{tabular}
\caption{Average persistence landscapes in degree $2$ of points sampled from $S^{d}$ for $d \in \{2,\ldots,10\}$. The spheres have been scaled so that the average distance between points is one.}
\label{fig:avLandDim2}
\end{figure}

\begin{figure}[h]
\begin{tabular}{ccc}
\includegraphics[scale=0.35]{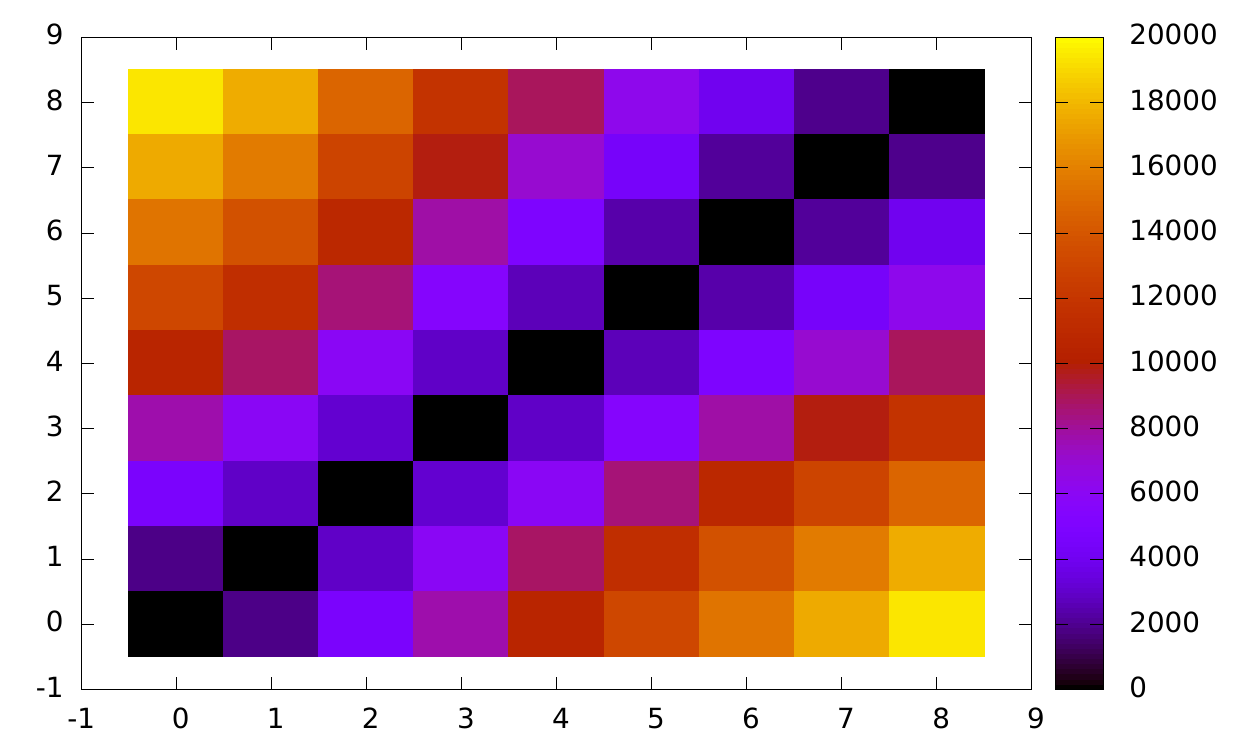}  & \includegraphics[scale=0.35]{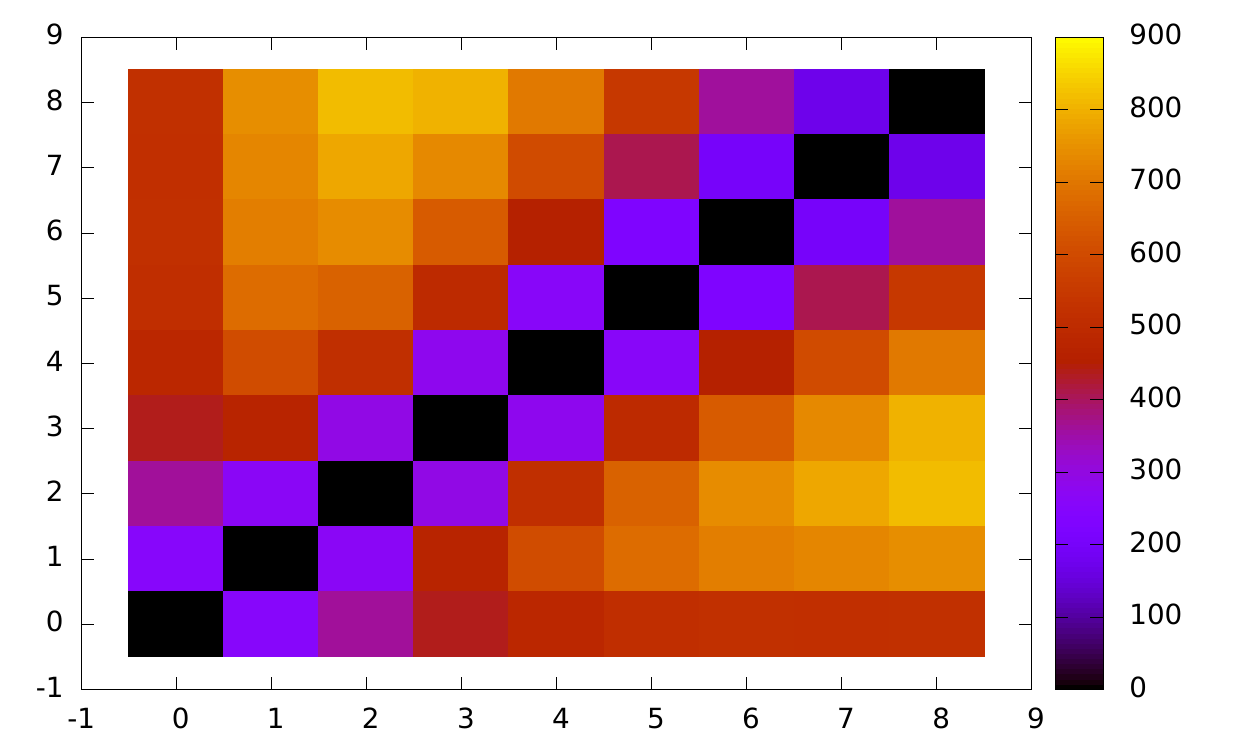} & \includegraphics[scale=0.35]{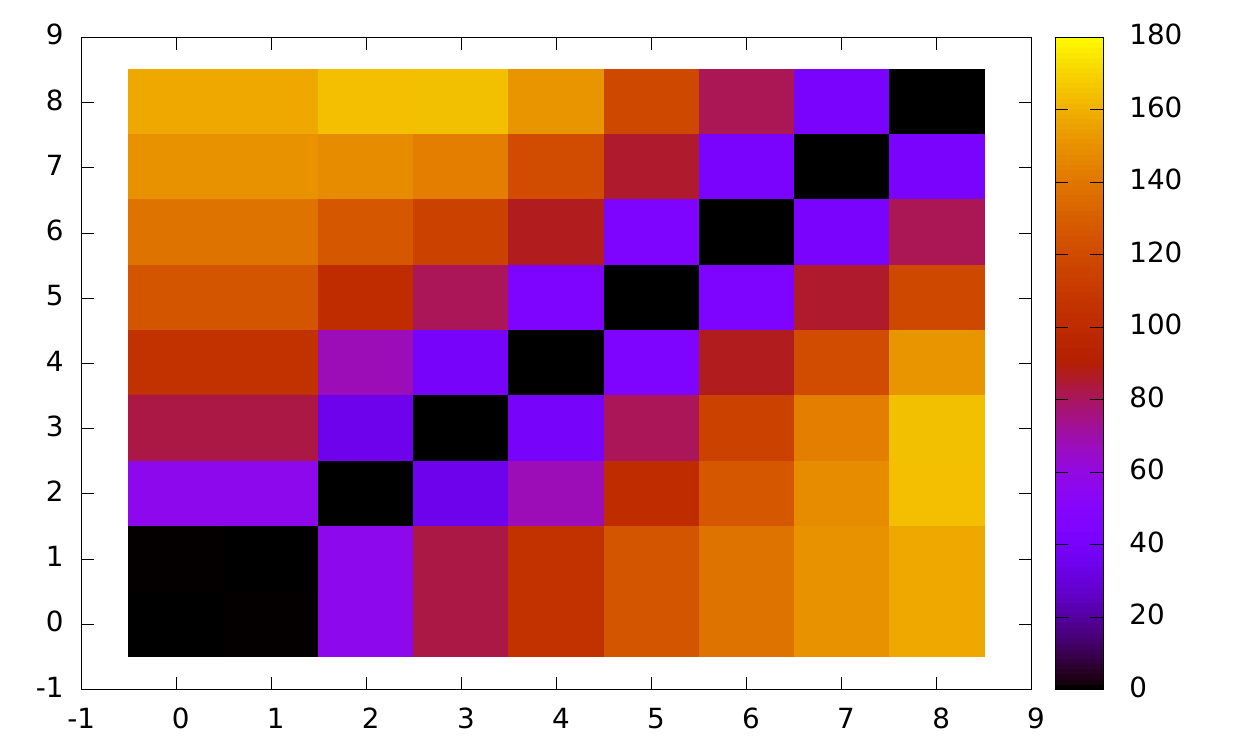} \\    
\includegraphics[scale=0.35]{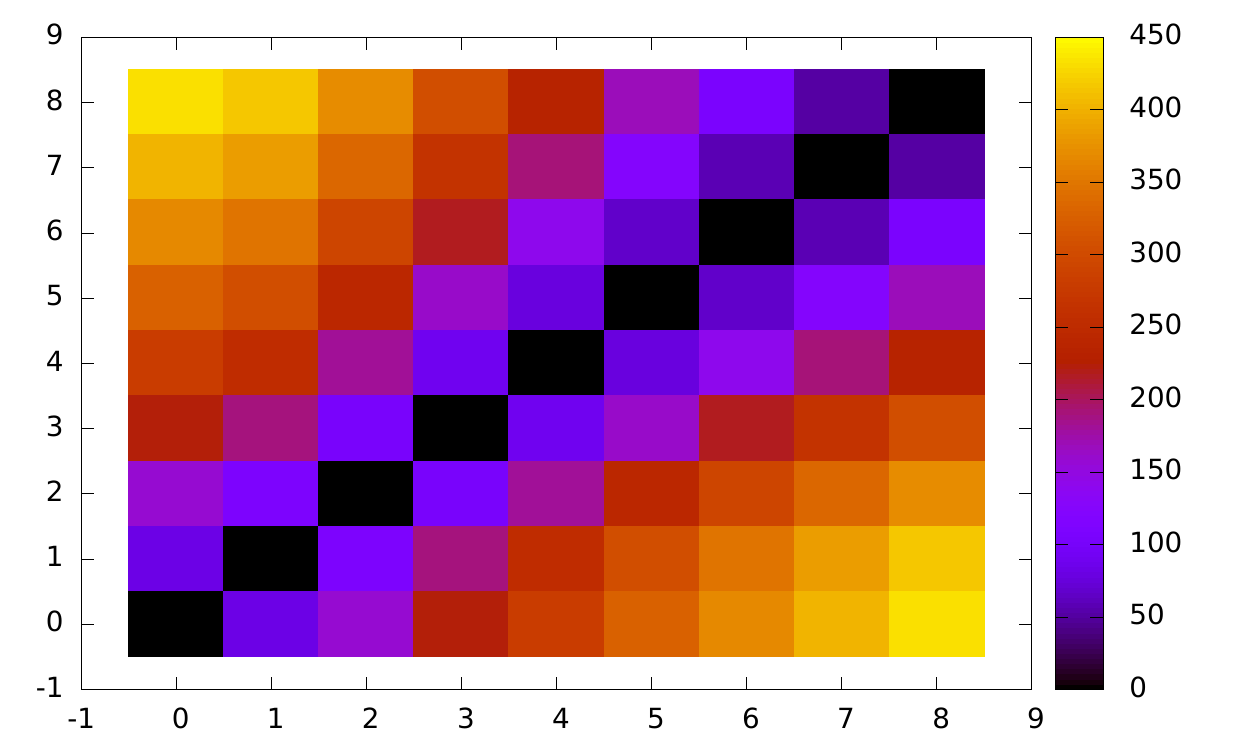} & \includegraphics[scale=0.35]{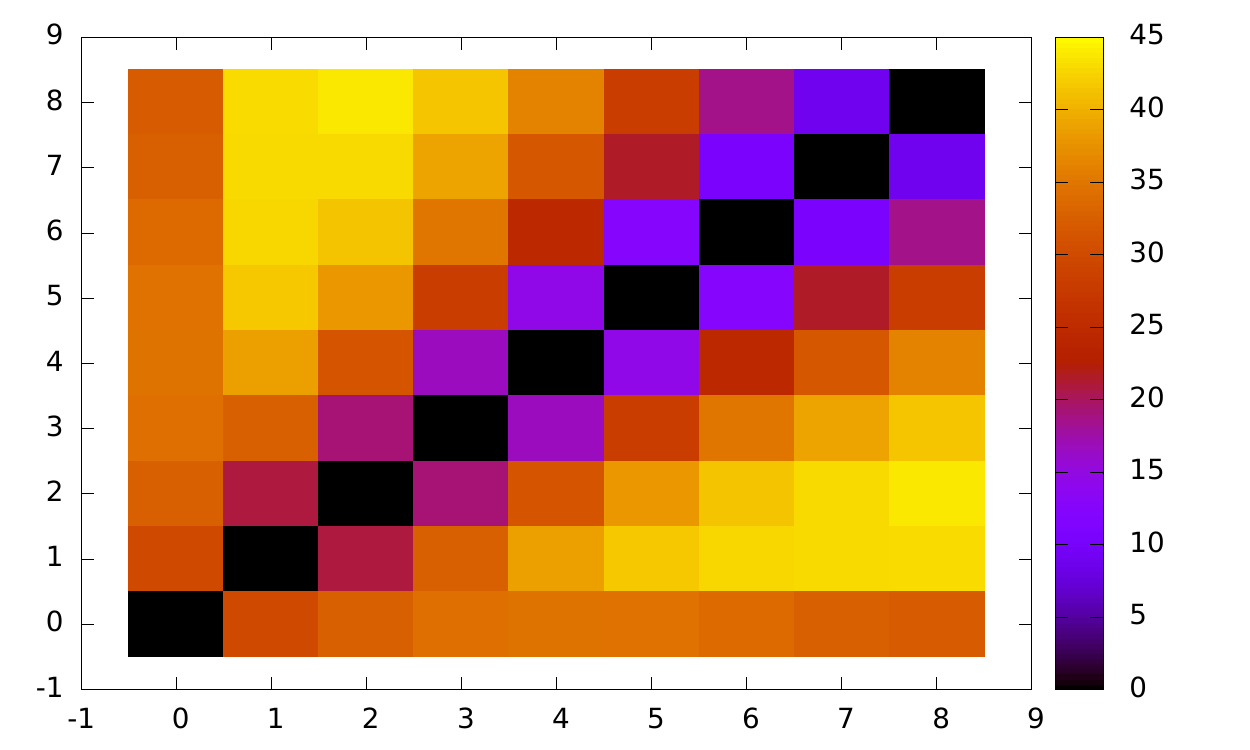} & \includegraphics[scale=0.35]{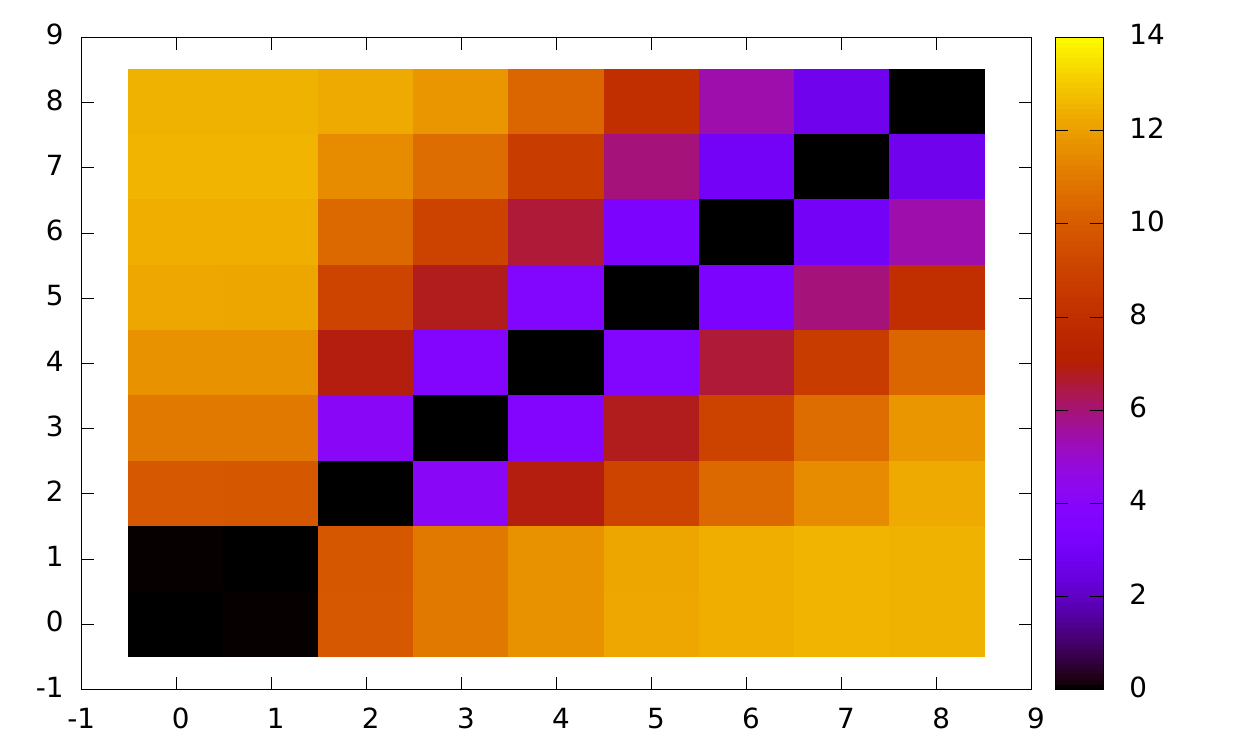} \\ 
\includegraphics[scale=0.35]{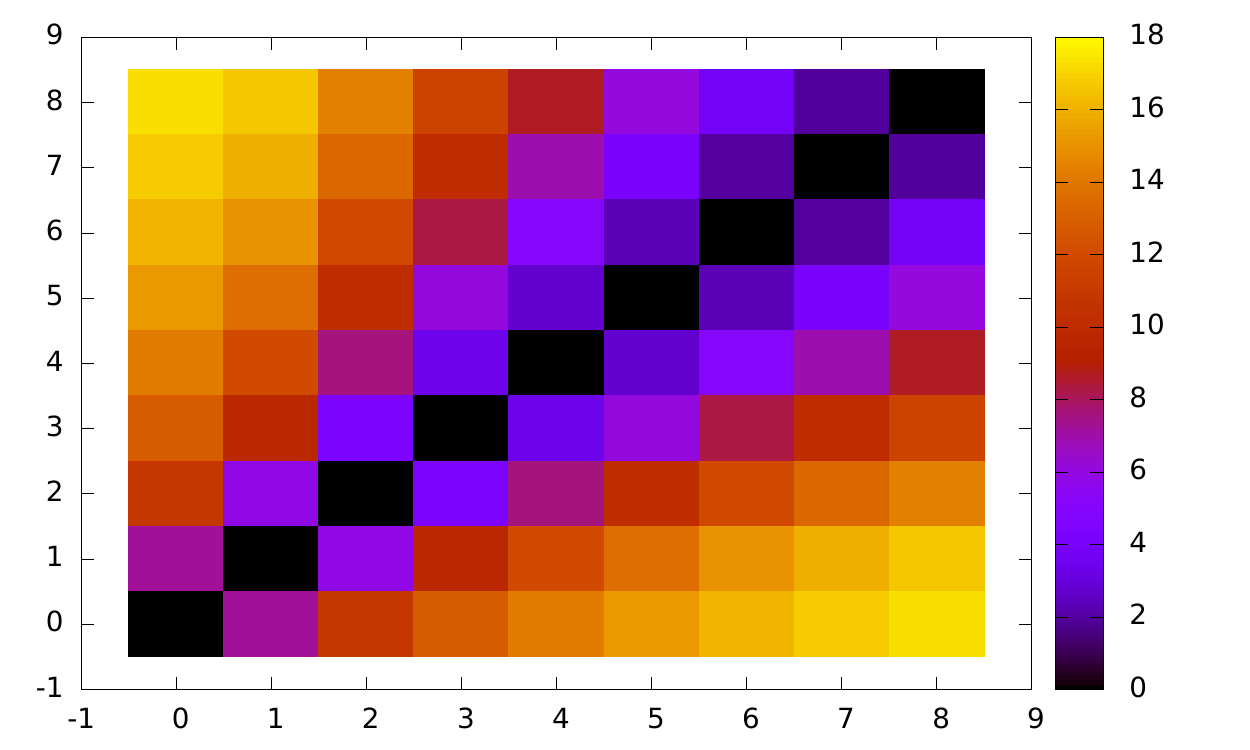} & \includegraphics[scale=0.35]{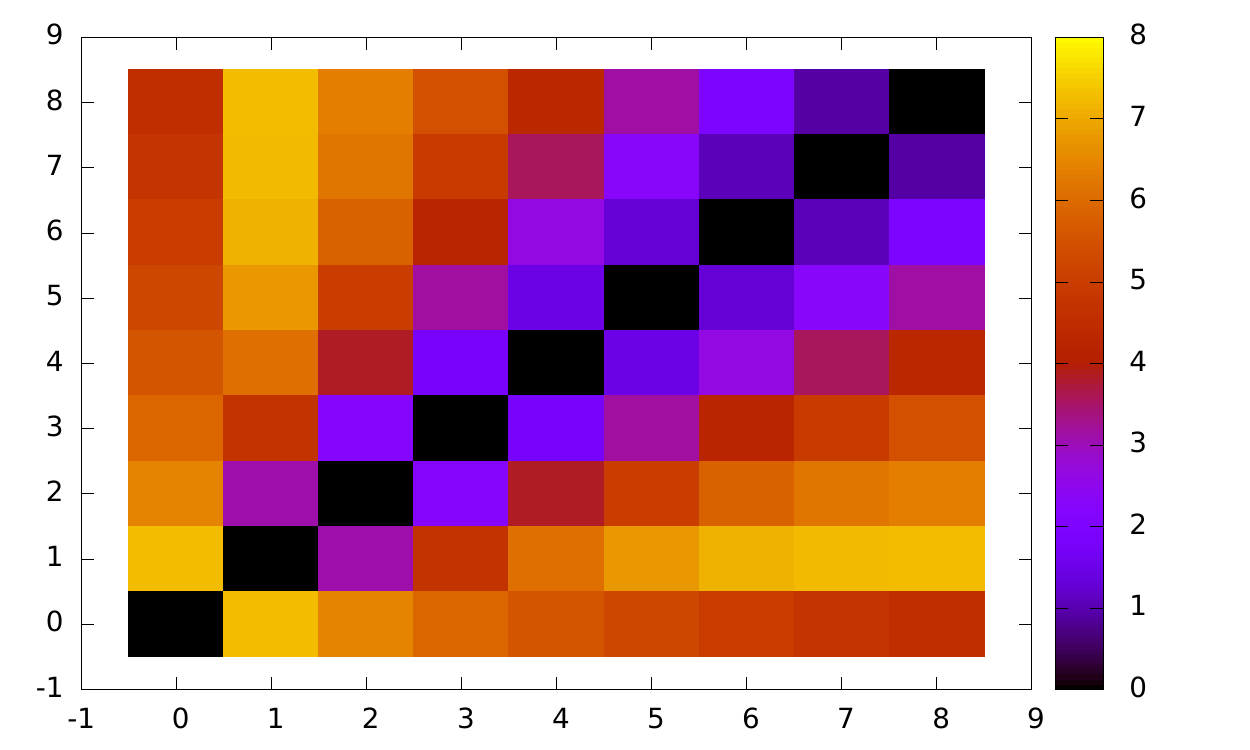} & \includegraphics[scale=0.35]{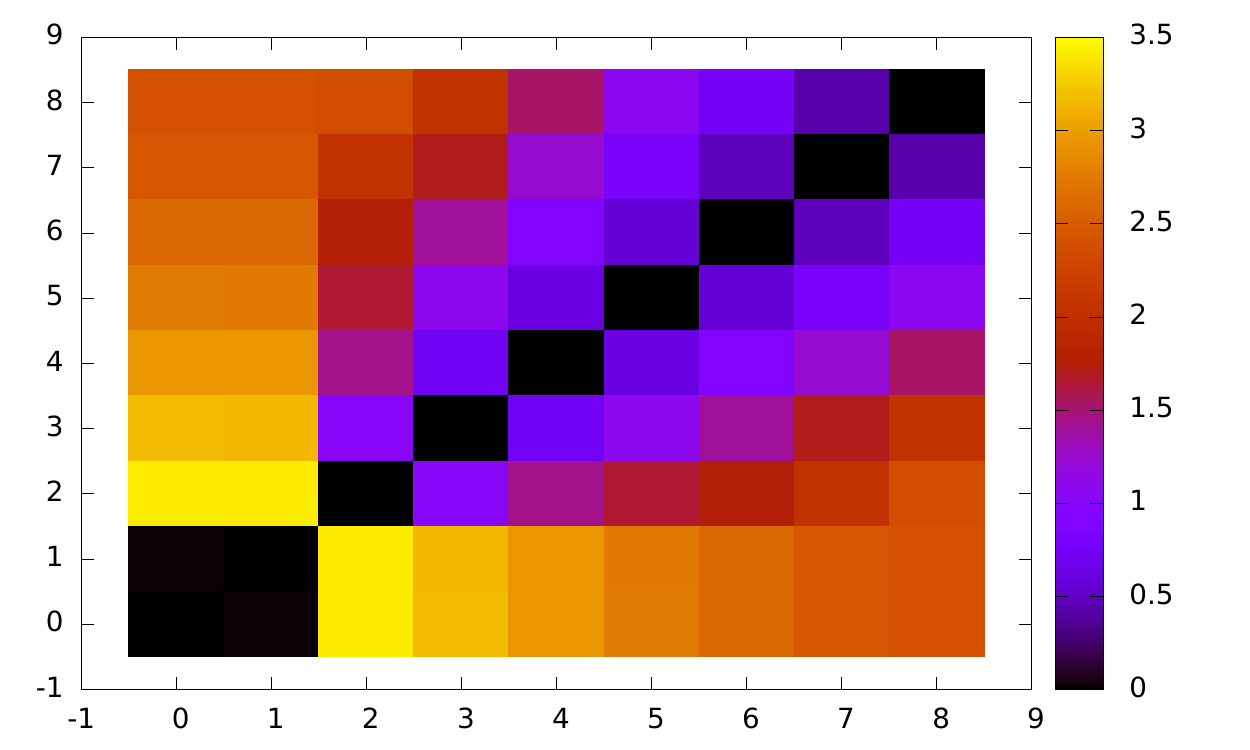} \\ 
\end{tabular}
\caption{Color plots of $L^1$ (top), $L^2$ (middle) and $L^{\infty}$ (bottom) distance matrices for average persistence landscapes for points sampled from spheres in dimensions $\{2,\ldots,10\}$ normalized so that average distance between points is $1$, for homological degree $0$ (left), $1$ (middle) and $2$ (right).}
\label{fig:dist-mat-color}
\end{figure}

The resulting
average persistence landscapes for degree zero, one and two are in
Figure~\ref{fig:avLandDim0}, Figure~\ref{fig:avLandDim1}, and
Figure~\ref{fig:avLandDim2}, respectively. The $L^1$, $L^2$ and
$L^{\infty}$ distances between these average landscapes given in
Figure~\ref{fig:dist-mat-color}.

To determine the significance of these distances between average landscapes we performed a permutation test~\cite{wasserman:book-statistics}, which we now explain.
First choose a significance level $\alpha=0.05$.
For
every pair $i,j \in \{2,\ldots,10\}$ with $i \neq j$, the two
corresponding sets of $1000$ persistence landscapes were combined in a
set of cardinality $2000$. Then this set was randomly split into
two subsets $A_1, A_2$ of cardinality $1000$ each. Next the average
landscapes $\lambda_1$ and $\lambda_2$ were computed based on
the landscapes in $A_1$ and $A_2$, respectively. 
Let $\delta$ denote the distance between the
original average landscapes in dimensions $i$ and $j$. The distance
between $\lambda_1$ and $\lambda_2$ is compared to $\delta$. 
This described process is repeated $10,000$ times.
The $p$ value equals the proportion of cases 
in which the distance between $\lambda_1$ and $\lambda_2$ is greater
than $\delta$.
For every pair $i \neq j$, it  never happened that the
distance between $\lambda_1$ and $\lambda_2$ was greater than the corresponding
$\delta$. Therefore we can conclude that there is very strong
statistical difference between the persistence landscapes in various
dimensions.

We believe that the ability to easily perform such calculations will be 
useful in topological data analysis.

\subsection{Points sampled from $[0,1]^d$}
\label{sec:pointsFromBoxes}
Given the success of the experiment presented in the Section~\ref{sec:pointsFromSpheres} it is natural to ask analogous question for the case of points sampled from a d-dimensional box $[0,1]^d$. Theorems on the random Vietoris-Rips and \v{C}ech complexes can be found in~\cite{KahleMeckes}. The results presented there indicate that Betti numbers of random complexes ``do not live together''; see Figure~1 in~\cite{KahleMeckes}. Here we focus not on the Betti numbers and their limit distribution, but on a task of dimension detection based on persistent homology. 

 For $d \in \{2,3,\ldots,10\}$, to get a single point cloud, we sampled $100$ points from $[0,1]^d$ by sampling each coordinate of each of the points independently from an interval $[0,1]$. For each dimension $d \in \{2,3,\ldots,10\}$, we sampled $1000$ point clouds. As in the previous section, to refute the charge that our detected differences are only due 
to the increase of the average distance between points in higher dimensions, we rescaled the boxes so that the average distance between points in each box is $1$. We computed persistent homology in dimension $0$ and $1$ of the resulting Rips complexes. We obtained $1000$ persistence diagrams in dimension $0$ and $1$ for each dimension $d \in \{2,3,\ldots,10\}$. 

The permutation test was run for $10,000$ permutations for the obtained persistence intervals separately in dimension $0$ and in dimension $1$. In all  cases, it never happened that the distance between shuffled sets was greater than between the original ones. 

This experiment, together with the one presented in the Section~\ref{sec:pointsFromSpheres}, show that low dimensional persistent homology can be a good tool for estimating the intrinsic dimension of data sets. 
Knowing the intrinsic dimension of the dataset is important, since when working with point clouds embedded in high dimensional spaces, it is typically assumed that the intrinsic dimension of the data is low and under this assumption one can overcome the curse of dimensionality. The presented techniques may eventually allow one to explicitly estimate the intrinsic dimension of the considered dataset. 

\subsection{Distance computations} \label{sec:distance-computations}

\begin{figure}[h!tb]
\centering
\begin{tabular}{ | c | c | c | }
\hline
\includegraphics[scale=0.35]{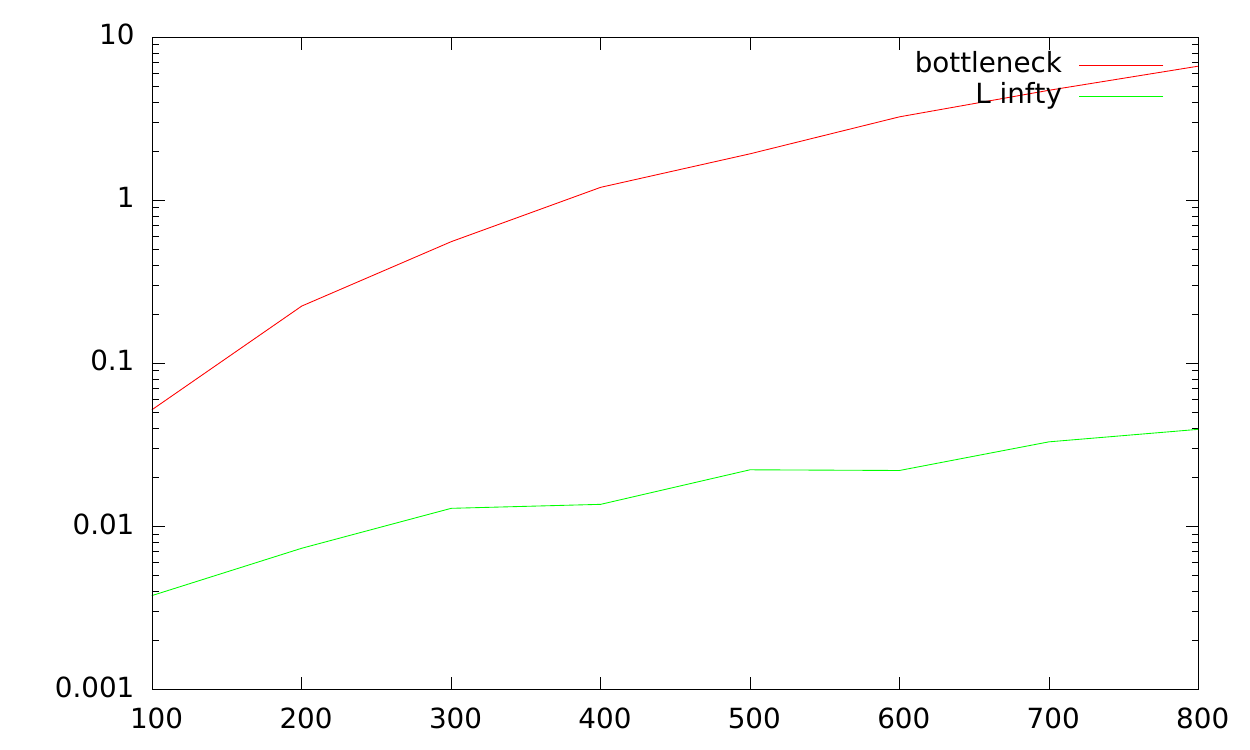}&
\includegraphics[scale=0.35]{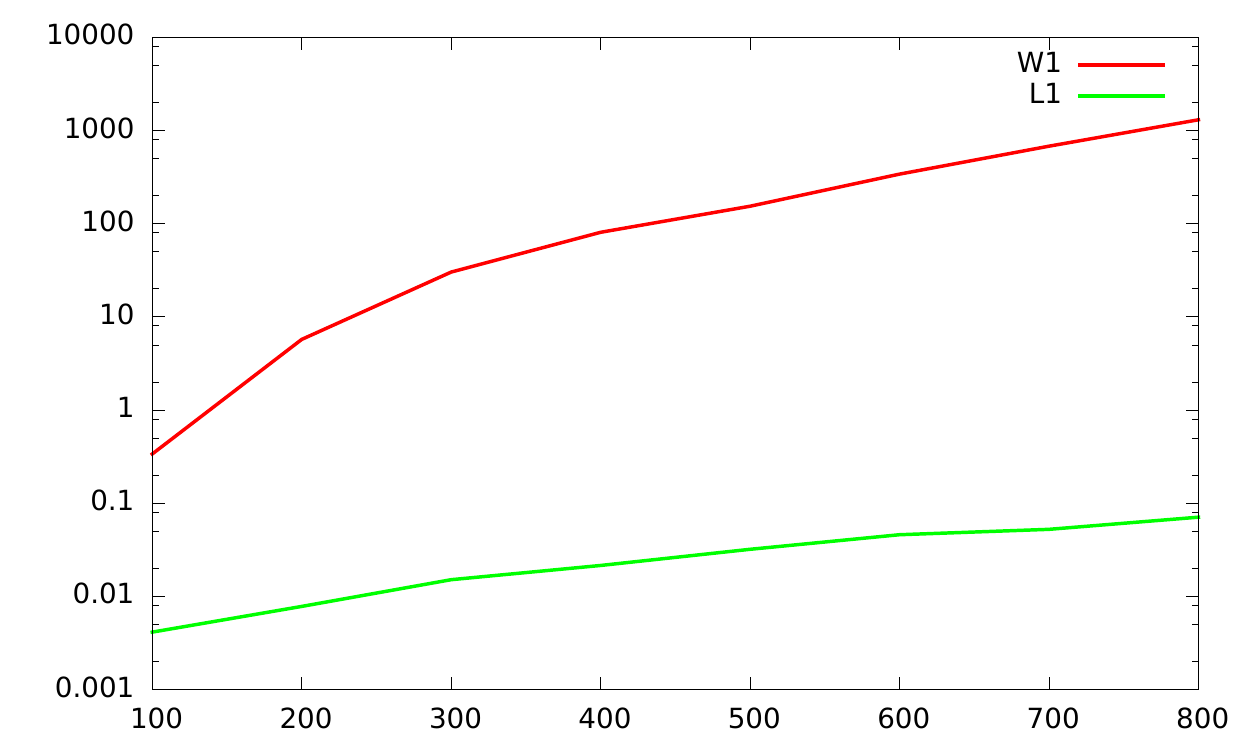} &
\includegraphics[scale=0.35]{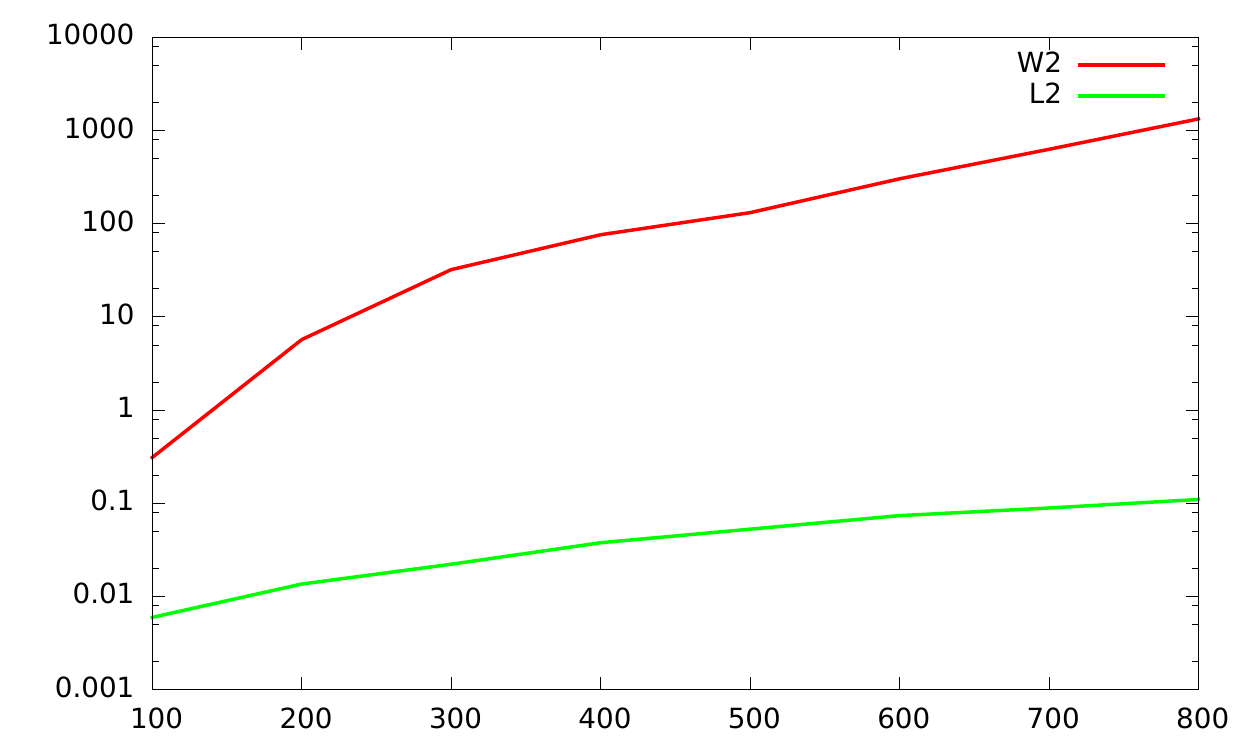} \\
\hline
\end{tabular}
\caption{Comparison of time of distance computations: (a) Bottleneck versus $L^{\infty}$, (b) $W^1$ versus $L^1$, (c) $W^2$ versus $L^2$. Note the logarithmic scale on the vertical axis.}
\label{fig:timeComparision}
\end{figure}

Here we compare our implementations of the persistence landscape distance algorithms with implementations of the bottleneck distance and Wasserstein distance in current use. For each $N \in \{100,200,\ldots,1000\}$  we sampled two random persistence diagrams consisting of $N$ birth-death pairs $(b,d)$ chosen independently from the uniform distribution on $\{0 \leq b \leq d \leq 1\}$. The pairs are chosen in the following way: we sample two points $a,b$ from $[0,1]$ interval. The smaller one became the birth time, the larger one, the death time. 
This was repeated 5 times.
For each $N$ the distance between all pairs of persistence diagrams was computed by using the Bottleneck, 1- and 2-Wasserstein metrics and the $L^{\infty}$, $L^1$ and $L^2$ persistence landscape metric. We used~\cite{dionysous} for the Bottleneck and Wasserstein distance computations. These implementations are known to use sub-optimal/slow algorithms. We are aware of efforts to implement the faster algorithms of~\cite{Efrat01geometryhelps,Agarwal99verticaldecomposition}, but we are unaware of any that are currently available.
We consider the average computation time for each $N$. The comparison of these times is presented in  Figure~\ref{fig:timeComparision}.

\subsection{Computing distance matrices on a random set of persistence intervals}
\label{sec:randomDiagramsDistanceMatrixComputations}

\begin{figure}[h]
\begin{tabular}{ | c | c |}
\hline
\includegraphics[scale=0.5]{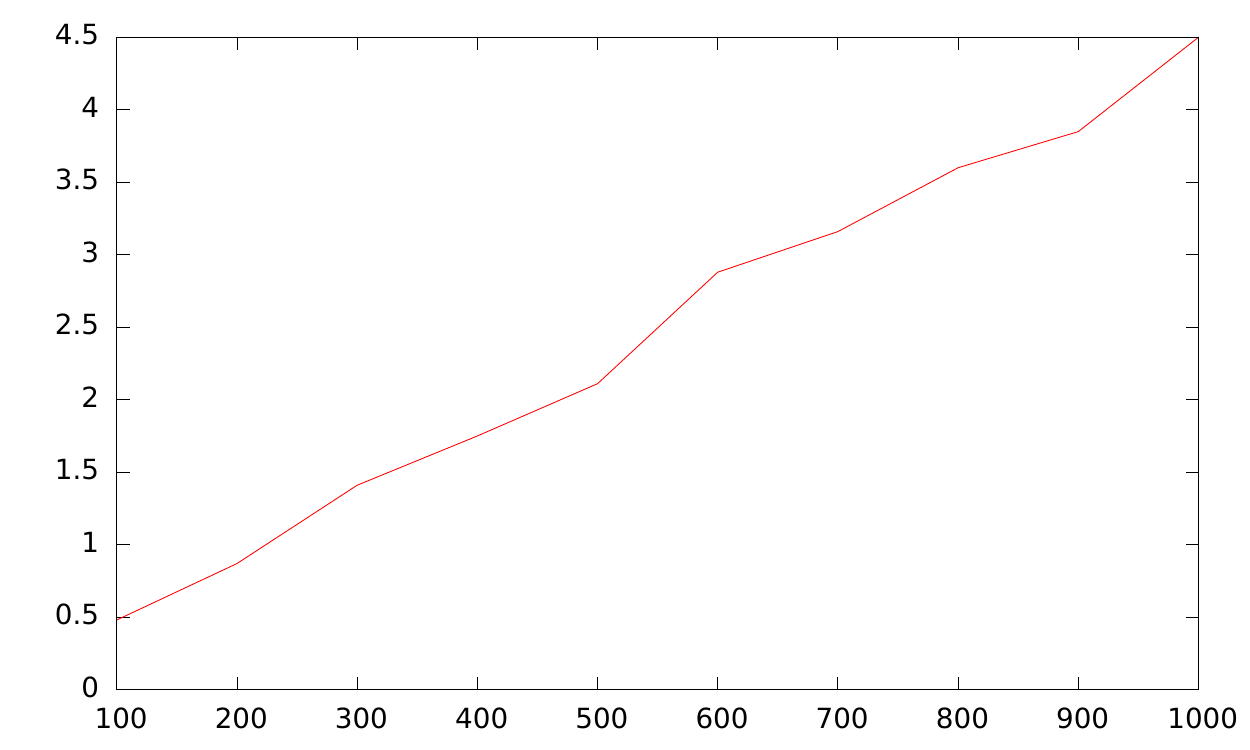}  & \includegraphics[scale=0.5]{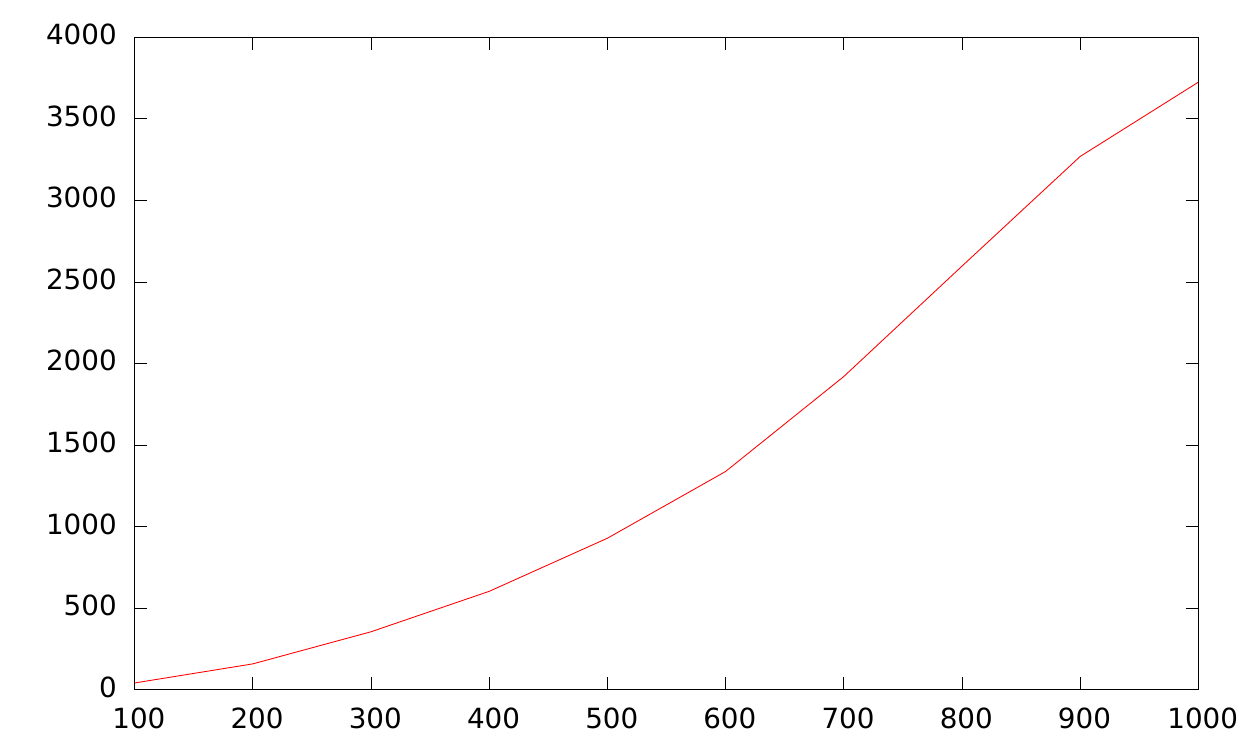}\\ 
\hline
Generation of a landscape & Computation of a distance matrix\\
\hline
\end{tabular}
\caption{Computation times for the distance matrix computations described in the Section~\ref{sec:randomDiagramsDistanceMatrixComputations}. Left: the average time to calculated the persistence landscape from a given number of birth-death pairs. Right: the time to calculate the distance matrix of $1000$ such persistence landscapes.}
\label{fig:randomDiagramsDistances}
\end{figure}

In this experiment,
for each $N \in \{100,200,\ldots,1000\}$, we generated $1000$ random collections of $N$ birth-death pairs, where the pairs $(b,d)$ chosen independently from the uniform distribution on $\{0 \leq b \leq d \leq 1\}$ as in Section~\ref{sec:distance-computations}.
Our aim was to compute the persistence landscape distance matrix for each of these and to see how the computation time scales with the number of birth-death pairs. The computation times are illustrated in the Table~\ref{fig:randomDiagramsDistances}.

\section{Specification of the implementation} \label{sec:specification}

An implementation of the presented procedures is available at the web page \url{http://hans.math.upenn.edu/~dlotko/persistenceLandscape.html}. 

The library is available as a programming tool and can be easily maintained by users who know C++. However our aim is also to provide easy to use tools for  users who are not familiar with programming. Therefore a number of programs that use the library have been created for the user's convenience. Also in the future, when there is such demand, we plan to add new programs to this collection. In this section, we describe the programs currently available and illustrate how to use them on a toy example. 

These programs have been compiled for Linux, Windows and OS X operating systems, and are also available in the package.
All the results described in this Section were obtained by using these programs.

\subsection{Input  and output files}
\label{sec:inputOutputFiles}

The input data to these programs has one of three forms.
It may be a file containing a \emph{persistence diagram} -- a list of birth-death pairs.
For example, the barcode $\{(1,4), (2, 3)\}$ is encoded as follows.
\begin{verbatim}
1 4
2 3
\end{verbatim}
It may also be a file containing a persistence landscape encoded as a sequence of critical points in the following form. The first integer denotes the degree of the persistence diagram from which the landscapes have been created. Then the sequence of critical points of $\lambda_i$ follows after a string $\verb+#lambda_i+$. Below, is an example of the file format for the persistence landscape corresponding to the persistence diagram $\{(1,4), (2,3)\}$  in degree zero.
\begin{verbatim}
0
#lambda_0
1 0 
2.5 1.5
4 0
#lambda_1
2 0
2.5 0.5
3 0

\end{verbatim}
More generally, such an input file may encode a linear combination of persistence landscapes.
Finally it may be a file containing a list of names of files containing either birth-death pairs or linear combinations of persistence landscapes. Due to its generality, this is a typical input for programs presented in this library. 

The output files consist of persistence landscapes and linear combinations of persistence landscapes as described above.

\subsection{Toy example} \label{sec:toy-example}

\begin{figure}[h!tb]
\centering
\includegraphics[scale=0.35]{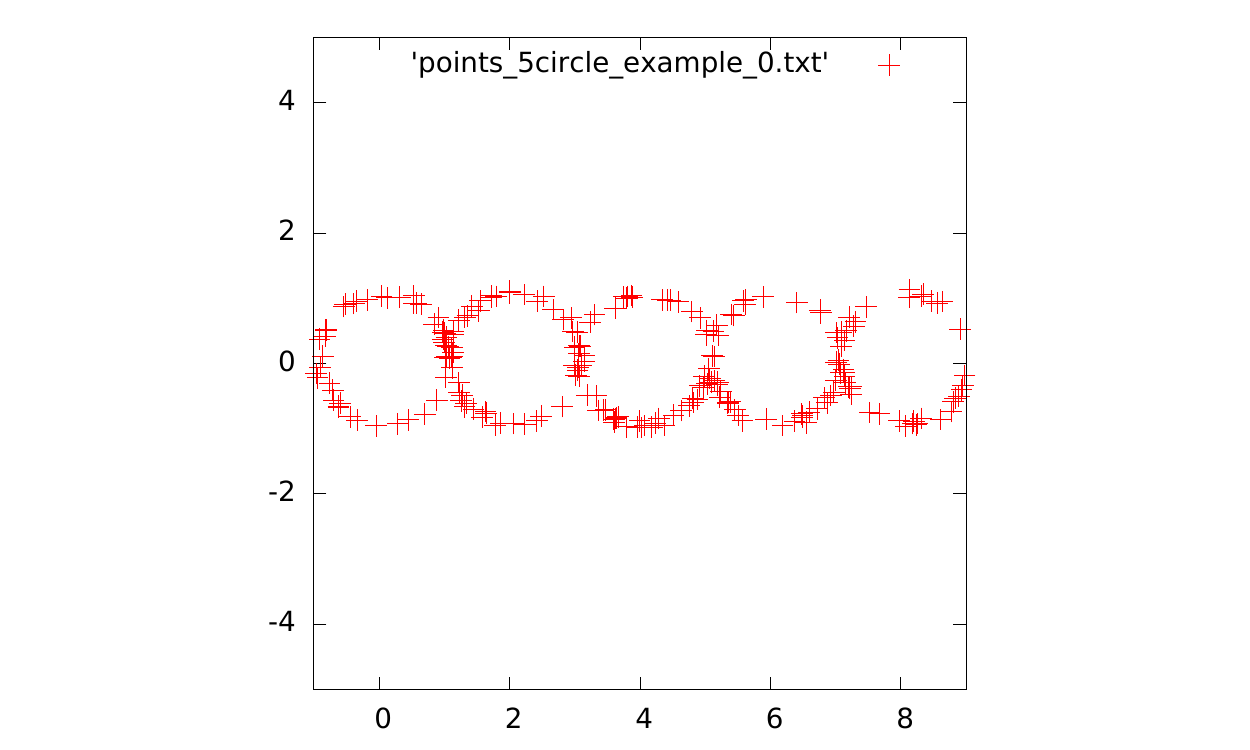}
\caption{Sample point cloud from 5-circle directory.}
\label{fig:5circle}
\end{figure}

Our toy test example for these programs is a set of files consisting of birth-death pairs, which we now describe.
For $n \in \{1,2,\ldots,5\}$, let $A_n$ denote the union of the circles of radius one centered at $(0,0), (2,0), \ldots, (2n,0)$.
From $A_n$ we sampled $50n$ points independently using the uniform measure. To each of these points we added a uniform error sampled from $[-0.15,0.15]^2$, see Figure~\ref{fig:5circle} for example.

From this set of points, we calculated the persistent homology in degrees zero and one of the corresponding Vietoris-Rips complex using Perseus~\cite{perseus}.
This was repeated $11$ times for each $n$. The results are encoded in $5$ directories (named \verb+1circle+,\ldots,\verb+5circle+) each of which contains $2\times 11=22$ files that each encode a persistence diagram of either degree zero or one.
For the further use in the programs the files containing homology in a particular degree are listed in a files \verb+1circle_dim1.txt,...,5circle_dim1.txt+ .
For example, the file \verb+1circle_dim1.txt+ lists the following files, each of which contains a persistence diagram.
\begin{verbatim}
1circle/points_1circle_example_0_persistence_1.txt
1circle/points_1circle_example_1_persistence_1.txt
1circle/points_1circle_example_2_persistence_1.txt
1circle/points_1circle_example_3_persistence_1.txt
1circle/points_1circle_example_4_persistence_1.txt
1circle/points_1circle_example_5_persistence_1.txt
1circle/points_1circle_example_6_persistence_1.txt
1circle/points_1circle_example_7_persistence_1.txt
1circle/points_1circle_example_8_persistence_1.txt
1circle/points_1circle_example_9_persistence_1.txt
1circle/points_1circle_example_10_persistence_1.txt
\end{verbatim}

\subsection{Average Persistence Landscapes}
\label{sec:averagesDemo}
Let us describe \verb+ComputeAverage+. The input parameter is a file containing a list of files containing either persistence diagrams or persistence landscapes from which an average landscape will be calculated. For example, 
one can call the program \verb+computeAverages+ \verb+1_circle_dim1.txt+. 
As a result, a file containing the average landscape is produced. The same procedure is tested for the remaining files and as a result we obtain ten files with average landscapes.

\subsection{Plots of landscape(s)}
\label{sec:plotsDemo}
In this section, we demonstrate how to generate output that can be used to plot landscapes. The presented software does not have a built-in graphical engine, so it instead creates gnuplot-readable files. To plot these, the user should install gnuplot~\cite{gnuplot}. The program \verb+PlotOfLandscape+
takes as its first parameter the file name of a file containing either a persistence diagram, a persistence landscape, or a linear combination of persistence landscapes. The remaining two parameters give the range of landscapes which should be plotted. For a range $a$, $b$, where $a,b \in \mathbb{N}$ and $a < b$ the functions $\lambda_a$, $\lambda_{a+1},\ldots,\lambda_{b-1}$ will be plotted. 

\begin{figure}[h]
\begin{tabular}{ccc}
\includegraphics[scale=0.3]{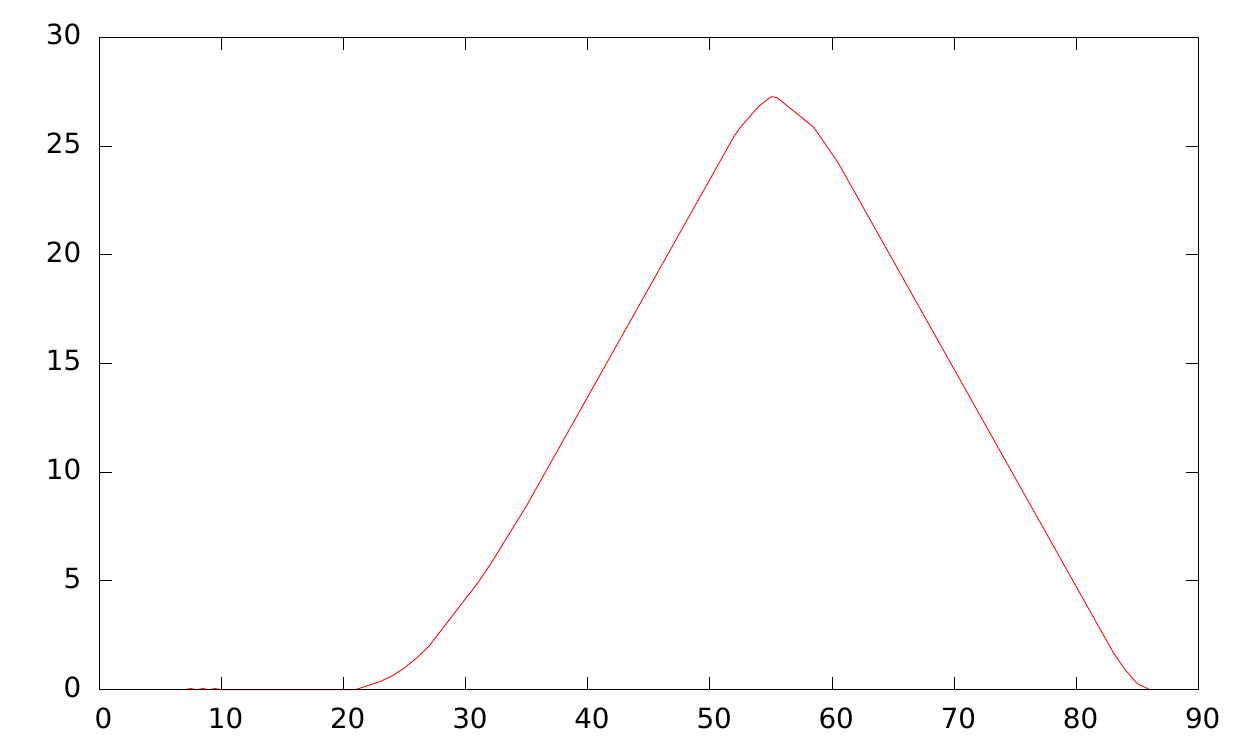} & \includegraphics[scale=0.3]{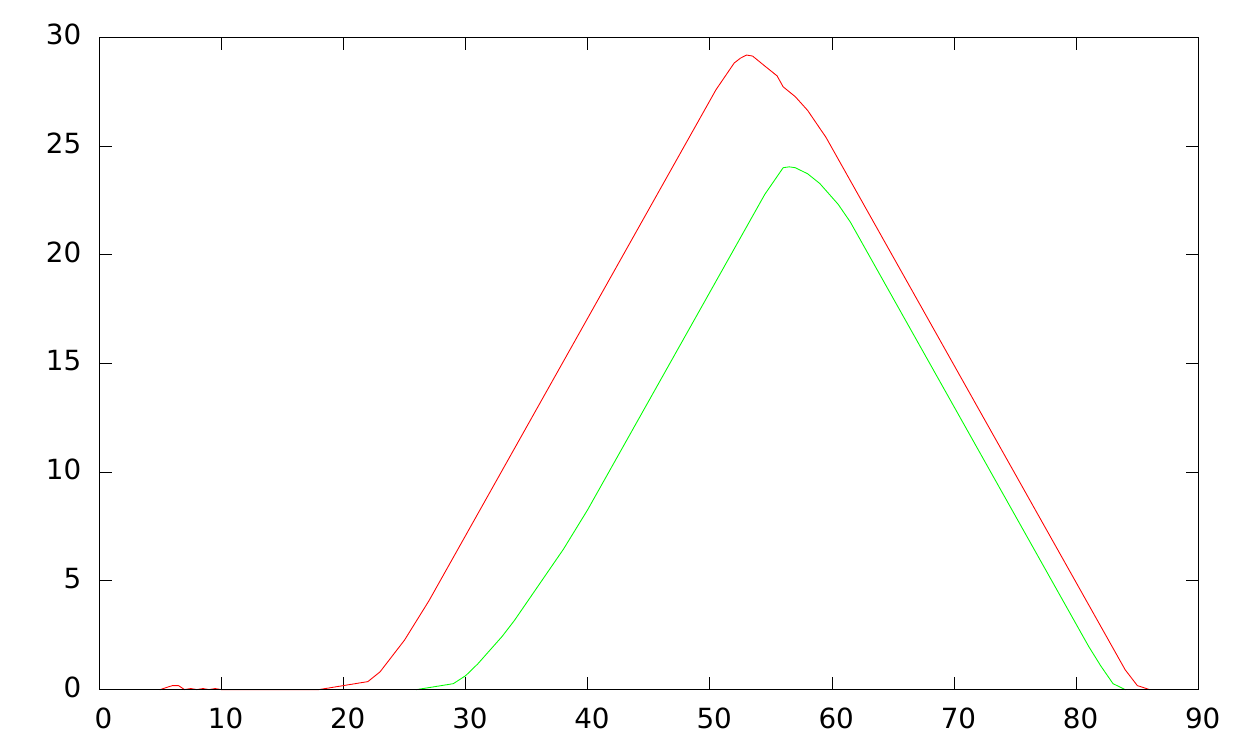} & \includegraphics[scale=0.3]{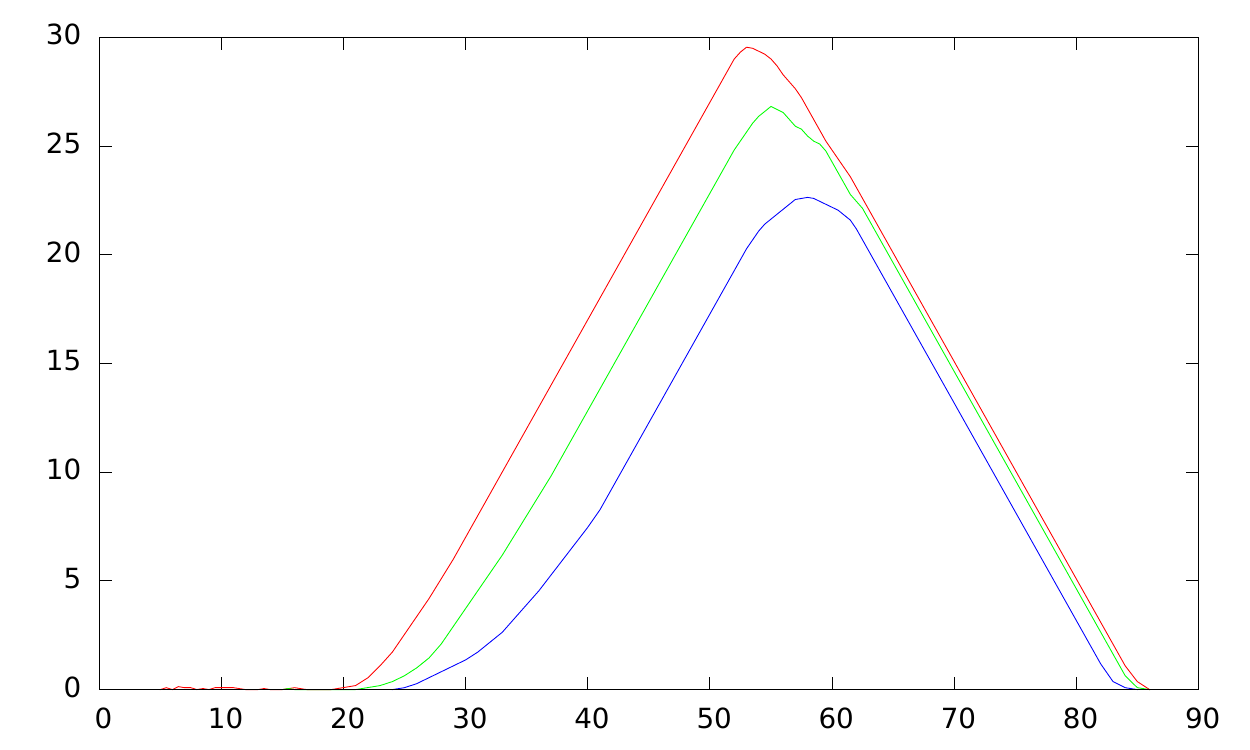} \\ 
1 circle & 2 circles & 3 circles \\   
\includegraphics[scale=0.3]{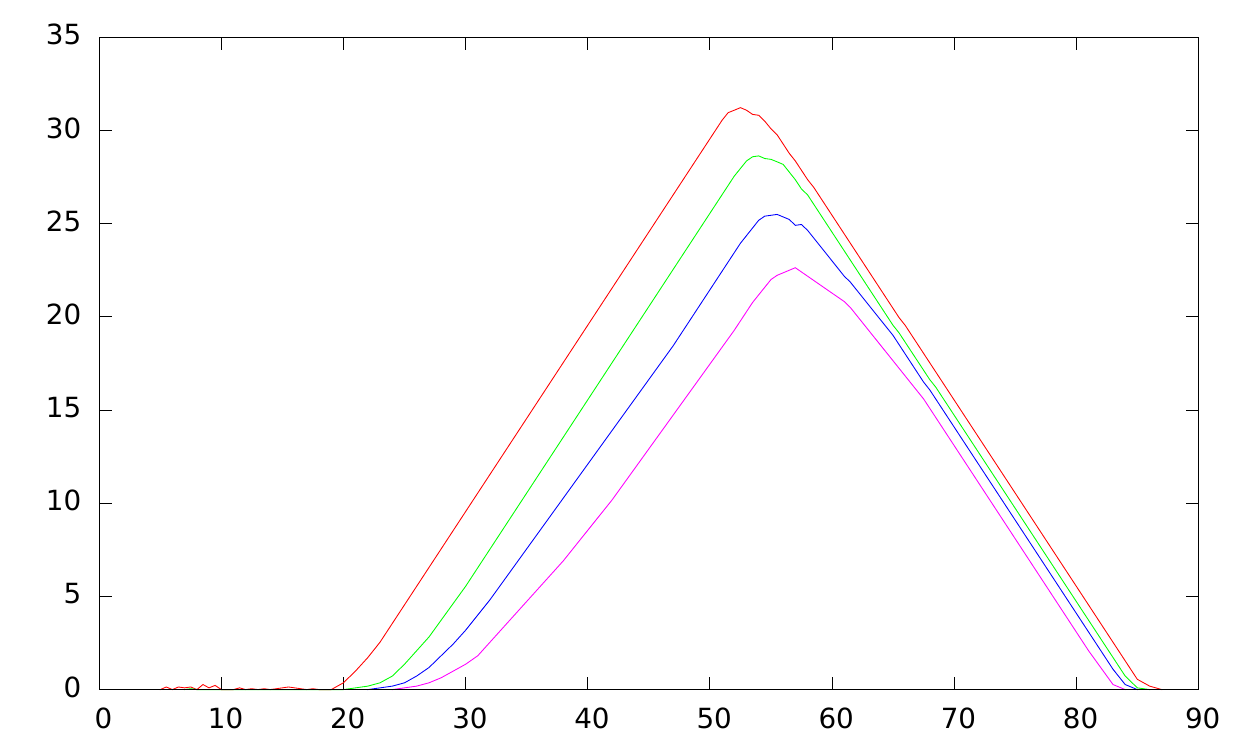} & \includegraphics[scale=0.3]{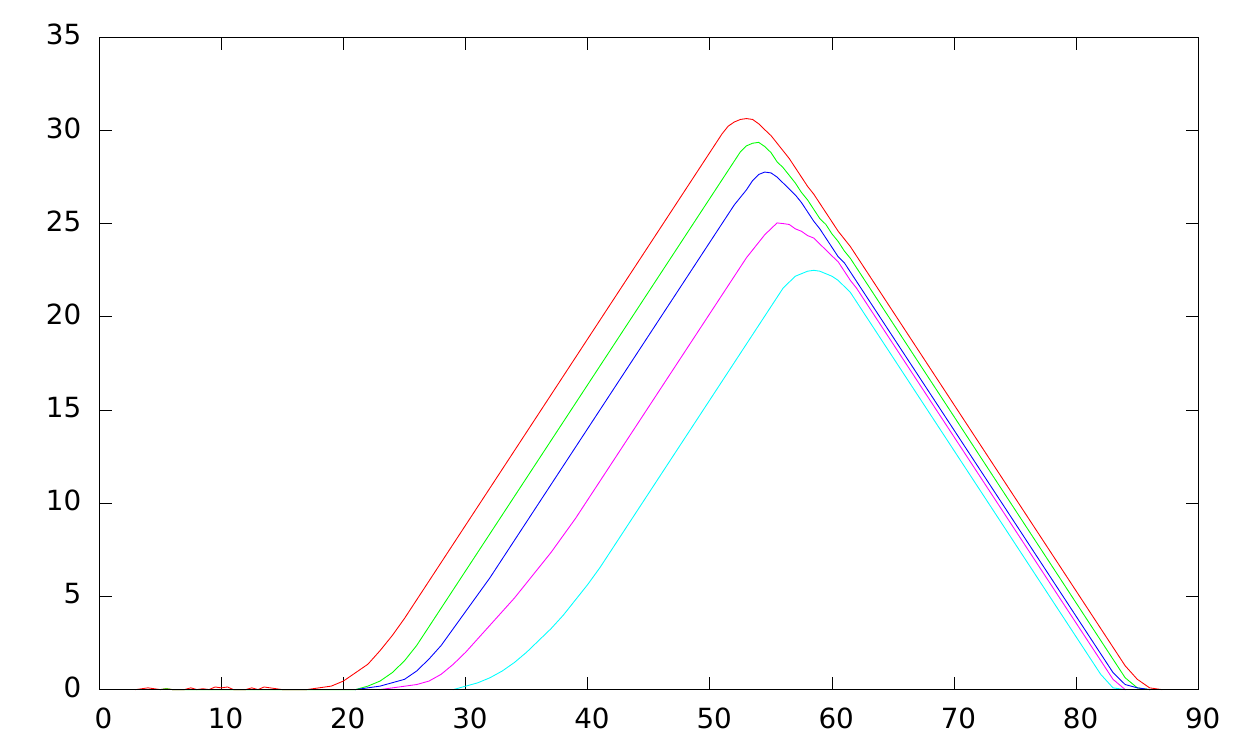} &  \\ 
4 circles & 5 circles & 
\end{tabular}
\caption{Average persistence landscapes in degree 1 from the toy example in Section~\ref{sec:toy-example}.}
\label{fig:averages}
\end{figure}

This program was used for each of the average landscapes computed in the Section~\ref{sec:averagesDemo} to obtain the plots in  Figure~\ref{fig:averages}. 

The program \verb+PlotsOfLandscapes+ has as its first parameter the name of a file listing files containing persistence landscapes, with the remaining parameters the same as \verb+PlotOfLandscape+. It was used to produce Figures \ref{fig:avLandDim0}, \ref{fig:avLandDim1} and \ref{fig:avLandDim2}.

\subsection{Norms of landscapes}
\label{sec:integralsDemo}
The program \verb+normsOfLandscapes+ computes norms of persistence landscapes and linear combinations of persistence landscapes. 
The parameters of the program are:
\begin{enumerate}
\item the name of a file containing names of files containing persistence diagrams, persistence landscapes, or linear combinations of persistence landscapes; and
\item a real $p\geq 1$, or $p=-1$ indicating which norm we want, where $p=-1$ represents the supremum norm.
\end{enumerate}
The output of the program is a list of the $p$-norms of the input landscapes. 

Here we compute the $p$-norms of the average landscapes in degree one from Section~\ref{sec:averagesDemo}.

\begin{center}
\begin{tabular}{| c | c | c | c |}
   \hline
    & $p=1$ & $p=2$ & $p = \infty$ \\
   \hline
  1-circle & $846.959$ & $126.715$ & $27.3333$ \\
  \hline
  2-circle & $1610.31$  & $174.102$ & $29.0833$ \\
  \hline
  3-circle & $2462.29$ & $214.864$ & $29.6667$ \\
  \hline
  4-circle & $3421.15$ & $256.394$ & $31.3333$ \\
  \hline
  5-circle & $4080.41$ & $278.459$ & $30.5833$ \\
  \hline
\end{tabular}
\end{center}

\subsection{Distance matrix}
\label{sec:distancesDemo}
In this section, we illustrate the usage of the program \verb+DistanceMatrix+. 
The parameters of the program are:
\begin{enumerate}
\item A file with names of files containing persistence diagrams or persistence landscapes or linear combinations persistence landscapes.
\item An integer p. If $p \geq 1$, then the $L^p$ distance between landscapes will be computed. If $p = -1$, the $L^{\infty}$ distance will be computed.
\end{enumerate}

The output is a distance matrix in a text file.

For example,
calling the program \verb+DistanceMatrix+ with the parameters \verb+1_circle_files.txt 2+ produces the following.
\begin{verbatim}
0	106.729	161.55	207.872	239.638
106.729	0	104.46	160.555	199.336
161.55	104.46	0	105.577	152.975
207.872	160.555	105.577	0	97.8873
239.638	199.336	152.975	97.8873	0
\end{verbatim}

\subsection{Permutation test}
\label{sec:permutationTestDemo}
The permutation test was explained in the Section~\ref{sec:pointsFromSpheres}. 
Our program \verb+PermutationTest+ performs a permutation test for each pair in a list of lists of persistence diagrams or persistence landscapes, and outputs a matrix of p-values.
Its parameters are
\begin{enumerate}
\item a positive integer $M$ being the number of files with the input persistence intervals or landscapes;
\item $M$ names of files with the input persistence diagrams or persistence landscapes. Each file is supposed to contain data from the same class, two different files are supposed to contain data from two different classes,
\item Positive integer $N$ indicating the number of tries in the permutation test,
\item An real number $p\geq 1$ indicating which distance is to be used in the procedure (or $p=-1$ for the $L^{\infty}$ distance). 
\end{enumerate}
Please note that this procedure usually takes a lot of time. Therefore, after each step a message is outputted to the screen so the user can verify that the program is progressing.

 As a sample test we have used all the files with persistence intervals for the $N$circle for $N \in \{1,\ldots,5\}$. In this case all p-values are $0$ which is what one should expect given how different the persistence landscapes of the different classes are.

\subsection{A classifier based on a single dimension}
\label{sec:classifierSingleDimDemo}
In this section, present a simple implementation of a nearest-neighbor classifier based on persistence landscapes. This is just one possible example of topological statistics in classification. There are many other ways to perform this task using the software described here. For example, one may apply a support vector machine (SVM)~\cite{svm} to our distance matrix. Our classifier is implemented in the program \verb+ClassifierBasedOnSingleDimension+.

The classifier proposed in this paper is based on the following idea. Suppose we are given a training set consisting of $N$ sets of persistence diagrams or persistence landscapes $T_1,\ldots,T_N$. To start, we calculate the average landscape $Av_1,\ldots,Av_N$ for each of these classes. Then a sequence of $M$ landscapes $l_1,\ldots,l_M$ are given to classification. We give two options:
\begin{enumerate}
\item for $l_j$ the program can return an index $i \in \{1,\ldots,N\}$ such that the $L^p$ distance, for a chosen $p$, between $l_j$ and $Av_i$ is the smallest one; or
\item for $l_j$ the program can return a vector of pairs $(distance,number\ of\ class)$ sorted according to the first coordinate.
\end{enumerate}

The usage of the program \verb+ClassifierBasedOnSingleDimension+ is  determined by the first parameter which is one of \verb+-construct+, \verb+-classify+, or \verb+-both+.

If the first parameter of the program is \verb+-construct+, the program will construct the average landscape of each class and write it to a \verb+.lan+ file in the same folder where the program is located. In this case the parameters of the program are:
\begin{enumerate}
\item An integer $N$ indicating the number of classes in the training set,
\item $N$ names of files with each listing files for one of the classes in the training set. 
\end{enumerate}
In this case the program will write $N$ \verb+.lan+ files with the average landscapes of the $N$ input classes. Those average landscapes can be later used by the program to perform classification when the \verb+-classify+ parameter is used.

If the first parameter of the program is \verb+-classify+, it is assumed that the average landscapes have already been created (by using \verb+-construct+ option). In this case the parameters of the program are:
\begin{enumerate}
\item A positive integer $N$ indicating how many classes there are in the considered data,
\item The name of a file listing with names of files which are to be classified,
\item A real number $p\geq 1$ indicating which norm is to be used in classification (with $p=-1$ for the supremum norm),
\item A parameter $q$ valued $0$ or $1$. If $q = 0$, then for each input persistence diagram or persistence landscape the best matching cluster will be computed. If q = 1, the distances to all averages will be calculated.
\end{enumerate}

If the first parameter of the program is \verb+-both+, then the program both computes the averages of the training set and classifies the test set in one run. In this case the parameters of the program are:
\begin{enumerate}
\item A positive integer $N$ indicating how many classes there are in the considered data,
\item $N$ names of files with each listing files for one of the classes in the training set,
\item The name of file listing the names of the files which are to be classified,
\item A real number $p\geq 1$ indicating which norm is to be used in classification, with $p=-1$ indicating the supremum norm,
\item A parameter $q$ valued $0$ or $1$ as above.
\end{enumerate}

In the latter two cases the result of the classification is written to the output file \verb+classification.txt+.

\medskip

In order to test the classifier we have used the persistence diagrams in degree $1$ 
calculated in Section~\ref{sec:toy-example}.
For each class, half of the persistence diagrams were used as a training set and the other half were later classified.

The file listing the files to classify contains the names of all the files which not used in the training set. The files come from \verb+1circle+, \ldots, \verb+5circle+, in order.
The results of the classification which outputs only the best matching element are presented below (the formatting has been adjusted  to make the interpretation easier).
\begin{verbatim}
1 1 1 1 1 1 
2 2 2 2 2 2 
3 3 3 3 3 3 
4 3 4 4 4 4 
5 5 5 5 5 5 
\end{verbatim}
Clearly the classification works very well except from the second case in $4circle$ where we get the wrong result. To understand the reason for the mismatch we went back to the data and it turned out that one of the intervals corresponding to a circle lived for relatively short time.

When we ask for all the matchings sorted from the best to the worst, we obtain the following.
\begin{verbatim}
(1,25.7761) (2,112.082) (3,157.937) (4,221.667) (5,245.942) 
(1,12.6717) (2,106.274) (3,153.505) (4,217.975) (5,242.595) 
(1,30.1331) (2,113.974) (3,159.168) (4,222.937) (5,246.981) 
(1,28.2563) (2,103.435) (3,151.521) (4,214.832) (5,240.057) 
(1,8.75305) (2,105.84) (3,153.331) (4,217.686) (5,242.4) 
(1,16.3392) (2,102.7) (3,151.279) (4,214.909) (5,240.174) 
(2,14.6244) (3,94.3967) (1,103.247) (4,171.567) (5,201.18) 
(2,15.1085) (3,92.8038) (1,110.353) (4,169.847) (5,199.583) 
(2,31.1131) (3,95.2895) (1,114.885) (4,172.342) (5,201.316) 
(2,18.2292) (3,91.73) (1,119.7) (4,168.167) (5,198.167) 
(2,18.1967) (3,97.6503) (1,102.252) (4,175.141) (5,204.504) 
(2,7.55125) (3,95.2824) (1,104.178) (4,172.385) (5,202.242) 
(3,32.8958) (2,111.348) (4,119.572) (5,159.077) (1,164.645) 
(3,47.7613) (2,105.823) (4,133.652) (1,147.63) (5,170.639) 
(3,24.6762) (2,108.863) (4,112.403) (5,153.737) (1,168.976) 
(3,30.7744) (4,111.813) (2,117.509) (5,153.46) (1,169.289) 
(3,51.5379) (4,106.492) (2,134.136) (5,148.542) (1,188.623) 
(3,56.3067) (4,107.379) (2,137.143) (5,148.599) (1,191.934) 
(4,48.5611) (3,93.7973) (5,110.772) (2,146.115) (1,194.061) 
(3,70.4748) (2,82.7482) (4,125.726) (1,141.654) (5,163.054) 
(4,28.5243) (5,100.564) (3,112.561) (2,161.836) (1,206.81) 
(4,42.9319) (5,102.437) (3,149.733) (2,199.963) (1,240.67) 
(4,29.0048) (5,98.0904) (3,134.157) (2,184.933) (1,231.01) 
(4,21.3686) (5,98.1081) (3,121.138) (2,171.799) (1,217.63) 
(5,40.7872) (4,100.327) (3,139.263) (2,180.146) (1,223.318) 
(5,47.0829) (4,119.938) (3,191.276) (2,234.445) (1,272.311) 
(5,28.9991) (4,91.9347) (3,154.484) (2,196.559) (1,234.49) 
(5,38.7583) (4,117.984) (3,172.077) (2,212.901) (1,249.565) 
(5,44.8902) (4,86.6282) (3,131.766) (2,180.086) (1,220.9) 
(5,24.9582) (4,105.167) (3,162.33) (2,203.913) (1,238.713) 
\end{verbatim}
So except for the one outlier, the results are exactly as one would expect. 

\subsection{A classifier based on all dimensions}
\label{sec:classifierAllDimDemo}
We have another classifier, \verb+ClassifierBasedOnAllDimensions+, similar to the one presented in Section~\ref{sec:classifierSingleDimDemo} that uses persistence data from more than one degree.

\section{Exact versus grid based computations.}
\label{sec:rigorousVsGrid}
In this section, we discuss the pros and cons of exact and grid-based computations of persistence landscapes. We do not provide comparison times, since it is not clear what dataset is the right benchmark for such comparisons. For a persistence diagram with a great number of points concentrated in a small region, then a grid-based implementation should be faster. On the other hand, for a persistence diagram in which points appears at a wide range of birth and death parameters, an exact implementation will be superior. This is because representing such a landscape with a reasonable grid spacing (i.e. reasonable accuracy) will require a very large grid. We encourage the users of the Persistence Landscape Toolbox to experiment with both strategies and to pick the better one for the data at hand. To change a software mode from exact computations to a grid-based computations, please modify the self-explanatory file \emph{configure} in the main folder of the library.

\begin{figure}[h!tb]
\centering
\includegraphics[scale=0.32]{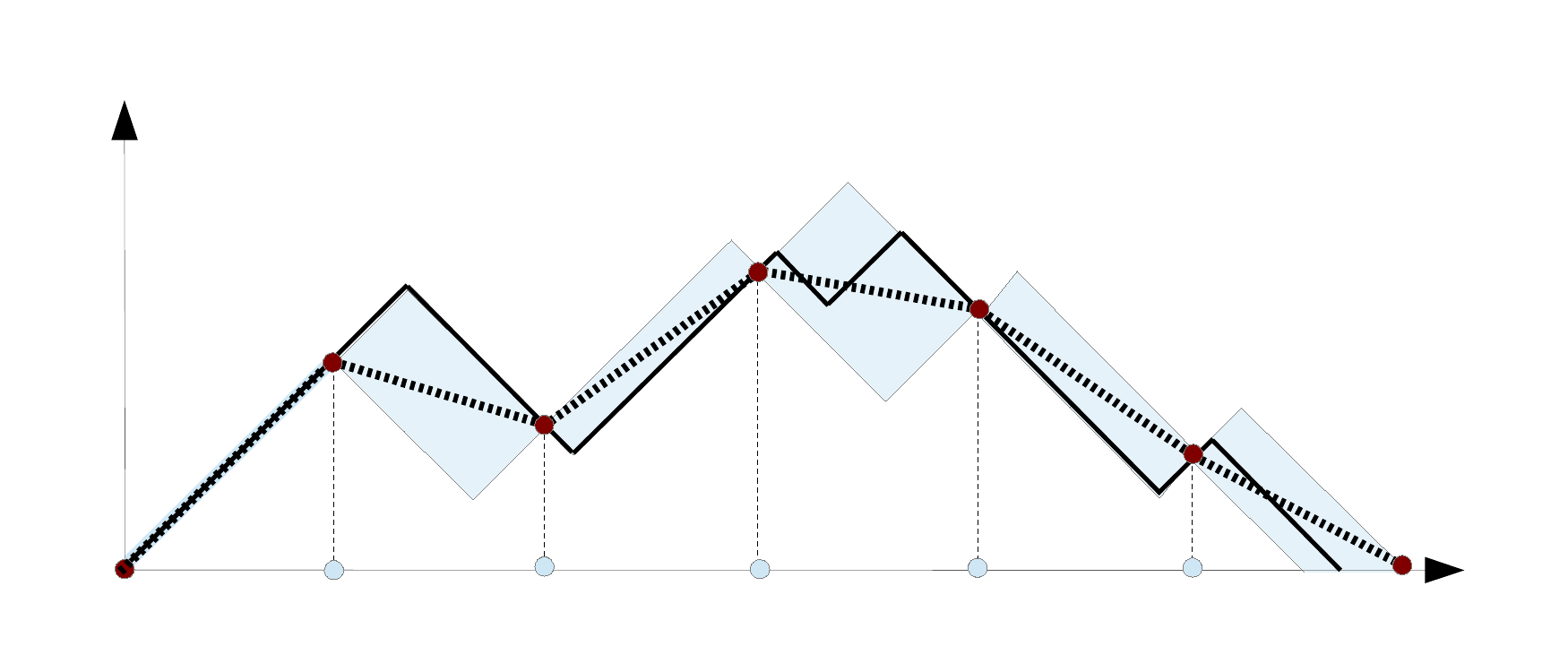}
\caption{Exact versus grid-based estimate of the persistence landscape. The underlying persistence landscape is given by the black lines. The blue dots in the x-axis indicate the grid points. The corresponding red dots in the plot of the persistence landscape give the values at the grid points. 
The dashed line is the estimated persistence landscape.
The blue regions indicate the intersection of the cones in which the persistence landscape
is guaranteed to be located.
}
\label{fig:gridErrorEstimate}
\end{figure}

We now discuss error bounds for grid-based estimates of persistence landscapes.
The persistence landscape and average persistence landscape are piecewise-linear functions with slopes between $-1$ and $1$. Therefore for each point $(x,y)$ in the plot of a landscape, the landscape will lie between the lines through $(x,y)$ having slope $\pm 1$. Now consider two points in the  landscape $(x_0,y_0)$ and $(x_1,y_1)$ such that $x_0$ and $x_1$ are consecutive points in the grid. Then the persistence landscape between those two points will lie in the intersection of the cones for $(x_0,y_0)$ and $(x_1,y_1)$. These intersecting cones are illustrated in Figure~\ref{fig:gridErrorEstimate}. 
We assume that the grid-based estimate of the persistence landscape is a piecewise linear function through the points obtained by evaluating the persistence landscape on the grid.

From the grid and the corresponding values of the persistence landscape one can bound the error. If the grid has spacing $\delta$, then the $L^{\infty}$ error is bounded by $\frac{\delta}{2}$. The $L^1$ error is bounded by half the sum of the areas of the rectangles described above. If the grid has size $m$ and there are $K$ nonzero persistence landscape functions, then this is bounded by $K(m-1)\frac{\delta^2}{4}$.

Even though the slope of an average landscape may be any real number between $-1$ and $1$, in our experience, the slope is often far from the extreme possible values. Therefore the error bounds above may be large overestimates.

\section{Configuration of the library.}
The Persistence Landscape Toolbox library is a collection of C++ programs. Each file with the $.cpp$ extension contains a ready-to use program. For a description of its functionality and required parameters, please run the program without any parameters. There is a \emph{configure} file which needs to be present in the folder in which a program is run. This file contains basic configuration parameters for the library. In this file one can set up the constant which represents infinity. 
The library assumes that the input file consists of a collection of numbers as described in the Section~\ref{sec:inputOutputFiles}. To encode infinite intervals, the infinities have to be changed to a ''magic number'' which is  defined in the \emph{configure} file. 
There one can find various options for what can be done with the infinite intervals. This file also allows one to switch the library from the default exact mode to a grid-based mode. If a grid-based mode is used, the user should set up the parameters of the grid in the configuration file.

\section{Conclusion}
\label{sec:conclusion}

To conclude, we have provided asymptotically optimal algorithms for computing persistence landscapes, averaging them and calculating distances between (average) persistence landscapes. We have implemented these algorithms and demonstrated how they may be used for hypothesis testing and classification. We hope that they will of use to the (topological) data scientist. 

Furthermore, this is not intended to be the end of the story. In future work, we aim to improve these algorithms and related software and increase their usefulness.

\appendix

\section{Persistence landscapes for barcodes with infinite intervals}

We can extend the definition of the persistence landscape in Section~\ref{sec:intro} to birth-death pairs $(b,d)$ with $-\infty \leq b < d \leq \infty$ using the following definitions.
\begin{gather*}
 f_{(-\infty,d)} =
  \begin{cases}
   0 & \text{if } x \not \in (-\infty,d) \\
   -x + d       & \text{if } x \in (-\infty , d) 
  \end{cases}\\
 f_{(b,\infty)} =
  \begin{cases}
   0 & \text{if } x \not \in (b,\infty) \\
   x - b & \text{if } x \in (b , \infty) 
  \end{cases}\\
  f_{(-\infty,\infty)}(x) = \infty
\end{gather*}

Algorithm~\ref{alg:landscapePointsInfinite} is a generalization of Algorithm~\ref{alg:landscapePoints} that constructs the persistence landscape from a barcode that may contain infinite intervals.

\begin{algorithm}[t]
\SetAlgoNoLine
\small
\KwIn{$A = \{(b_i,d_i)\}_{i=1}^n$ -- a list of birth-death pairs, $-\infty \leq b_i < d_i \leq \infty$.}
\KwOut{$\{\mathbb{L}_k\}$ -- the persistence landscape, a list of lists of critical points $(x,y)$.}
Sort $A$ first according to increasing $b$ and second according to decreasing $d$\;
$k \leftarrow 1$\;
\While {$A \neq \emptyset$}
{
  Initialize $\bbL_k$\;
  Pop first $(b,d)$ from $A$; Let $p$ point to the next term\;
  \eIf { $(b,d)=(-\infty,\infty)$ }
  {
    Add $(-\infty,\infty), (\infty,\infty)$ to $\bbL_k$\;
  }
  {
    \eIf { $d=\infty$ }
    {
	Add $(-\infty,0), (b,0), (\infty,\infty)$ to $\bbL_k$\;
    }
    {
    \eIf { $b=- \infty$ }
    {
	Add $(-\infty,\infty)$ to $\bbL_k$\;
    }
    { 
        Add $(-\infty,0),(b,0),(\frac{b+d}{2},\frac{d-b}{2})$ to $\bbL_k$\;
    }
    }
  }
  \While {$\bbL_k.\last \neq (0,\infty) \text{ or } (\infty,\infty)$} 
  {
    \eIf { $d$ maximal among remaining terms in $A$ starting at $p$}
    {
	Add $(d,0),(\infty,0)$ to $\bbL_k$\;
    }
    {
        Let $(b',d')$ be the first of the terms starting at $p$ with $d'>d$\;
	Pop $(b',d')$ from $A$; Let $p$ point to the next term\;
	\If { $b'>d$ }
	{
          Add $(d,0)$ to $\bbL_k$\;
	}
	\eIf { $b'\geq d$ }
	{
          Add $(b',0)$ to $\bbL_k$\;
	}
	{
          Add $(\frac{b'+d}{2},\frac{d-b'}{2})$ to $\bbL_k$\;
          Push $(b',d)$ into $A$ in order, starting at $p$; Let $p$ point to the next term\;
        }
	\eIf { $d'=\infty$ }
	{
          Add $(\infty,\infty)$ to $\bbL_k$\;
	}
        {
	  Add $(\frac{b'+d'}{2},\frac{d'-b'}{2})$ to $\bbL_k$\;
	  $(b,d) \leftarrow (b',d')$\;
        }
    }
  }
  ++k\;
}
Return $\{\bbL_k\}$\;
\caption{Compute the persistence landscape.}
\label{alg:landscapePointsInfinite}
\end{algorithm}

\section*{Acknowledgments}
The authors would like to thank Tamal K. Day, Brittany T. Fasy, Michael Kerber, Miro Kramar, Donald Sheehy and the anonymous referees for their valuable suggestions.

\bibliographystyle{plain}
\bibliography{bibfile}

\begin{thebibliography}{10}

\bibitem{adcock:2013}
A.~Adcock, E.~Carlsson, and G.~Carlsson.
\newblock The ring of algebraic functions on persistence bar codes.
\newblock {\em arXiv:1304.0530}, 04 2013.

\bibitem{Agarwal99verticaldecomposition}
Pankaj~K. Agarwal, Alon Efrat, and Micha Sharir.
\newblock Vertical decomposition of shallow levels in 3-dimensional
  arrangements and its applications.
\newblock {\em SIAM J. Comput}, 29:39--50, 1999.

\bibitem{phat}
U.~Bauer, M.~Kerber, and J.~Reininghaus.
\newblock {\em PHAT (Persistent Homology Algorithm Toolbox), accessed
  11/15/2013.}, 2014.

\bibitem{bendich:tracking}
P.~Bendich, S.~Chin, J.~Clarke, J.~deSena, J.~Harer, E.~Munch, A.~Newman,
  D.~Porter, D.~Rouse, N.~Strawn, and A.~Watkins.
\newblock Topological and statistical behavior classifiers for tracking
  applications.
\newblock arXiv:1406.0214 [cs.SY], 2014.

\bibitem{bendich:brainArtery}
P.~Bendich, J.S. Marron, E.~Miller, A.~Pieloch, and S.~Skwerer.
\newblock Persistent homology analysis of brain artery trees.
\newblock arXiv:1411.6652 [stat.AP], 2014.

\bibitem{peter}
P.~Bubenik.
\newblock Statistical topological data analysis using persistence landscapes.
\newblock {\em Journal of Machine Learning Research}, 16:77--102, 2015.

\bibitem{bubenikScott:1}
P.~Bubenik and J.~A. Scott.
\newblock Categorification of persistent homology.
\newblock {\em Discrete Comput. Geom.}, 51(3):600--627, 2014.

\bibitem{carlsson:topologyAndData}
G.~Carlsson.
\newblock Topology and data.
\newblock {\em Bull. Amer. Math. Soc. (N.S.)}, 46(2):255--308, 2009.

\bibitem{cidsz:mumford}
G.~Carlsson, T.~Ishkhanov, V.~de~Silva, and A.~Zomorodian.
\newblock On the local behavior of spaces of natural images.
\newblock {\em Int. J. Comput. Vision}, 76:1--12, 2008.

\bibitem{carriereOudotOvsjanikov}
M.~Carrière, S.~Y. Oudot, and M.~Ovsjanikov.
\newblock Stable topological signatures for points on 3d shapes.
\newblock {\em Eurographics Symposium on Geometry Processing 2015, Volume 34
  (2015), Number 5}, 2015.

\bibitem{Chazal:2014c}
F.~Chazal, B.~T. Fasy, F.~Lecci, B.~Michel, A.~Rinaldo, and L.~Wasserman.
\newblock Subsampling methods for persistent homology.
\newblock {\em arXiv:1406.1901}, 06 2014.

\bibitem{Chazal:2014a}
F.~Chazal, B.T. Fasy, F.~Lecci, A.~Rinaldo, A.~Singh, and L.~Wasserman.
\newblock On the bootstrap for persistence diagrams and landscapes.
\newblock {\em Modeling and Analysis of Information Systems}, 20(6):96--105,
  2014.

\bibitem{csgo:persistenceModules}
Frederic Chazal, Vin de~Silva, Marc Glisse, and Steve Oudot.
\newblock The structure and stability of persistence modules.
\newblock arXiv:1207.3674 [math.AT], 2012.

\bibitem{chepushtanova:2015}
S.~Chepushtanova, T.~Emerson, E.~Hanson, M.~Kirby, F.~Motta, R.~Neville,
  C.~Peterson, P.~Shipman, and L.~Ziegelmeier.
\newblock Persistence images: An alternative persistent homology
  representation.
\newblock {\em arXiv:1507.06217}, 07 2015.

\bibitem{cbk:ipmi2009}
M.~K. Chung, P.~Bubenik, and P.~T. Kim.
\newblock Persistence diagrams in cortical surface data.
\newblock In {\em Information Processing in Medical Imaging (IPMI) 2009},
  volume 5636 of {\em Lecture Notes in Computer Science}, pages 386--397, 2009.

\bibitem{stability}
D.~Cohen-Steiner, H.~Edelsbrunner, and J.~Harer.
\newblock Stability of persistence diagrams.
\newblock {\em Discrete \& Computational Geometry}, 37:103--120, 2007.

\bibitem{extendedPersistence}
D.~Cohen-Steiner, H.~Edelsbrunner, and J.~Harer.
\newblock Extending persistence using poincare and lefschetz duality.
\newblock {\em Found. Comput. Math.}, 9:79--103, 2009.

\bibitem{csehm:lipschitz}
D.~Cohen-Steiner, H.~Edelsbrunner, J.~Harer, and Y.~Mileyko.
\newblock Lipschitz functions have $l_p$-stable persistence.
\newblock {\em Found. Comput. Math.}, 10:127--139, 2010.

\bibitem{svm}
C.~Cortes and V.~Vapnik.
\newblock Support-vector networks.
\newblock {\em Machine Learning}, 20, 1995.

\bibitem{deSilvaGhrist:coverageInSNvPH}
V.~de~Silva and R.~Ghrist.
\newblock Coverage in sensor networks via persistent homology.
\newblock {\em Algebr. Geom. Topol.}, 7:339--358, 2007.

\bibitem{diFabioFerri}
B.~Di~Fabio and M.~Ferri.
\newblock Comparing persistence diagrams through complex vectors.
\newblock {\em http://arxiv.org/abs/1505.01335}, 2015.

\bibitem{donatini:1998}
P.~Donatini, P.~Frosini, and A.~Lovato.
\newblock Size functions for signature recognition.
\newblock In {\em Proceedings of the SPIE's Workshop ``Vision Geometry VII''},
  volume 3454 of {\em SPIE}, pages 178--183, 1998.

\bibitem{herbert}
H.~Edelsbrunner and J.~Harer.
\newblock {\em Computational Topology}.
\newblock American Mathematical Society, 2010.

\bibitem{elz:tPaS}
H.~Edelsbrunner, D.~Letscher, and A.~Zomorodian.
\newblock Topological persistence and simplification.
\newblock {\em Discrete \& Computational Geometry}, (28):511--533, 2002.

\bibitem{Efrat01geometryhelps}
A.~Efrat, A.~Itai, and M.~J. Katz.
\newblock Geometry helps in bottleneck matching and related problems.
\newblock {\em Algorithmica}, 31:2001, 2001.

\bibitem{fasy:tda}
B.~T. Fasy, J.~Kim, F.~Lecci, and C.~Maria.
\newblock Introduction to the r package tda.
\newblock arXiv:1411.1830 [cs.MS], 2014.

\bibitem{Fasy:2014}
B.~T. Fasy, F.~Lecci, A.~Rinaldo, L.~Wasserman, S.~Balakrishnan, and A.~Singh.
\newblock Confidence sets for persistence diagrams.
\newblock {\em Ann. Statist.}, 42(6):2301--2339, 2014.

\bibitem{ferri:1998}
M.~Ferri, P.~Frosini, A.~Lovato, and C.~Zambelli.
\newblock Point selection: a new comparison scheme for size functions (with an
  application to monogram recognition).
\newblock In T.~Pong R.~Chin, editor, {\em Proceedings Third Asian Conference
  on Computer Vision}, volume 1351 of {\em Lecture Notes in Computer Science},
  pages 329--337, Berlin Heidelberg, 1998. Springer-Verlag.

\bibitem{gambleHeo}
J.~Gamble and G.~Heo.
\newblock Exploring uses of persistent homology for statistical analysis of
  landmark-based shape data.
\newblock {\em J. Multivariate Anal.}, 101(9):2184--2199, 2010.

\bibitem{ghrist:survey}
R.~Ghrist.
\newblock Barcodes: the persistent topology of data.
\newblock {\em Bull. Amer. Math. Soc. (N.S.)}, 45(1):61--75, 2008.

\bibitem{Hershberger2}
J.~Hershberger.
\newblock Finding the upper envelope of n line segments in o(n log n) time.
\newblock {\em Information Processing Letters}, 33:169–174, 1989.

\bibitem{Hershberger}
J.~Hershberger.
\newblock Upper envelope onion peeling.
\newblock {\em Lecture Notes in Computer Science}, 447:368--379, 1990.

\bibitem{KahleMeckes}
M.~KAHLE and E~Meckes.
\newblock Limit theorems for betti numbers of random simplicial complexes.
\newblock {\em Unknown Journal}, 2015.

\bibitem{gudhi}
C.~Maria.
\newblock {\em GUDHI, Simplicial Complexes and Persistent Homology Packages.},
  2014.

\bibitem{perseus}
K.~Mischaikow and V.~Nanda.
\newblock Morse theory for filtrations and efficient computation of persistent
  homology.
\newblock {\em Discrete \& Computational Geometry}, 50:330--353, 2013.

\bibitem{dionysous}
D.~Morozov.
\newblock {\em The Dionysus software project, accessed 11/15/2013.}, 2014.

\bibitem{munch:probabilistic-f}
E.~Munch, P.~Bendich, K.~Turner, S.~Mukherjee, J.~Mattingly, and J.~Harer.
\newblock Probabilistic fr{\'e}chet means and statistics on vineyards.
\newblock {\em arXiv:1307.6530}, 2014.

\bibitem{perseus1}
V.~Nanda.
\newblock {\em The Perseus software project, accessed 11/15/2013.}, 2014.

\bibitem{nlc:topologyBreastCancer}
M.~Nicolau, Ar.~J. Levine, and G.~Carlsson.
\newblock Topology based data analysis identifies a subgroup of breast cancers
  with a unique mutational profile and excellent survival.
\newblock {\em Proc. Nat. Acad. Sci.}, 108(17):7265--7270, 2011.

\bibitem{giseon:maltose}
V.~K. Nikolic, G.~Heo, D.~Nikoli\'c, and P.~Bubenik.
\newblock Using cycles in high dimensional data to analyze protein binding.
\newblock {\em arXiv:1412.1394}, 2014.
\newblock arXiv:1412.1394 [stat.ME].

\bibitem{pereaHarer:swipers}
J.~Perea and J.~Harer.
\newblock Sliding windows and persistence: An application of topological
  methods to signal analysis.
\newblock arXiv:1307.6188 [math.AT], 2013.

\bibitem{reininghaus:2014}
J.~Reininghaus, S.~Huber, U.~Bauer, and R.~Kwitt.
\newblock A stable multi-scale kernel for topological machine learning.
\newblock arXiv:1412.6821 [stat.ML], 2014.

\bibitem{robins:2015}
V.~Robins and K.~Turner.
\newblock Principal component analysis of persistent homology rank functions
  with case studies of spatial point patterns, sphere packing and colloids.
\newblock {\em arXiv:1507.01454}, 07 2015.

\bibitem{robinsonTurner:2013}
A.~Robinson and K.~Turner.
\newblock Hypothesis testing for topological data analysis.
\newblock {\em arXiv:1310.7467}, 2013.
\newblock arXiv:1310.7467 [stat.AP].

\bibitem{plex}
M~Sexton and M.~Vejdemo-Johansson.
\newblock {\em PLEX libdary, accessed 11/15/2013}, 2014.

\bibitem{tmmh:frechet-means}
K.~Turner, Y.~Mileyko, S.~Mukherjee, and J.~Harer.
\newblock Fr{\'e}chet means for distributions of persistence diagrams.
\newblock {\em arXiv:1206.2790}, 2014.

\bibitem{wasserman:book-statistics}
L.~Wasserman.
\newblock {\em All of statistics}.
\newblock Springer Texts in Statistics. Springer-Verlag, New York, 2004.
\newblock A concise course in statistical inference.

\bibitem{gnuplot}
T.~Williams and C.~Kelley.
\newblock {\em Gnuplot 4.5: an interactive plotting program}.
\newblock 2011.

\bibitem{zomorodianCarlsson:computingPH}
A.~Zomorodian and G.~Carlsson.
\newblock Computing persistent homology.
\newblock {\em Discrete \& Computational Geometry}, (33):249--274, 2005.

\end{thebibliography}

\end{document}